\documentclass[a4paper,11pt]{article}
\pdfoutput=1

\usepackage[a4paper,left=2.73cm,right=2.7cm,top=3cm,bottom=3.5cm]{geometry}

\usepackage{amsmath,amssymb,graphicx,wrapfig,caption,subcaption,enumerate,color,tikz}
\usetikzlibrary{arrows}
\usepackage{colortbl}
\usepackage{booktabs}
\usepackage{cite}
\usepackage{wrapfig}
\usepackage[mathscr]{euscript}

\usepackage[colorlinks=true,linktocpage=true,linkcolor=blue,citecolor=blue]{hyperref}

\definecolor{gray}{rgb}{.9,.9,.9}
\tikzstyle{block} = [rectangle, text width=5em, text  centered, rounded corners]
\tikzstyle{line} = [draw, very thick, -latex']
\tikzstyle{RGflow}= [->, shorten <=1pt, thick, dashed, color=black!70]

\addtocontents{toc}{\protect\setcounter{tocdepth}{2}}
\numberwithin{equation}{section}

\newcommand{\Sec}[1]{Sec.~\ref{#1}}

\newcommand{\bea}{\begin{eqnarray}}
\newcommand{\beal}[1]{\begin{eqnarray}\label{#1}}
\newcommand{\eea}{\end{eqnarray}}
\newcommand{\be}{\begin{equation}}
\newcommand{\bel}[1]{\begin{equation}\label{#1}}
\newcommand{\ee}{\end{equation}}

\newcommand{\bit}{\begin{itemize}}
\newcommand{\eit}{\end{itemize}}
\newcommand{\ben}{\begin{enumerate}}
\newcommand{\een}{\end{enumerate}}

\def\del{\partial}

\newcommand{\mt}[1]{\textrm{\tiny #1}}
\newcommand{\nc}{N_\mt{c}}

\newcommand{\sac}{\, , \qquad}

\newcommand{\eqq}[1]{(\ref{#1})}
\newcommand{\fig}[1]{Fig.~\ref{#1}}
\newcommand{\figs}[1]{Figs.~\ref{#1}}
\newcommand{\sect}[1]{Sec.~\ref{#1}}

\newcommand{\bal}{\begin{align}}
\newcommand{\eal}{\end{align}}
\newcommand{\bse}{\begin{subequations}}
\newcommand{\ese}{\end{subequations}}

\newcommand{\Ehigh}{\mathcal{E}_{\text{high}}}
\newcommand{\Elow}{\mathcal{E}_{\text{low}}}

\def\E{{\mathcal{E}}}
\def\Ezero{{\mathcal{E}\left(t=0\right)}}

\def\Pt{{{P}}_T}
\def\Pl{{{P}}_L}

\begin{document}

\begin{titlepage}

\thispagestyle{empty}

\begin{flushright}
\hfill{ICCUB-19-004}
\end{flushright}

\vspace{40pt}  

\begin{center}
	 
\title{Dynamics of Phase Separation from Holography}
{\LARGE \textbf{\mbox{Dynamics of Phase Separation from Holography}}}
	\vspace{30pt}
		
{\large \bf  Maximilian Attems,$^{1,\,2}$ Yago Bea,$^{2}$ Jorge Casalderrey-Solana,$^{2}$ \\ David Mateos,$^{2,\,3}$ and Miguel Zilh\~ao$^{4,\,2}$}

\vspace{25pt}

{\normalsize $^{1}$Instituto Galego de F\'\i sica de Altas Enerx\'\i as (IGFAE), Universidade de Santiago de Compostela, 15782 Galicia, Spain.}\\
\vspace{15pt}

{$^{2}$ Departament de F\'\i sica Qu\`antica i Astrof\'\i sica and Institut de Ci\`encies del Cosmos (ICC),\\  Universitat de Barcelona, Mart\'\i\  i Franqu\`es 1, ES-08028, Barcelona, Spain.}\\

\vspace{15pt}
{ $^{3}$Instituci\'o Catalana de Recerca i Estudis Avan\c cats (ICREA), \\ Passeig Llu\'\i s Companys 23, ES-08010, Barcelona, Spain.}\\
\vspace{15pt}

{ $^{4}$CENTRA, Departamento de F\'\i sica, Instituto Superior T\'ecnico, Universidade de Lisboa, Avenida Rovisco Pais 1, 1049 Lisboa, Portugal.}



\vspace{40pt}
				
\abstract{
We use holography to develop a physical picture of the real-time evolution of the spinodal instability of a four-dimensional, strongly-coupled gauge theory with a first-order, thermal phase transition. 
We numerically solve Einstein's equations to follow the evolution, in which we identify four generic stages: A first, linear stage in which the instability grows exponentially; a second, non-linear stage in which peaks and/or phase domains are formed; a third stage in which these structures  merge; and a fourth stage in which the system finally relaxes to a static, phase-separated configuration. On the gravity side the latter is described by a static, stable, inhomogeneous horizon. We conjecture and provide evidence that all static, non-phase separated configurations in large enough boxes are dynamically unstable. We show that all four stages are well described by the constitutive relations of second-order hydrodynamics that include all second-order gradients that are purely spatial in the local rest frame. In contrast, a M\"uller-Israel-Stewart-type formulation of hydrodynamics fails to provide a good description for two reasons. First, it misses some large, purely-spatial gradient corrections. Second, several second-order transport coefficients in this formulation, including the relaxation times $\tau_\pi$ and $\tau_\Pi$,  diverge at the points where the speed of sound vanishes. 
}


\end{center}
\end{titlepage}

\tableofcontents  
\hrulefill 
\vspace{10pt}

\section{Introduction}
\label{intro}
Holography has provided numerous insights into the out-of-equilibrium properties of hot, strongly-coupled, non-Abelian plasmas. Examples include \cite{Chesler:2008hg,Chesler:2009cy,Heller:2011ju,Chesler:2010bi,Heller:2012km,Heller:2013oxa,Casalderrey-Solana:2013aba,Casalderrey-Solana:2013sxa,Chesler:2015wra,Chesler:2015bba,Chesler:2015lsa,Buchel:2015saa,Chesler:2016ceu,Attems:2016tby,Casalderrey-Solana:2016xfq,Attems:2016ugt,Attems:2017zam,Rougemont:2017tlu,Gursoy:2016ggq,Critelli:2018osu,Buchel:2018ttd,Czajka:2018bod,Czajka:2018egm,Attems:2018gou} (see e.g.~\cite{CasalderreySolana:2011us} for a review). In most of these cases the final state of the plasma at asymptotically late times is a homogeneous state. The purpose of this paper is to analyse a case in which the final configuration is expected to be an inhomogeneous state exhibiting phase separation. Specifically, we will study a four-dimensional gauge theory  
with a first-order, thermal phase transition. We will place the theory in an initial homogeneous state with an energy density in the unstable, spinodal region (see \fig{fig:energydensity}). If this state is perturbed, the system will evolve to a final state that will necessarily be inhomogeneous. Following this real-time evolution with conventional quantum-field theoretical methods for an interacting gauge theory  is challenging. Therefore we will use holography, in which case following the evolution can be done by solving the time-dependent Einstein's equations on the gravity side. A first study of this system was presented in  \cite{Attems:2017ezz}. Subsequently, \cite{Janik:2017ykj} analysed a case in which the gauge theory is three-dimensional. Here we will extend the analysis of \cite{Attems:2017ezz} in several directions and we will develop a detailed physical picture of the entire evolution of the system. 

The real-time dynamics of phase separation in a strongly-coupled, non-Abelian gauge theory might be relevant to understand the physics of future heavy ion collision (HIC) experiments such as  the beam energy scan at RHIC, the compressed baryonic matter experiment at FAIR and other experiments at NICA. These experiments will open an unprecedented window into the properties of the phase diagram of Quantum Chromodynamics (QCD) at large baryon chemical potential, which is expected to contain a line of first order phase transitions ending at a critical point \cite{Stephanov:1998dy,Stephanov:1999zu,Stephanov:2017ghc}. If this scenario is realised then the real-time dynamics  of the spinodal instability may play an important role, which provides one motivation for our work. 

Hydrodynamics has been extremely successful at describing the quark-gluon plasma formed in HICs so far. If the future HIC experiments  do explore  a region with phase transitions, the applicability of the   formulation of hydrodynamics used in most numerical codes might need to be reconsidered. In \cite{Attems:2017ezz,Attems:2018gou} we initiated a study in this direction. We found that near a critical point, due to the slowing down of the dynamics, the system is accurately described by the constitutive relations of a formulation of hydrodynamics that includes all second-order gradients that are purely spatial in the local rest frame. However, this formulation is an acausal theory for which the initial-value problem is not well posed. A cure that is vastly used in hydrodynamic codes consists of   exchanging the terms  with second-order purely spatial derivatives in the local rest frame for terms with one time and one spatial derivative (see \cite{Romatschke:2017ejr} for a review). We found in \cite{Attems:2017ezz,Attems:2018gou} that the resulting theory, which we will call a M\"uller-Israel-Stewart-type (MIS) formulation, failed to provide a good description in the region near the phase transition. In this paper we find similar conclusions and we elaborate on the reasons for the failure of the MIS-type formulation. In particular, we show that some second-order transport coefficients, including the relaxation times 
$\tau_\pi$ and $\tau_\Pi$, diverge at the points where the speed of sound vanishes.


\section{Model and instability}

Our gravity model is described by the 
Einstein-scalar action
\begin{equation}
\label{eq:action}
S=\frac{2}{\kappa_5^2} \int d^5 x \sqrt{-g} \left[ \frac{1}{4} {\cal R}  - \frac{1}{2} \left( \nabla \phi \right) ^2 - V(\phi) \right ] ~,
\end{equation}
with potential 
\begin{equation}
\ell^2 V(\phi)=-3-\frac{3\phi^2}{2}-\frac{\phi^4}{3}-\frac{\phi^6}{3\phi_M^2}+\frac{\phi^6}{2\phi_M^4}-\frac{\phi^8}{12\phi_M^4}~,
\label{potential}
\end{equation}
where $\ell$ is the asymptotic curvature radius of the corresponding AdS geometry and $\phi_M$ is a parameter that we will set to $\phi_M=2.3$. This  potential can be derived from the superpotential
\be
\ell W(\phi)=-\frac{3}{2}-\frac{\phi^2}{2}-\frac{\phi^4}{4\phi_M^2}  \, ,
\ee
via the usual relation 
\be
V(\phi) = -\frac{4}{3} W(\phi)^2 + \frac{1}{2} W' (\phi)^2 \,.
\ee
The potential (\ref{potential}) is the same as in \cite{Attems:2017ezz,Attems:2018gou}. The dual gauge theory is  a Conformal Field Theory (CFT) deformed with a dimension-three scalar operator with source $\Lambda$. On the gravity side this scale  appears as a boundary condition for the scalar $\phi$. 

\begin{figure}[t]
\begin{center}
\begin{tabular}{cc}
\includegraphics[width=.46\textwidth]{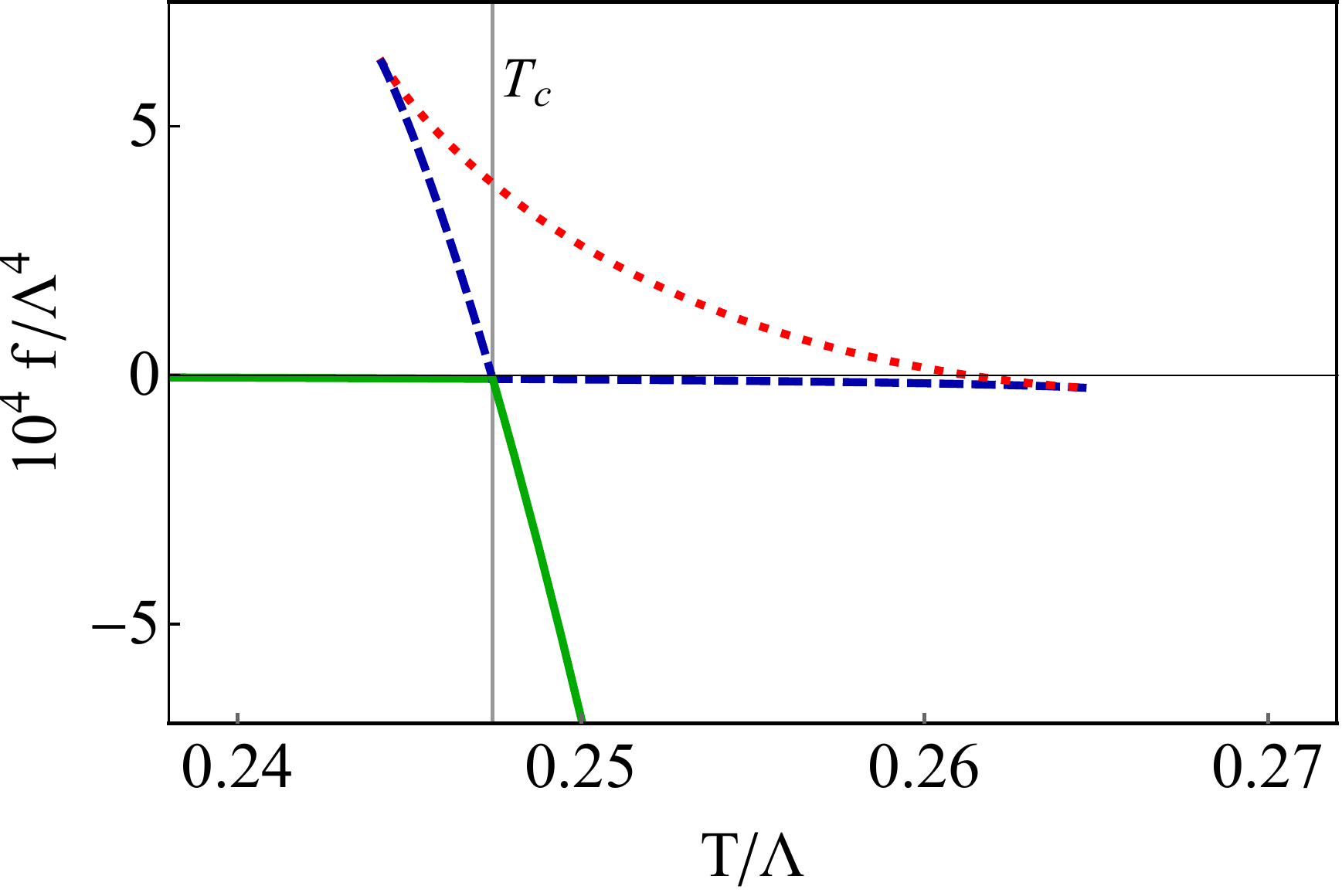}
 \quad&\quad
\includegraphics[width=.46\textwidth]{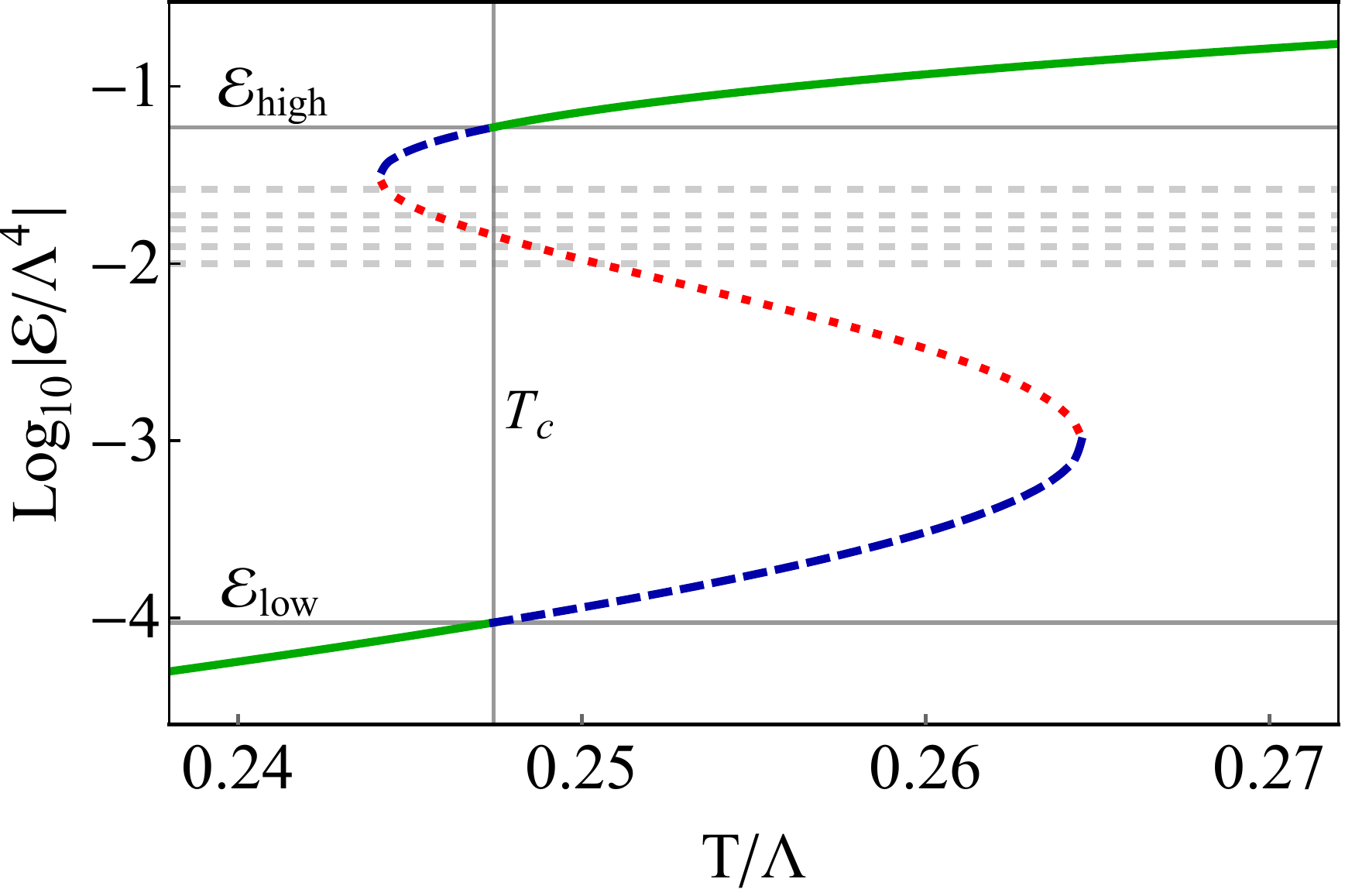}
\end{tabular}
	\caption{\label{fig:energydensity} 	 \small Free energy density (left) and energy density (right)  versus temperature for the gauge theory dual to \eqq{eq:action}.  The thermodynamically preferred phase is shown in solid green. In dashed blue we show the locally stable but globally unstable region. The locally unstable region, the so-called spinodal region, is shown in dotted red. The grey vertical lines indicate $T_c= 0.247 \Lambda$. The solid horizontal grey lines correspond (from top to bottom) to the energy densities \eqq{energiesphase transition}. The dashed horizontal grey  lines correspond (from top to bottom) to the energy densities $\{ \mathcal{E}_1, \mathcal{E}_2, \mathcal{E}_3, \mathcal{E}_4, \mathcal{E}_5\}$ given in \eqq{energies}.}
\end{center}
\end{figure}

Our motivation to choose this model is simplicity. The presence of the scalar breaks conformal invariance. The first two terms in the superpotential are fixed by the asymptotic AdS radius and by the dimension of the dual scalar operator. The quartic term in the superpotential is the simplest addition that results in a thermal first-order phase transition in the gauge theory (for $\phi_M \leq 2.521$), which we extract  from the homogeneous black brane solutions on the gravity side. In particular, the gauge theory possesses  a first-order phase transition at a critical temperature $T_c= 0.247 \Lambda$, as illustrated by the multivalued plot of the free energy density as a function of the temperature  in \fig{fig:energydensity}(left). \fig{fig:energydensity}(right) shows the corresponding energy density, where the high- and low-energy phases at $T_c$ have energy densities
\be
\label{energiesphase transition}
\mathcal{E}_{\text{high}} \simeq 5.9 \times \frac{\Lambda^4}{10^2} \,, \quad  \mathcal{E}_{\text{low}} \simeq 9.4 \times \frac{\Lambda^4}{10^5} \,.
\ee
Notice that we work with the rescaled quantities 
\be
({\cal E}, \Pl, \Pt)=\frac{\kappa_5^2}{2\ell^3} (-T^t_t, T^z_z, T^{x_\perp}_{x_\perp}) \,,
\ee
where $\Pl$ and $\Pt$ are the longitudinal and transverse pressures with respect to the $z$-spatial direction along which the dynamics will take place. For an $SU(\nc)$ gauge theory the prefactor on the right-hand side typically scales as $\nc^2$. Note the large hierarchy between the energy densities:
\be
\label{hi}
\mathcal{E}_{\text{high}}/\mathcal{E}_{\text{low}} \simeq 0.6 \times 10^3 \,.
\ee
\begin{figure}[t]
	\begin{center}
		\begin{tabular}{cc}
			\includegraphics[width=.465\textwidth]{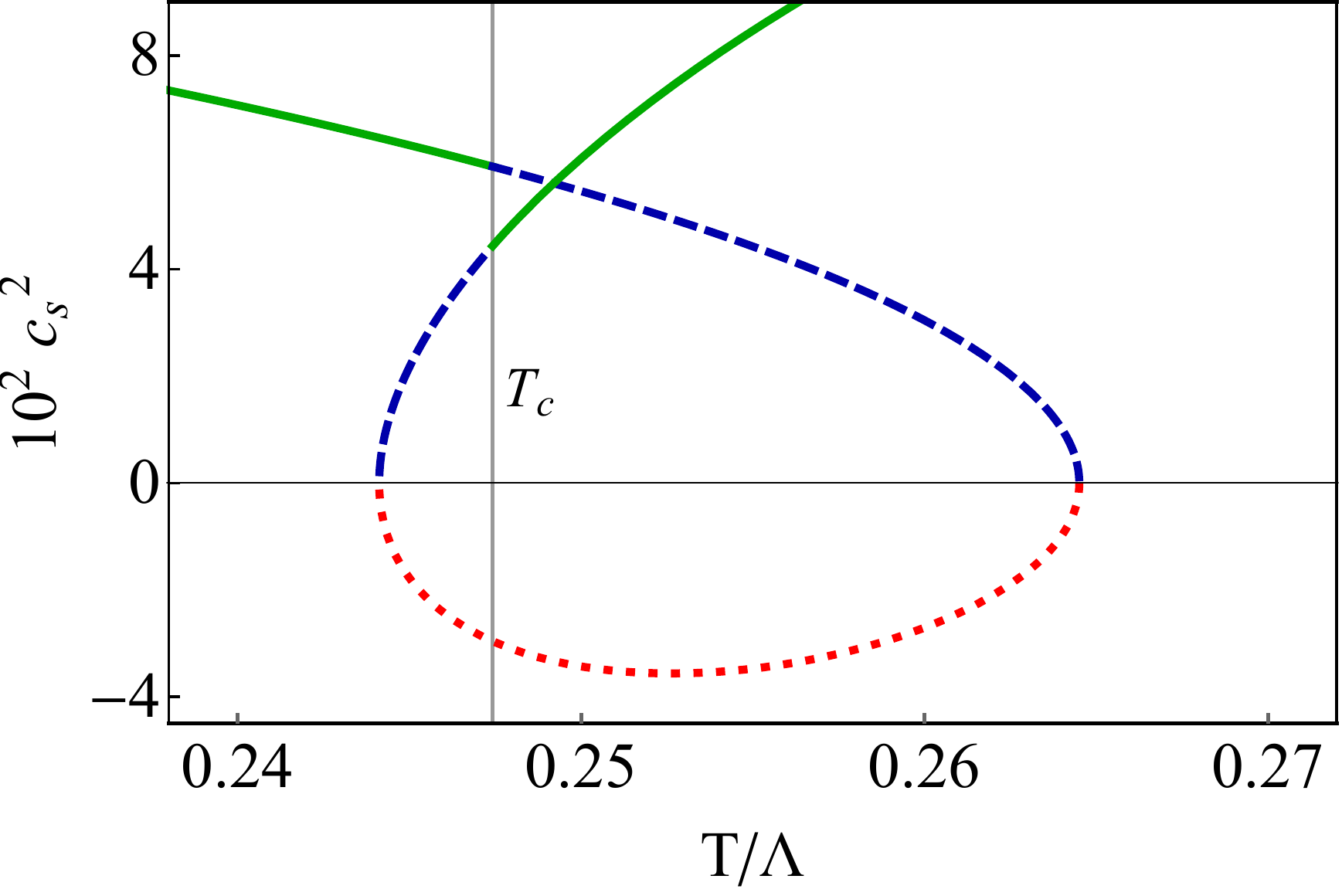}
			\quad&\quad
			\includegraphics[width=.465\textwidth]{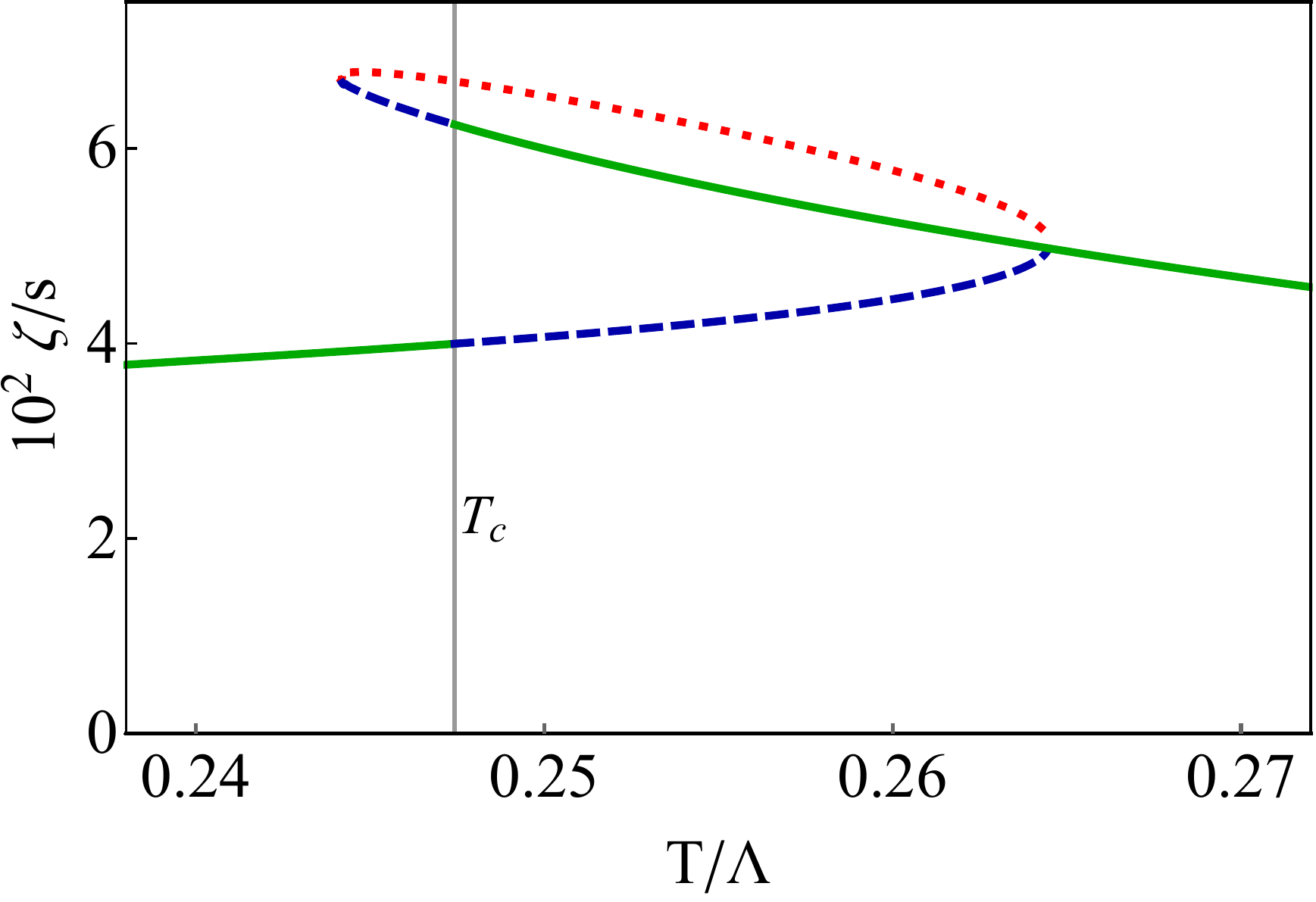}
		\end{tabular}
		\caption{\label{fig:cs2} 	 \small (Left) Speed of sound squared versus temperature for the gauge theory dual to \eqq{eq:action}. The color coding is as in Fig. \ref{fig:energydensity}. (Right) The ratio of the bulk viscosity over entropy $\zeta /s$ versus temperature. The ratio has a maximum value of $max(\zeta/s)\simeq 0.0678$ at $T/\Lambda \simeq 0.245$. 
		}
	\end{center}
\end{figure}

States on the dotted red curves of \fig{fig:energydensity} are locally thermodynamically unstable since the specific heat  is negative, $c_v =d \E / dT <0 $. As we now explain, these states are also dynamically unstable. This is in agreement with the Gubser-Mitra conjecture \cite{Gubser:2000ec,Gubser:2000mm}, which states that a black brane with a non-compact translational symmetry is classically stable if, and only if, it is locally thermodynamically stable. We will comment again on this conjecture at the end of \Sec{unstable}.

The connection with the dynamic instability was pointed out in an analogous context in \cite{Buchel:2005nt} and it arises as follows. The speed of sound is related to the specific heat $c_v$ and the entropy density $s$ through 
\be
c_s^2 = \frac{s}{c_v} \,.
\ee
Therefore $c_s^2$ is negative on the dashed red curves of \fig{fig:energydensity}, as shown in \fig{fig:cs2}(left), and consequently $c_s$ is purely imaginary. The amplitude of long-wave length, small-amplitude sound modes behaves as 
\be
\mathcal{A} \sim e^{-i \omega(k) t } \, , 
\label{amplitude0}
\ee
with a dispersion relation given by
 (see Appendix \ref{appendixA} for the derivation)
\be
\label{disp}
\omega_{\pm} (k) =\pm  \sqrt{ c_s^2 k^2 - k^4 f_L - k^4 \frac{1}{4} \Gamma^2 } - \frac{i}{2} \Gamma k^2 \,.
\ee 
The plus sign corresponds to an unstable mode, while the minus sign leads to a stable mode. In this expression 
\be
\label{ate}
\Gamma = \frac{1}{T} \left(\frac{4}{3} \frac{\eta}{s}  + \frac{\zeta}{s} \right)   
\ee
is the sound attenuation constant, $\eta$ and $\zeta$  are the shear and  bulk viscosities, and $f_L$ is a second-order transport coefficient related to the coefficients $\tau_{\pi}$ and $\tau_{\Pi}$ in \cite{Romatschke:2009kr} through (see Sec.~\ref{sectionhydro})
\be
f_L=-\frac{c_s^2}{\mathcal{E}+P} \left( \zeta \tau_{\Pi} +\frac{4}{3}\, \eta\,  \tau_{\pi} \right) \,. 
\ee	
In our model $\eta/s=1/4\pi$ \cite{Kovtun:2004de}, we compute $\zeta$ numerically following \cite{Eling:2011ms}, and we obtained $f_L$ in \cite{Attems:2017ezz,Attems:2018gou}. 

\begin{figure}[t]
\begin{center}
\includegraphics[width=.455\textwidth]{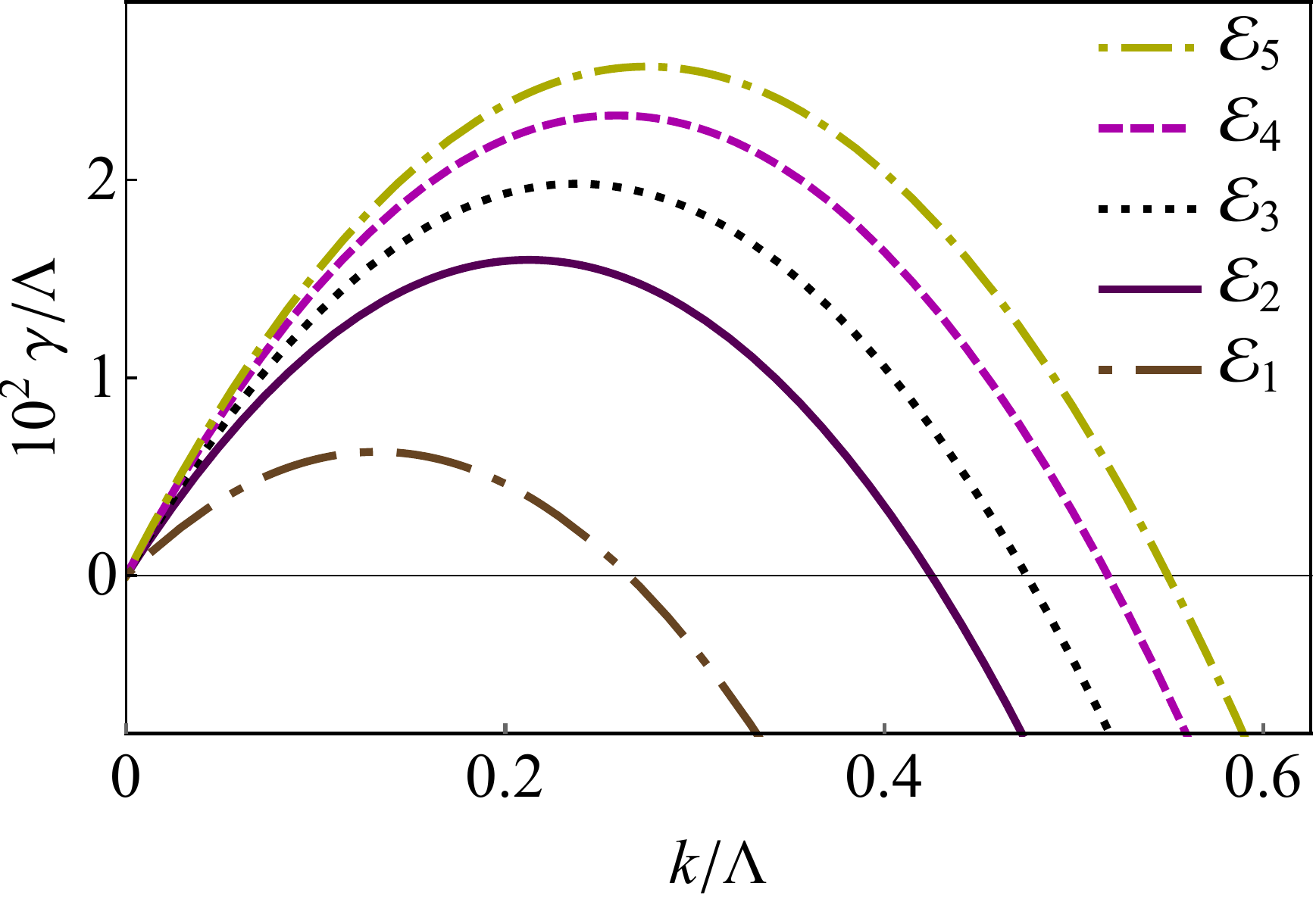}
\caption{\label{fig:parabola}  \small
		Growth rates $\gamma(k)$ given by the first two terms in (\ref{small}) for the energy densities 
		$\{ \mathcal{E}_1, \mathcal{E}_2, \mathcal{E}_3, \mathcal{E}_4, \mathcal{E}_5\}$ given in \eqq{energies}. 
}
\end{center}
\end{figure}

An imaginary value of $c_s$ leads to a purely real value of the growth rate
\be
\label{exponents}
\gamma(k) \equiv -i \omega (k) \,. 
\ee
For small momenta \eqq{disp} yields for the unstable mode
\be
\label{small}
\gamma (k) =  \left |c_s \right | k - \frac{1}{2} \Gamma k^2 + k^3 \frac{4 f_L + \Gamma^2 }{8 \left |c_s \right | } + O(k^5)\,.
\ee
The first two terms alone give the familiar parabolas corresponding to the curves in \fig{fig:parabola}. Note that these curves depend on the energy density $\mathcal{E}$ of the state under consideration because both $c_s$ and $\Gamma$ depend on $\mathcal{E}$.
We see that the growth rate is positive for momenta in the range 
$0 \leq k < k_*$ with 
\be
k_* \simeq \frac{2\left | c_s \right |}{\Gamma} \,.
\ee
The corrections to these parabolas coming from the inclusion of the $k^3$ term in \eqq{small} or from evaluation of the full square root in \eqq{disp} are small if we use the values of $f_L$ obtained in \cite{Attems:2017ezz,Attems:2018gou}. Nevertheless, because $4f_L$ is very close to $\Gamma^2$ in magnitude but has the  opposite sign, fitting these corrections provide an accurate method to extract this coefficient. In subsection \ref{exponentialgrowth} we will use this method to obtain a value of $f_L$ in good agreement with \cite{Attems:2017ezz,Attems:2018gou}.

In conclusion, states on the red dashed curves of \fig{fig:energydensity} are afflicted by a dynamical instability, known as spinodal instability, whereby long-wave length, small-amplitude perturbations in the sound channel grow exponentially in time. The corresponding statement on the gravity side is that the  black branes dual to the states on the dashed red curve are afflicted by a long-wave length  instability. Although this is similar \cite{Buchel:2005nt,Emparan:2009cs,Emparan:2009at} to the  Gregory-Laflamme  (GL) instability of black strings \cite{Gregory:1993vy}, there is an important difference: In the GL case all strings below a certain mass density are unstable, whereas in our case only states on the red dashed curves of \fig{fig:energydensity} are unstable.


\section{Time evolution}
\label{time}

\subsection{Initial state}
\label{init}
To investigate  the fate of the spinodal instability we compactify the 
$z$-direction on a circle of length $L$  in the range $L \Lambda \in (107, 213)$. For comparison, Ref.~\cite{Attems:2017ezz} considered  $L\Lambda\simeq 57$. This infrared cut-off reduces the number of unstable sound modes to a finite number, since modes along the  $z$-direction must have quantized momenta 
\be
k_n = \frac{2\pi n}{L} \,.
\ee
For simplicity, we impose homogeneity along the other two gauge theory directions $x_\perp$.
We then consider a set of homogeneous, unstable initial states with energy densities  $\{ \mathcal{E}_1, \mathcal{E}_2, \mathcal{E}_3, \mathcal{E}_4, \mathcal{E}_5 \}$  in the spinodal region given by 
\be
\label{energies}
\{ \mathcal{E}_1, \mathcal{E}_2, \mathcal{E}_3, \mathcal{E}_4, \mathcal{E}_5 \} \simeq
\{ 2.6, 1.9, 1.6, 1.2, 1.0 \} \times \frac{\Lambda^4}{10^2}  \,.
\ee
For comparison, we have also listed the energy density 
$\mathcal{E}_5$ of the state considered in \cite{Attems:2017ezz}. These  states are indicated by the dashed, horizontal  grey lines in \fig{fig:energydensity}(right). To trigger the instability, Ref.~\cite{Attems:2017ezz} introduced a small $z$-dependent perturbation in the energy density corresponding to a specific Fourier mode on the circle. However, this is not indispensable since numerical noise alone is enough to trigger the instability. Therefore in this paper we will consider both cases. Note that for the instability to play a role the size of the box must be large enough to fit at least one unstable mode. In other words we must have $L > 2\pi/k_*$.

\begin{figure*}[h!!!] 
\begin{center}
\begin{tabular}{cc}
\hspace{-6mm}
\includegraphics[width=.51\textwidth]{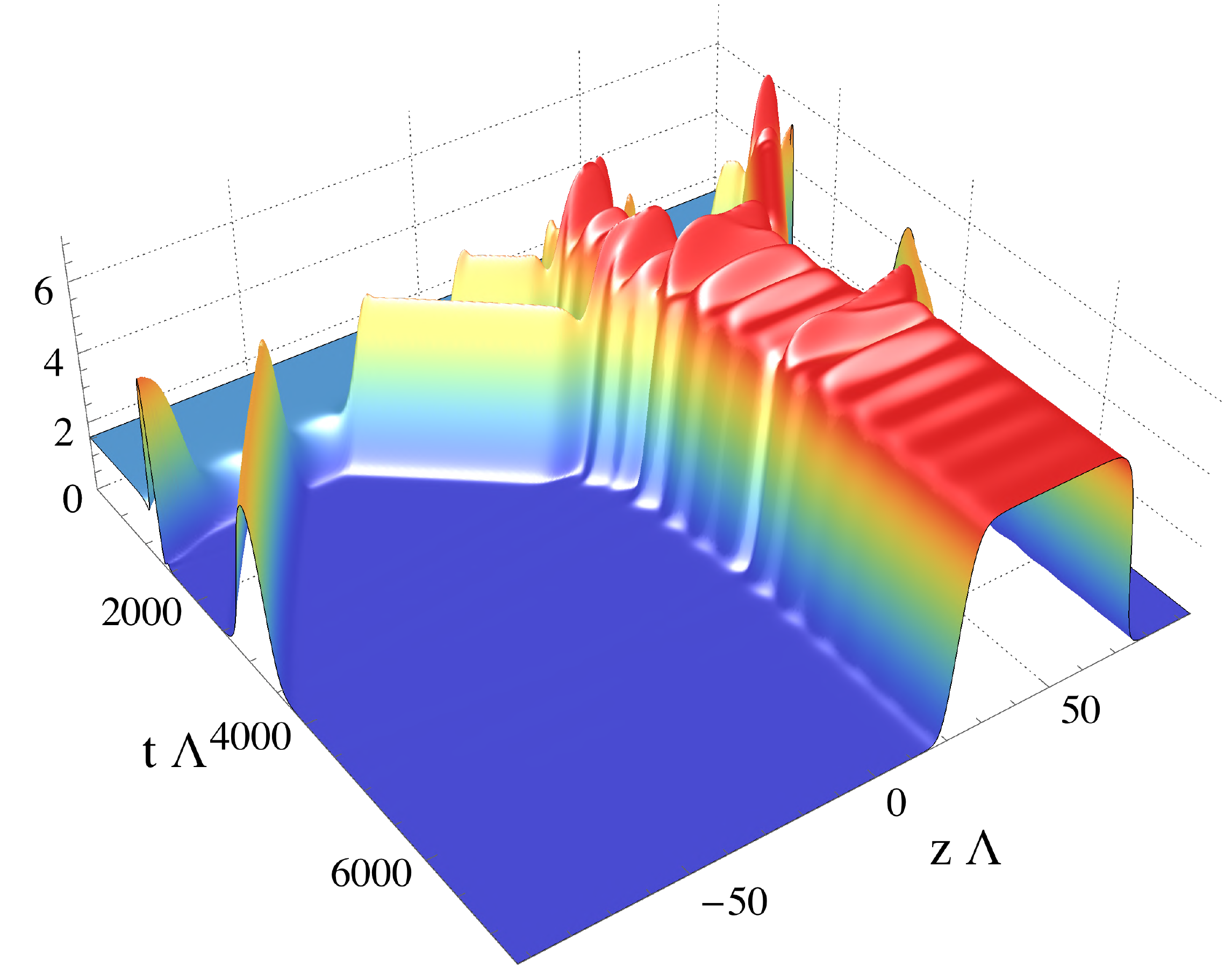} 
\put(-231,141){\mbox{\small{$10^2$ $\mathcal{E}/\Lambda^4$}}}
\quad \hspace{-2mm}
\includegraphics[width=.51\textwidth]{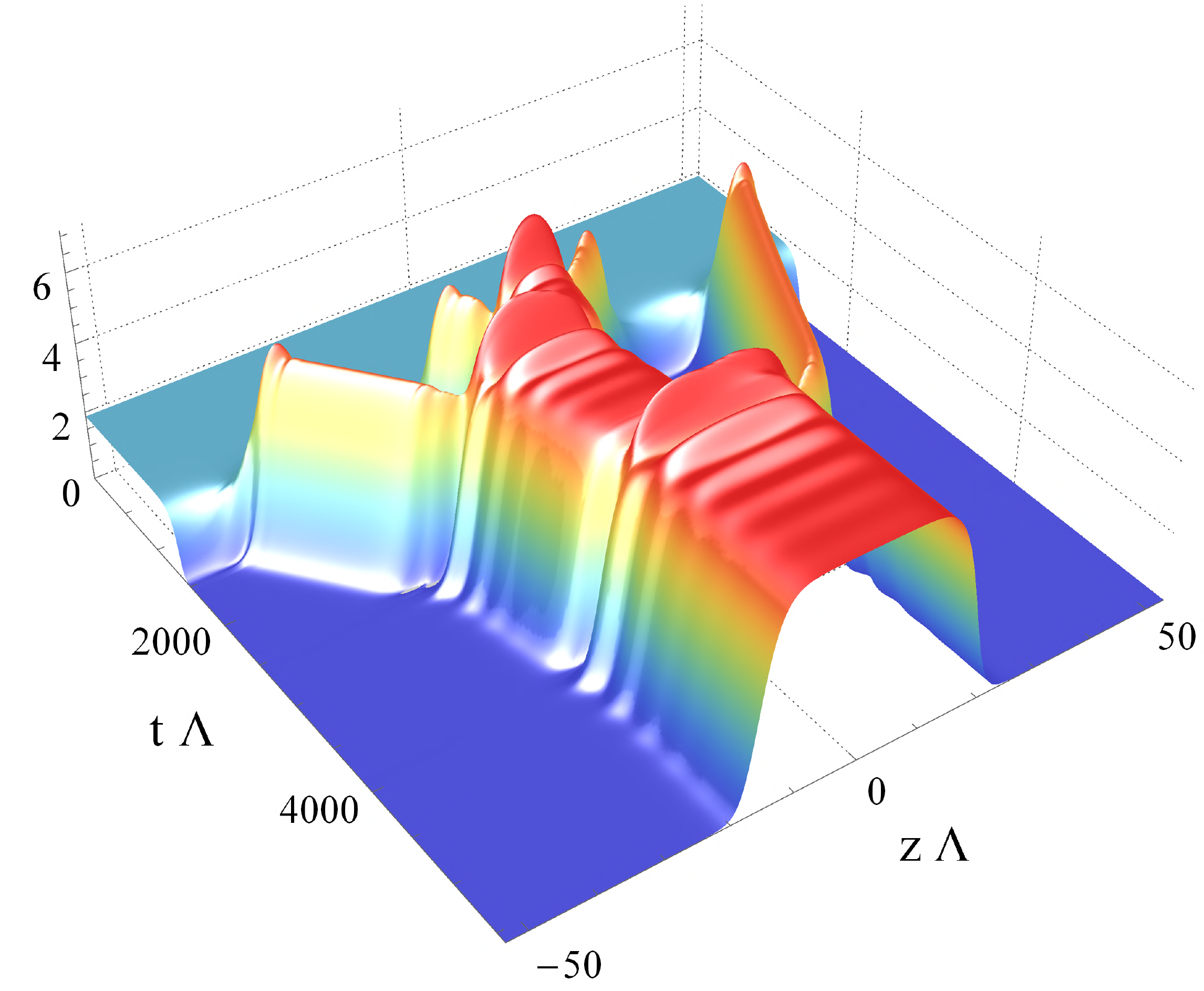} 
\put(-232,148){\mbox{\small{$10^2$ $\mathcal{E}/\Lambda^4$}}}
\end{tabular}
\end{center}
\vspace{-5mm}
\caption{
\label{3Denergy}
	\small  Spacetime evolution of the energy densities of initial homogeneous states with energies and size boxes $\Ezero=\E_3, L\Lambda \simeq 187$ (left) and $\Ezero=\E_2, L\Lambda \simeq 107$ (right). The instability is triggered by an initial $n=1$ perturbation (left) and by pure numerical noise (right). 
}
\vspace{7mm}
\end{figure*}

We follow the instability by numerically evolving the Einstein-plus-scalar equations as in \cite{Attems:2016tby,Attems:2017zam}. From the dynamical metric  we extract the boundary stress tensor. We have performed 15 runs in which we observe phase separation at late times. We have published the corresponding boundary data extracted from the evolutions as open data \cite{zenodo:2019a} with a script to visualize each of them. The results for the energy density for two representative runs are shown in \fig{3Denergy}.
\begin{figure*}[h!!]
	\begin{center}
	\begin{tabular}{c}
			\hspace{-4mm}
			\includegraphics[width=.9\textwidth]{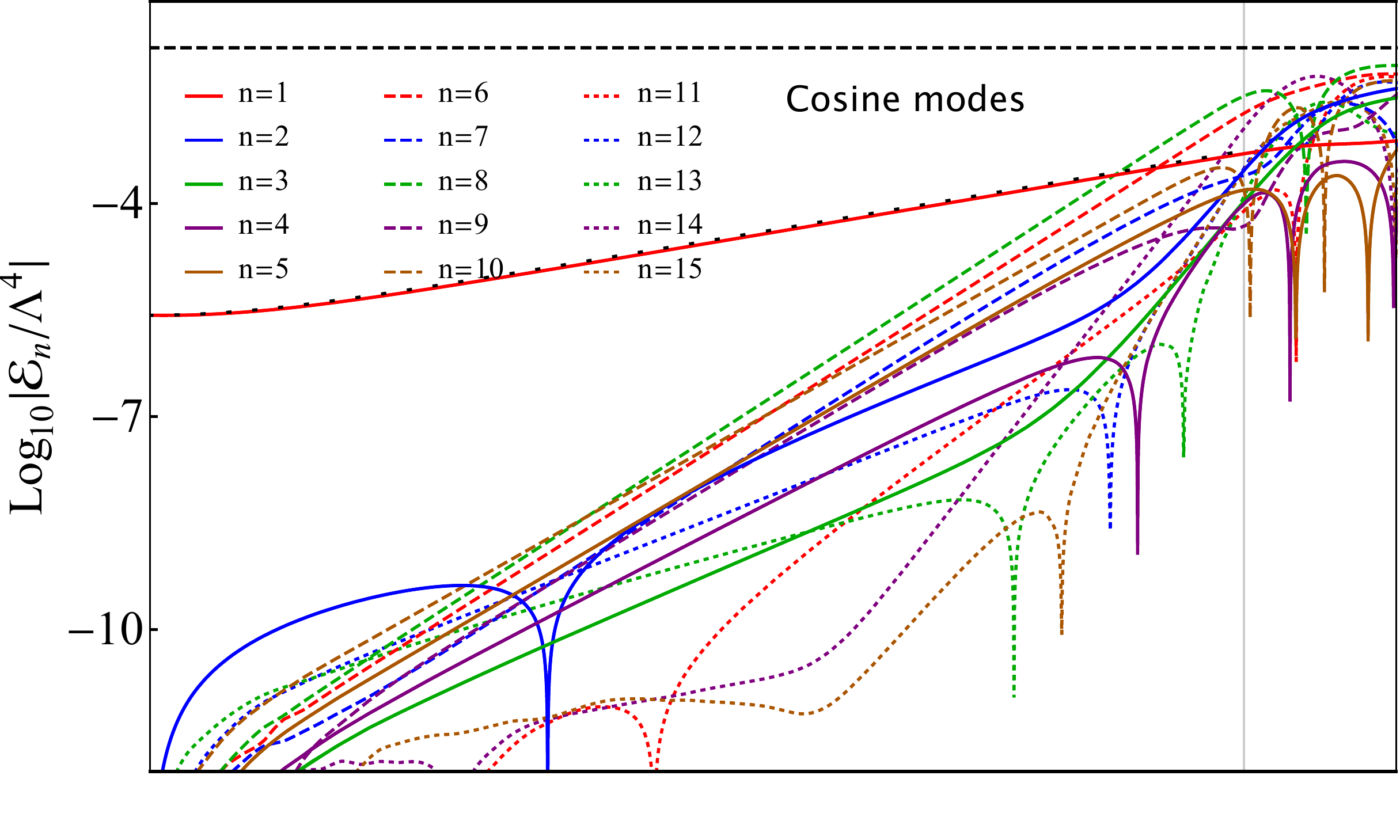} 
			\\[-0.041\textwidth]
			\includegraphics[width=.9\textwidth]{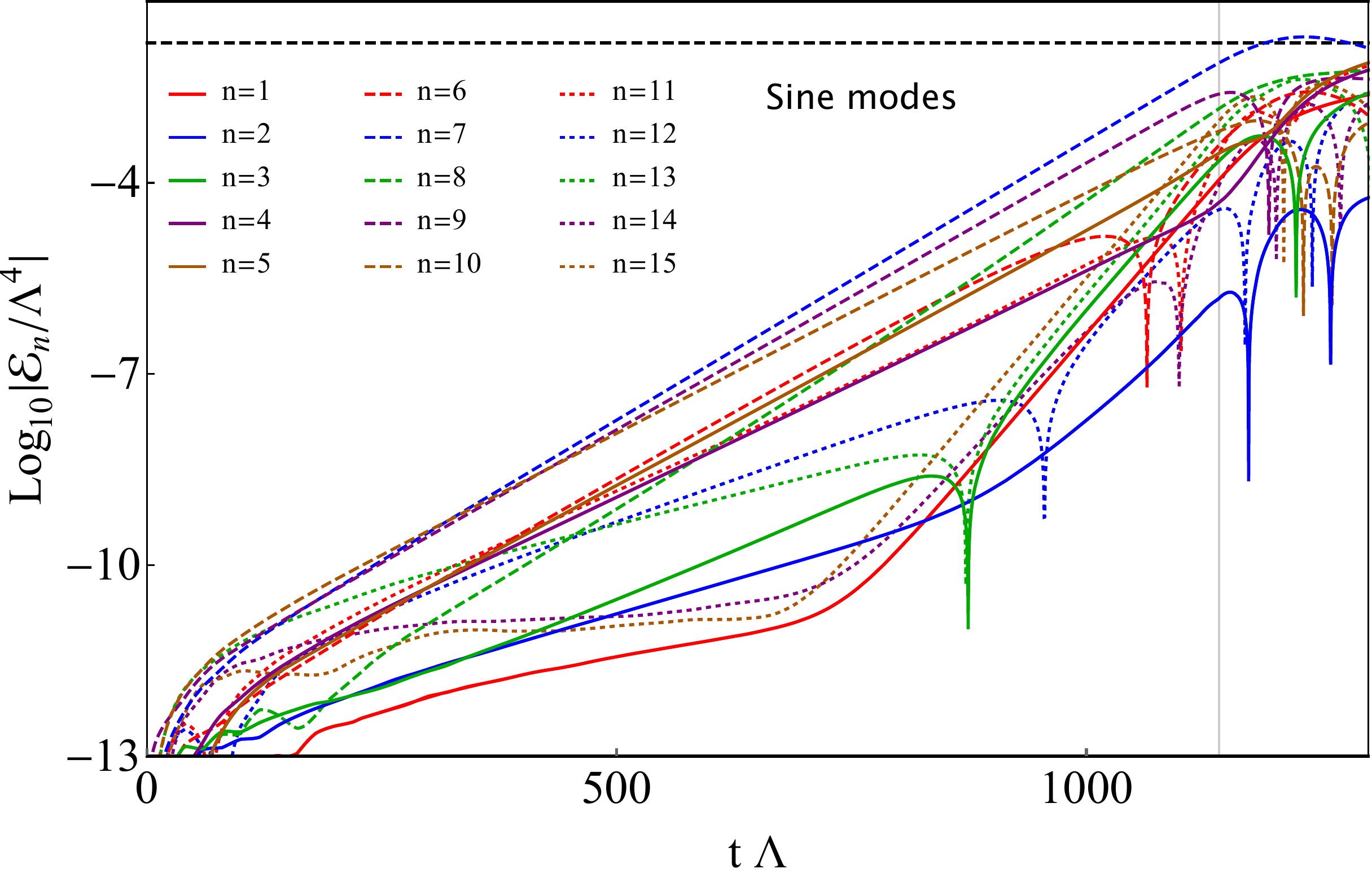} 
			\hspace{0.0093\textwidth}
	\end{tabular}
	\end{center}
	\vspace{-5mm}
	\caption{
		\label{modes} 
		\small 
		Time evolution of some cosine (top) and sine (bottom) Fourier modes of the energy density shown in \fig{3Denergy}(left). The dashed horizontal lines at the top indicate the average energy density or, equivalently, the $n=0$, constant mode. The grey vertical lines at $t\Lambda =1141$ indicate the end of the linear regime. A snapshot of the energy profile at this time is shown in \fig{snap}(left) in solid blue. The black dots on top of the $n=1$ line in (top) correspond to the analytical solution for the sound mode in the linear regime, namely to the sum of one growing and one decaying exponential with exponents given by \eqq{exponents}.
	}
	\vspace{7mm}
\end{figure*} 
\begin{figure*}[h!!!]
	\begin{center}
		\begin{tabular}{c}
			\hspace{-4mm}
			\includegraphics[width=.9\textwidth]{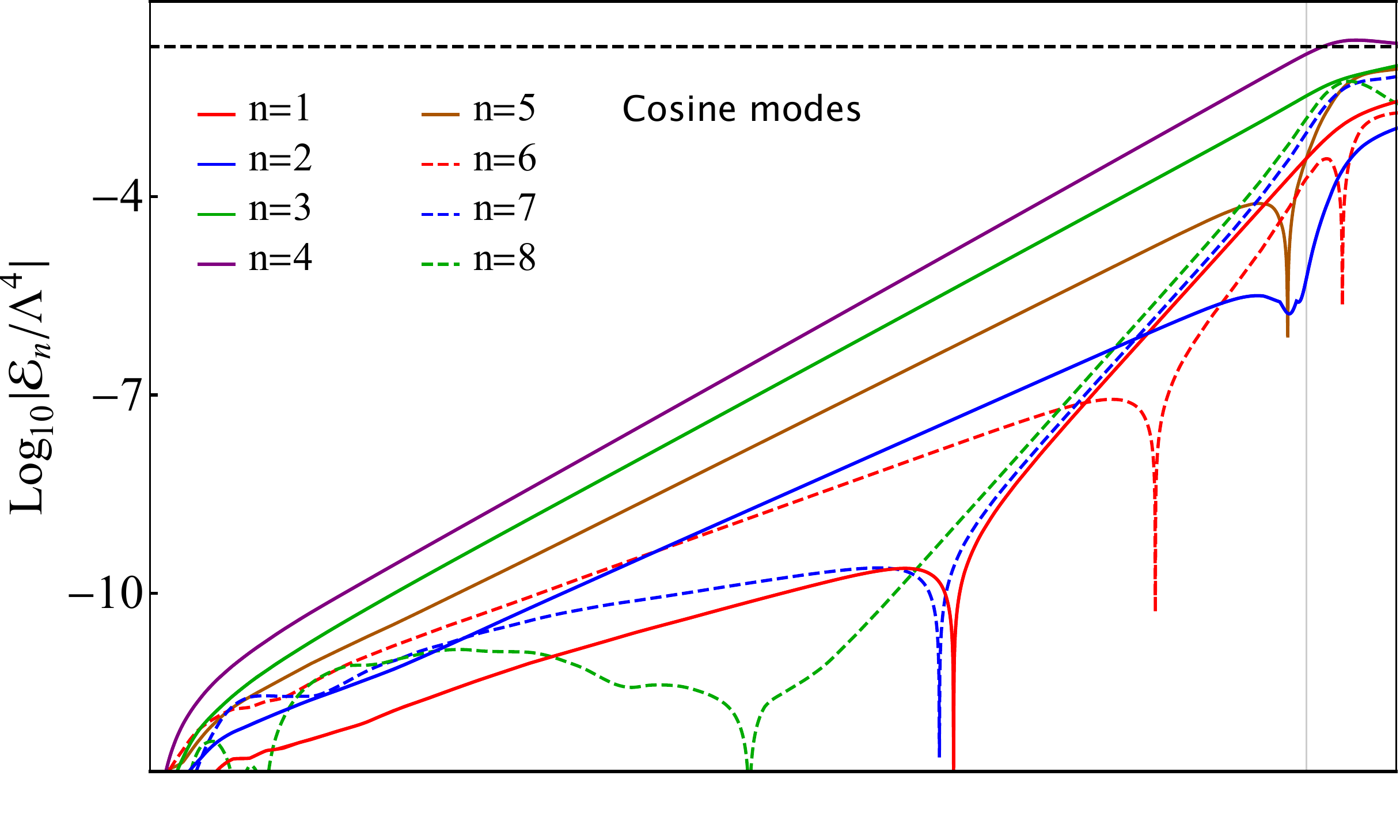} 				\\[-0.041\textwidth]
			\includegraphics[width=.9\textwidth]{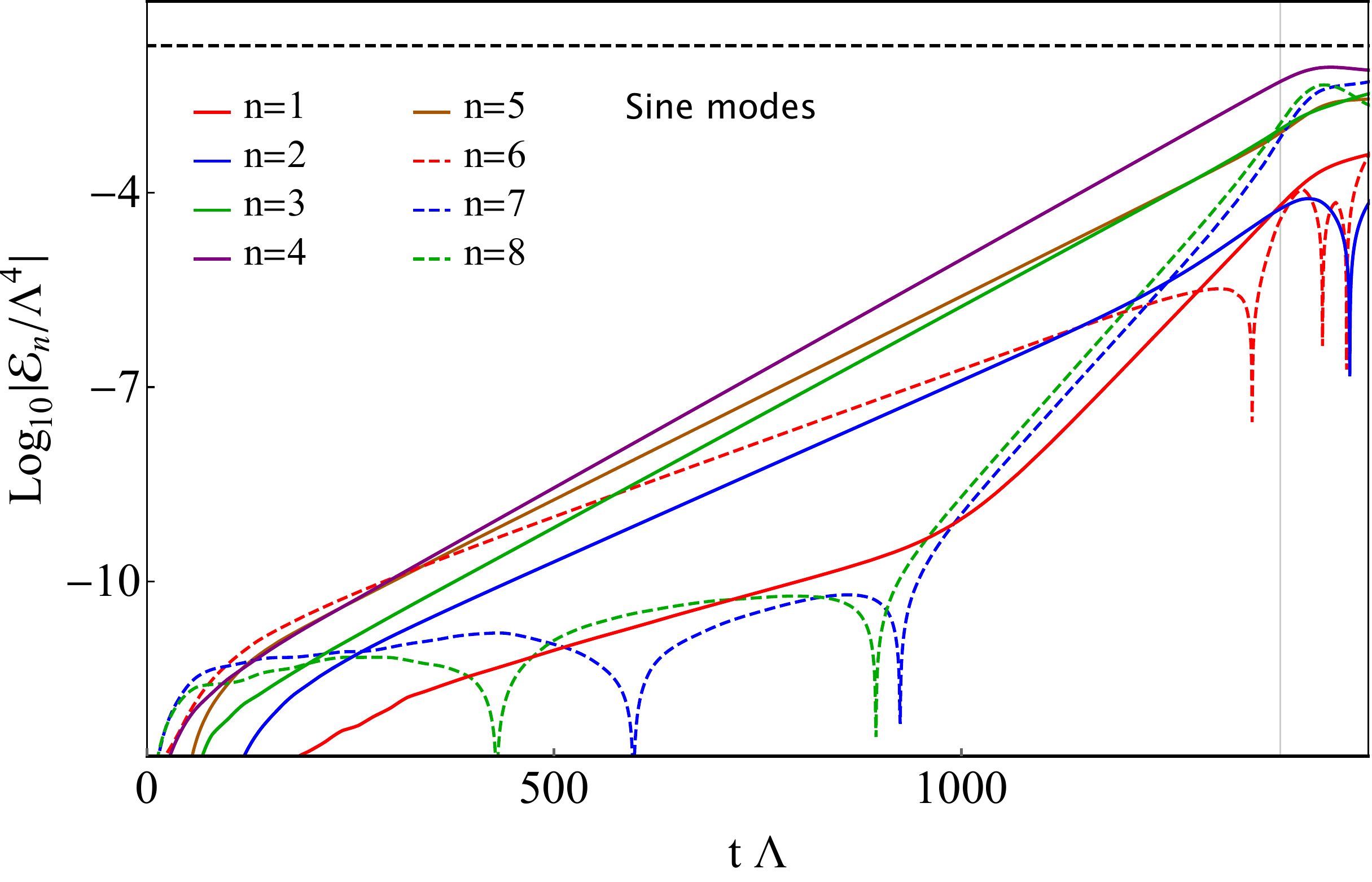} 
			\hspace{0.0093\textwidth}
				\end{tabular}
	\end{center}
	\vspace{-5mm}
	\caption{
		\label{modes2} 
		\small 
		Time evolution of some cosine (top) and sine (bottom) Fourier modes of the energy density shown in \fig{3Denergy}(right). The dashed horizontal lines at the top indicate the average energy density or, equivalently, the $n=0$, constant mode. The grey vertical lines at $t\Lambda =1392$ indicate the end of the linear regime. A snapshot of the energy profile at this time is shown in \fig{snap}(right) in solid blue.
	}
	\vspace{7mm}
\end{figure*}

\subsection{The role of the box}
\label{box}
In order to avoid confusion in the discussion below, it is important to realize from the beginning in which ways the physics of the system can  depend on the size of the box or, more precisely, on the dimensionless quantity $L\Lambda$. As we said above, the finite value of $L$ implements and IR cut-off on the allowed modes. Thus a first potential effect is that, if the box has size $L < 2\pi/k_*$, then homogeneous states that would  be dynamically unstable in bigger  boxes become dynamically stable because no unstable modes fit in the box. 

The dynamics of inhomogeneous states can also be crucially affected. Suppose for example that we have a configuration with a single domain (to be defined more precisely in \sect{reshaping}) like that at late times in \figs{3Denergy}(left) or (right). Here we must clarify an issue of terminology. Because of our periodic boundary conditions, the number of domains is only unambiguously defined in relation to the size of the box. In other words, given a static configuration with one domain in a box of size $L$ we can take $n$ copies of it and generate a new configuration with $n$ domains in a box of size $nL$. The crucial point is that, although these two configurations are equivalent as static configurations, their dynamics once slightly perturbed may be radically different. Physically, the reason is that the configurations in \figs{3Denergy}(left) or (right) are phase-separated and hence stable, whereas $n$ copies of them in a box of size $nL$ are unstable towards merging of the different domains into a single one, as we will see in \sect{unstable}. Technically, the reason is again that some modes that are unstable in the box of size $nL$ do not fit in a box of size $L$. In fact, even the final stable domain in the $nL$-sized  box will not fit in the smaller box if $n$ is large enough. This illustrates that, ultimately, the  different dynamics is due to the fact that, while any configuration in a box of size $L$ can be viewed as a configuration in a box of size $nL$, the reverse is not true. The space of possible configuration  and the dynamics in a bigger box are richer.

In summary, in each simulation we will specify and keep fixed the size of the box. We will also refer to the number of domains in the system, or speak of multi-peak configurations, or count the number of maxima of the energy profile, etc. with the understanding that these are meaningful, unambiguous concepts because we have a fixed, specific box size in mind.

\subsection{Linear regime}
\label{exponentialgrowth}
Since the initial perturbation is small, the first stage of the evolution is well described by a linear analysis around the initial homogeneous state. Linear theory predicts a behavior which is the sum of two exponentials, precisely the two solutions of the sound mode (\ref{disp}).
In the spinodal region, one of these modes decays with time while the other one grows. After some time the latter dominates. \figs{modes} and \ref{modes2} show the time evolution of the amplitudes of several Fourier modes corresponding to the runs in \fig{3Denergy}. The straight lines in the logarithmic plots of \figs{modes} and \ref{modes2} correspond to the regime of exponential growth. 

\begin{figure*}[h] 
	\begin{center}
		\begin{tabular}{cc}
			\includegraphics[width=.46\textwidth]{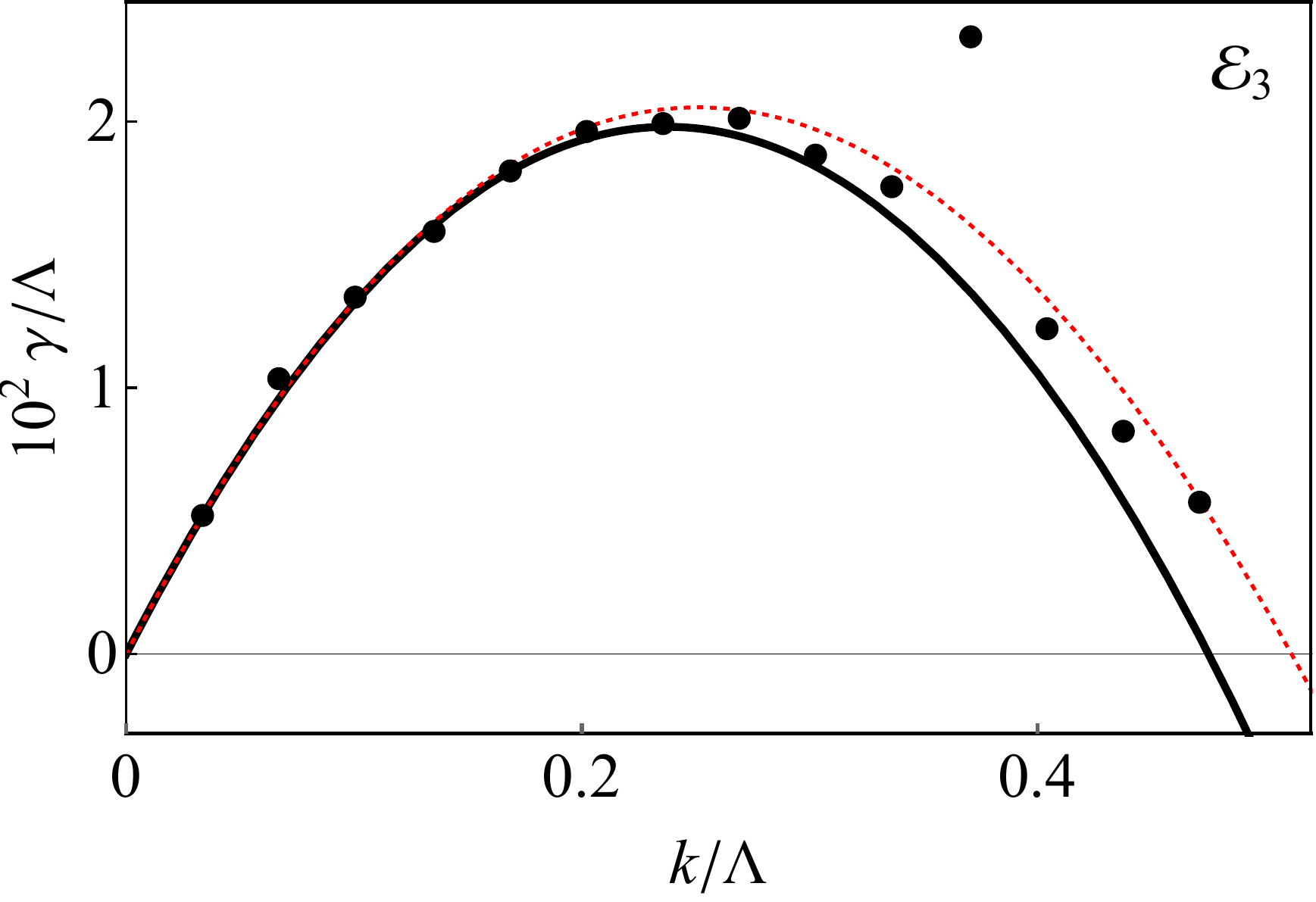} 
			\quad&\quad
			\includegraphics[width=.46\textwidth]{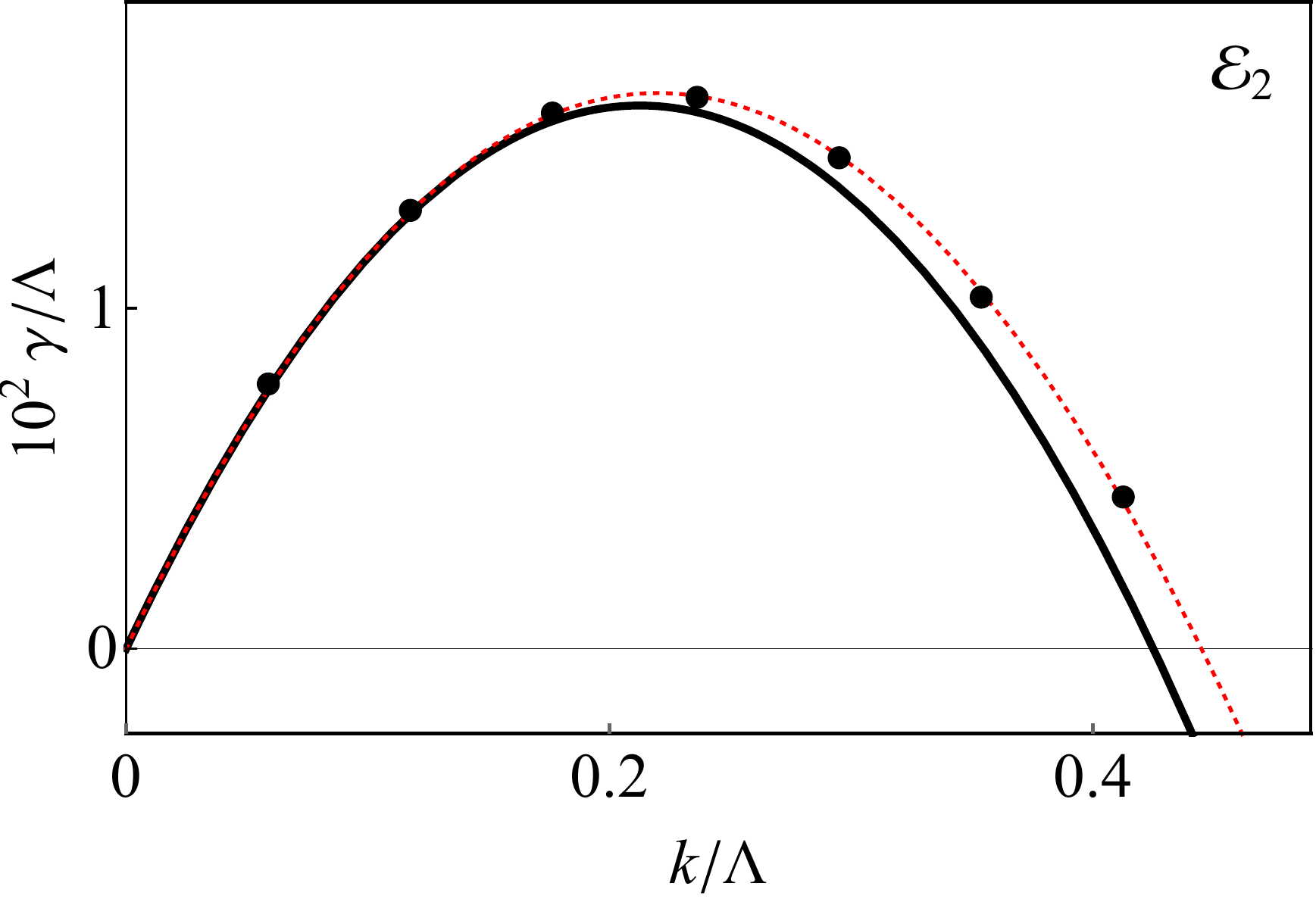} 
		\end{tabular}
	\end{center}
	\vspace{-5mm}
	\caption{
		\label{comparison}
		\small 
		The continuous black curves show the growth rates predicted by the linear approximation (\ref{small}) to order $k^2$ for the energies corresponding to the evolutions shown in \fig{3Denergy}. The black dots show the growth rates extracted from a fit to the slopes in \figs{modes} and \ref{modes2}. The dotted red curves corresponds to a fit to the expression (\ref{disp}), with the result \eqq{fitresult}.
	}
	\vspace{7mm}
\end{figure*}

In \fig{comparison} we compare the growth rates predicted by linear analysis up to order $k^2$ with those extracted from a fit to the slopes of the straight lines in \figs{modes} and \ref{modes2}. We obtain good agreement, except for some particular cases. These correspond to resonant behavior, i.e.~to the coupling between two modes which contributes to the growth of a third mode. For example, in \fig{comparison}(left) the 11th mode corresponds to a resonant behavior of the 1st and 10th modes. Other modes are also affected by resonant behavior, for example the  14th mode in \fig{modes}(top) changes its slope at $t \Lambda \sim 650$ from the growth rate given by the sound mode to a new growth rate given by a resonance. As these may change in time, we have obtained the dots of \fig{comparison} from early times, when the resonant behaviors are minimal. 

The continuous black curves in \fig{comparison} show the prediction of linear analysis to order $k^2$. As explained above, we can consider the full non-linear expression (\ref{disp}) to extract a value for $f_L$ by performing a fit. The dotted red curves in  \fig{comparison} show the results of these fits, from which we obtain the values 
\be
\label{fitresult}
f_L (\E_2) \simeq f_L (\E_3) \simeq   -0.136 / \Lambda^2 \,.
\ee
These agree well with the values obtained in \cite{Attems:2017ezz,Attems:2018gou}.

Eqs.~(\ref{amplitude0})-(\ref{disp}) determine the time evolution of each Fourier mode once two initial conditions are specified, for example its amplitude and its derivative. We have illustrated this with the dotted black curve shown in \fig{modes}(top), which we obtained by fitting the initial conditions at $t=0$ for the $n=1$ mode, and which falls on top of the 
exact $n=1$ red curve.  In principle, this could be done for every Fourier mode and obtain the full description of the system along the linear evolution. 
In practice, it is not possible to specify the precise initial conditions for the modes that are excited from the noise, but an estimate can be given by recalling that it is white noise. For example, in \fig{modes2} a reasonable estimate would be obtained by assuming that the initial amplitudes are equal for all modes.

\begin{figure*}[t]
\begin{center}
\begin{tabular}{cc}
\includegraphics[width=.46\textwidth]{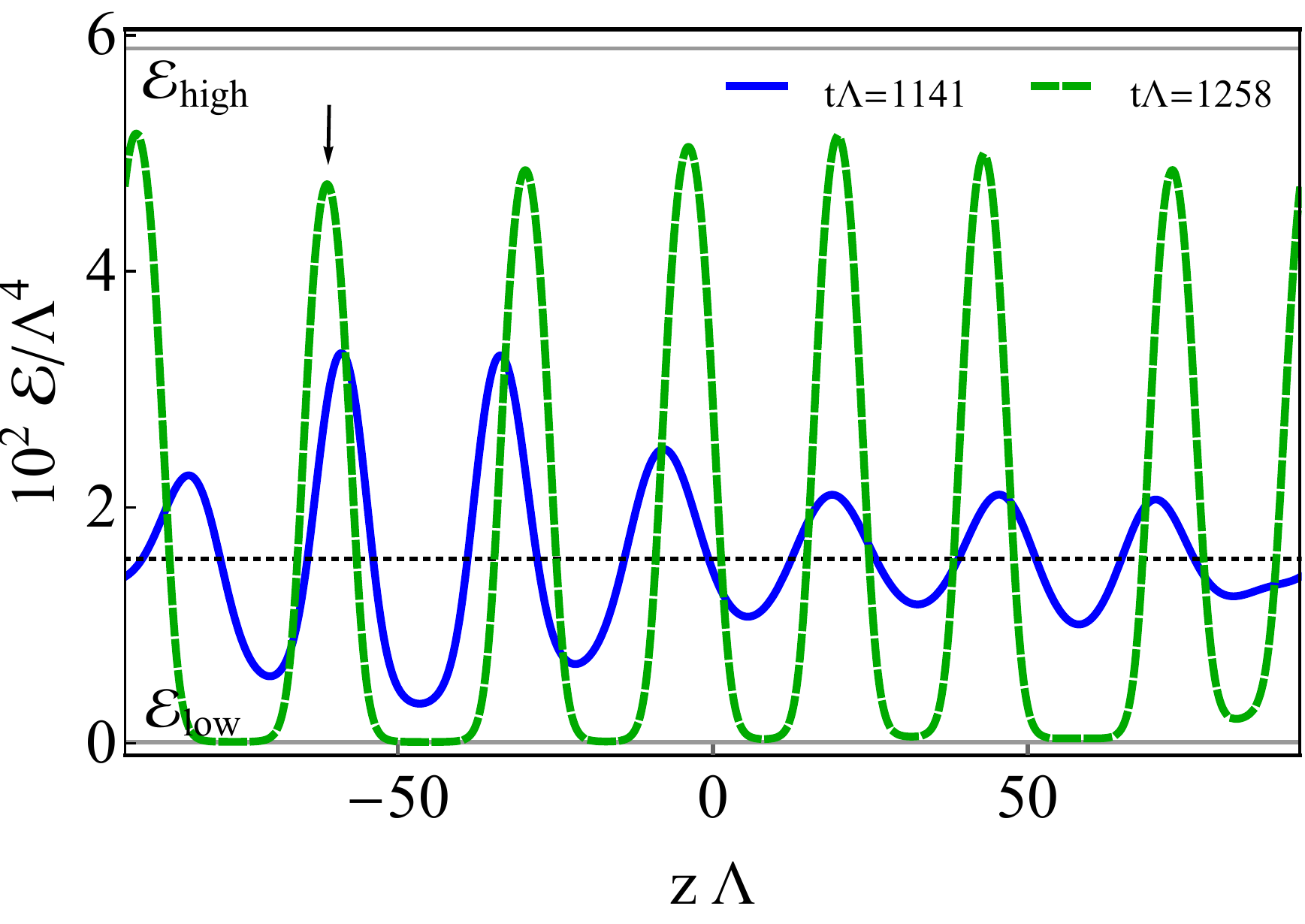} 
\quad&\quad
\includegraphics[width=.46\textwidth]{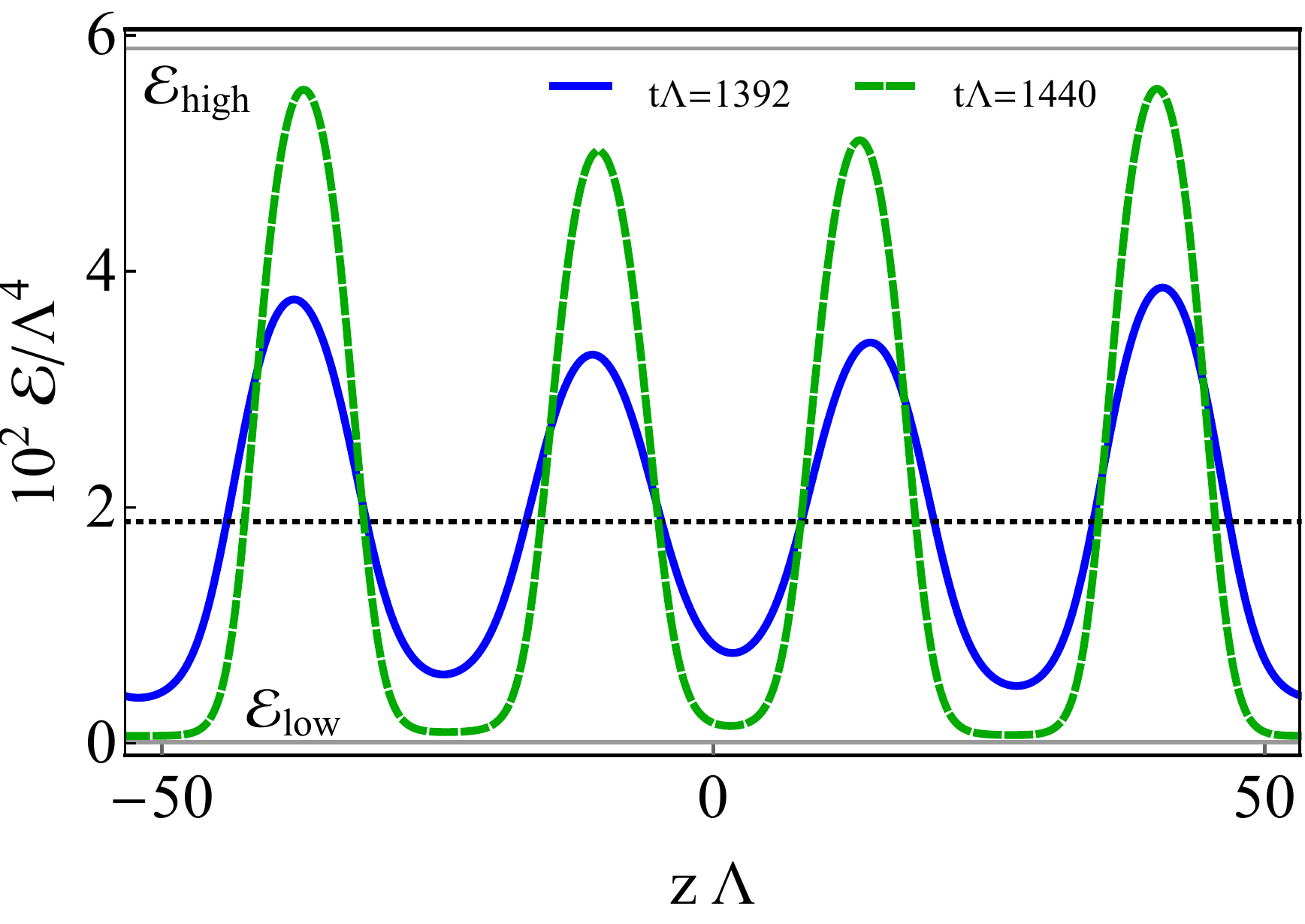} 
\end{tabular}
\end{center}
\vspace{-5mm}
\caption{\label{snap} \small 
	Snapshots of the energy densities of \fig{3Denergy} at the end of the linear regime (solid blue) and at the end of reshaping period (dashed green). The arrow on the left plot indicates the peak that corresponds to \fig{peakundeformedfig}. The green dashed curves precisely correspond to the peaks observed at early times in \fig{ZOOM1}. Notice that the maxima of the profiles have some velocity which will translate into the initial velocity of the structures formed.
}
\vspace{7mm}
\end{figure*}

\subsection{End of the linear regime}

After the initial period of linear evolution, eventually the system enters into the non-linear regime. Of course, this time is not sharply defined. We choose to define it as the time at which the slope of the leading Fourier mode in the log plots of \figs{modes} and \ref{modes2} deviates from the straight line predicted by the linear analysis by more than 10\%. The resulting times  are indicated by the grey vertical lines in \figs{modes} and \ref{modes2}. The corresponding energy profiles at those times are shown in \fig{snap} in solid blue. As can be seen in \figs{modes} and \ref{modes2}, the subleading modes can deviate earlier from the  exponential growth predicted by the linear analysis due to resonant behavior.  

A priori one may think that the linear regime ends simply when the amplitude of the inhomogeneous modes becomes a significant fraction of, but is still smaller than, the amplitude of the initial, homogeneous zero mode. However, our analysis indicates that this is too simplistic. On the one hand, in some cases the linear regime can persist until the amplitude in certain modes is so large that it is actually larger than that of the homogeneous mode. For example, this is the case  in \fig{modes2}(top). Note that this does not mean that the energy density  becomes negative in some regions, since the large negative contribution of the leading mode in these regions is compensated  by the positive contributions of the other modes, as is clear from \fig{snap}(right).

On the other hand, in some other cases we expect the linear regime to end when the amplitude of the inhomogeneous modes is still arbitrarily small. The reason for this is that, in generic circumstances, we expect the final state of the evolution to be a phase-separated configuration that should maximise the total entropy given the total energy available in the box. This implies that, at the latest, the exponential growth of the inhomogeneous modes should cease  around the time when  the energy density profile reaches $\Ehigh$ or $\Elow$, since in the final state no region should have energy higher than $\Ehigh$ or lower than 
$\Elow$. Note that this is a global condition in the sense that it does not apply to an individual mode but to the full energy density, which is the sum of all the modes. In some cases this condition should cut off the growth of the inhomogeneous modes at arbitrarily small amplitudes. For example, in the case of a  gauge  theory with a first order phase transition with arbitrarily small latent heat the growth stops because the energy profile quickly reaches both $\Ehigh$ and $\Elow$. Similarly, in a generic theory in a homogeneous initial state with energy very close to the upper or to the lower endpoints of the unstable branch, the growth stops because the profile of the energy density quickly reaches $\Ehigh$ or $\Elow$, respectively. We leave a more detailed investigation of the precise mechanism that cuts off the exponential growth for future work. 

At the end of the linear regime, the exact number  of maxima and minima of the energy profile  is given by the leading Fourier modes at that time. These depend on the initial amplitude of the modes and on the growth rates, and can be determined from the initial conditions. 
Consider first the case in which the initial modes correspond to numerical noise. Since this is assumed to be white noise all modes start with similar initial amplitudes. Therefore in this case the modes with the largest growth rates will dominate. 
For example, we can see this in \fig{modes2}(top), where the $n=4$ mode clearly dominates, which is why we have 4 maxima and minima in \fig{snap}(right). Now consider instead the case in which some initial mode is excited by hand with a large amplitude. If this amplitude is large enough then  this mode will be  the dominant one at the end of the linear regime. 
However, in some cases other modes with larger growth rates may still overtake this mode and become dominant at the end of the linear regime. 
This is illustrated in Fig.~\ref{modes}, in which the $n=1$ cosine mode is overtaken by the faster $n=6,8$ cosine modes and also by the $n=7$ sine mode. The latter is actually the dominant mode at the end of the linear regime, which is the reason why there are 7 peaks in \fig{snap}(left).

\subsection{Reshaping}
\label{reshaping}

By ``reshaping'' we mean a stage of non-linear evolution immediately after the end of the linear regime in which energy keeps being redistributed in the system in such a way that the structures formed during the linear regime keep adjusting their shape. This adjustment may or may not include a change in the number of maxima and minima. As we will see in \sect{mergers}, the reshaping period results in the formation of some structures that are either static or move with respect to one another with slowly varying velocity and almost constant shape. 

\begin{figure*}[t]
\begin{center}
\begin{tabular}{cc}
\hspace{-6mm}
\includegraphics[width=.48\textwidth]{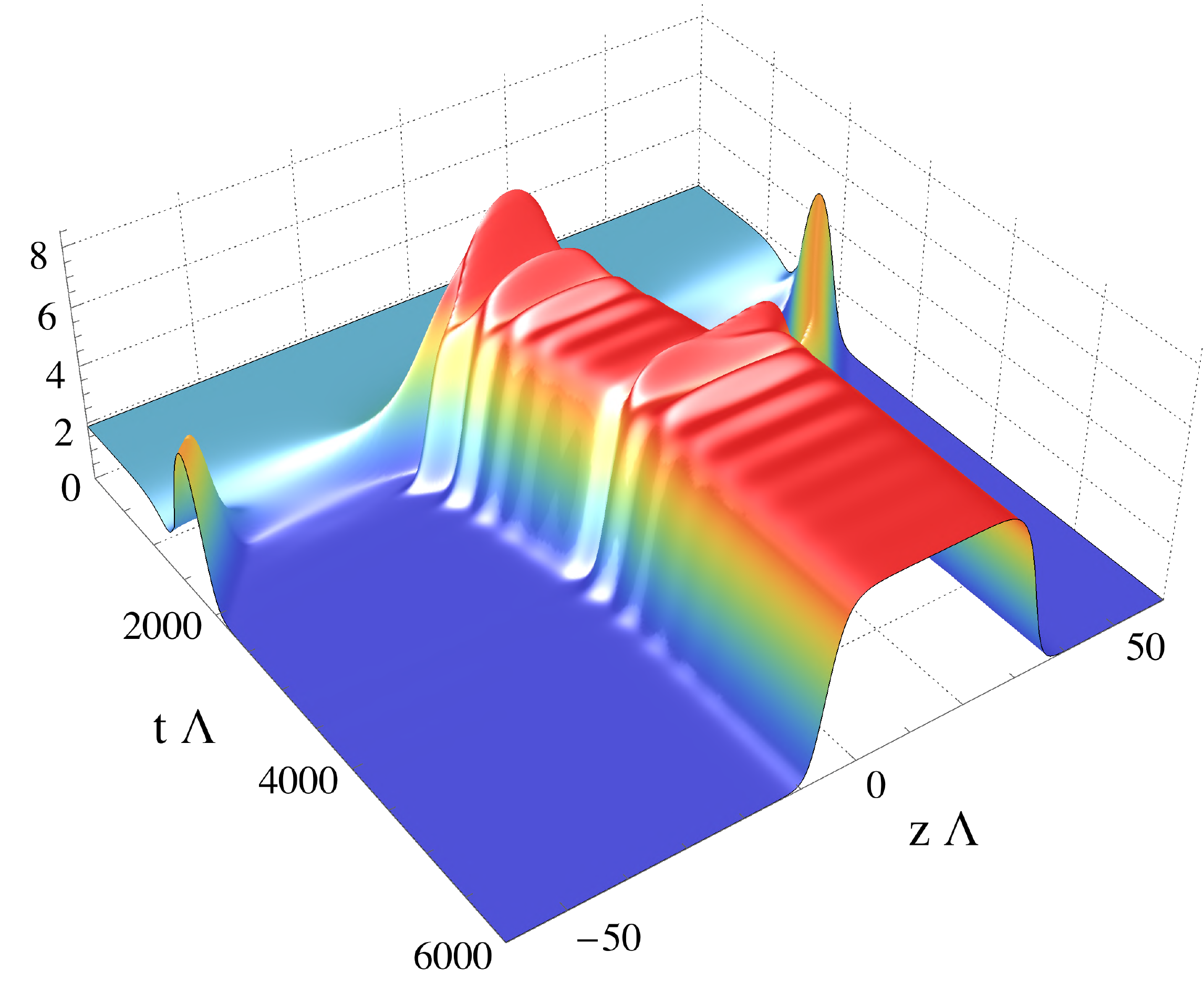} 
\put(-230,140){\mbox{\small{$10^2$ $\mathcal{E}/\Lambda^4$}}}
\qquad
\includegraphics[width=.46\textwidth]{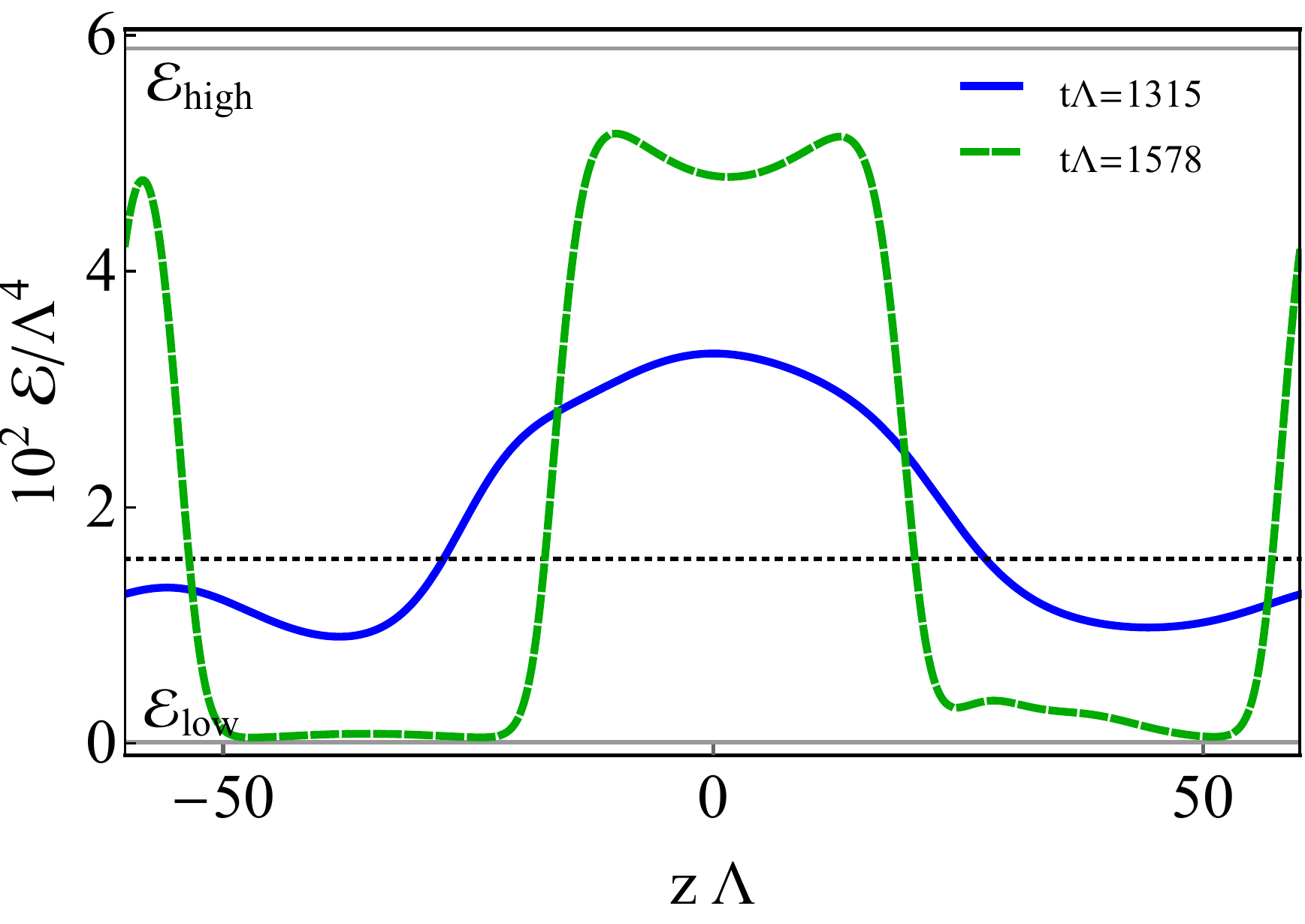}
\end{tabular}
\end{center}
\vspace{-5mm}
\caption{
\label{second} 
\small  (Left) Energy density of the evolution of an initial state with $\Ezero=\E_2, L\Lambda \simeq 120$. The central structure formed after the reshaping period is a phase domain. (Right) Snapshots of the energy density of 
the plot on the left at the end of the linear regime (solid blue) and at the end of reshaping period (dashed green). 
}
\vspace{7mm}
\end{figure*}

There are two qualitative possibilities for the type of  structures that can be formed at the end of the reshaping period: peaks and domains. By peaks we mean Gaussian-looking profiles as those around the maxima of the energy density in \fig{snap}. By domains we mean plateaus in which the energy density is approximately constant and equal to either $\Ehigh$  or $\Elow$. If we need to distinguish we will refer to these as ``high-energy domains" and ``low-energy domains'', respectively. If we simply use the term ``domain'' we mean a high-energy domain. Some low-energy domains are present in \fig{snap}, and  examples of both types are shown in \fig{second}.  
\begin{figure*}[h!!!] 
	\begin{center}
		\begin{tabular}{cc}
			\includegraphics[width=.46\textwidth]{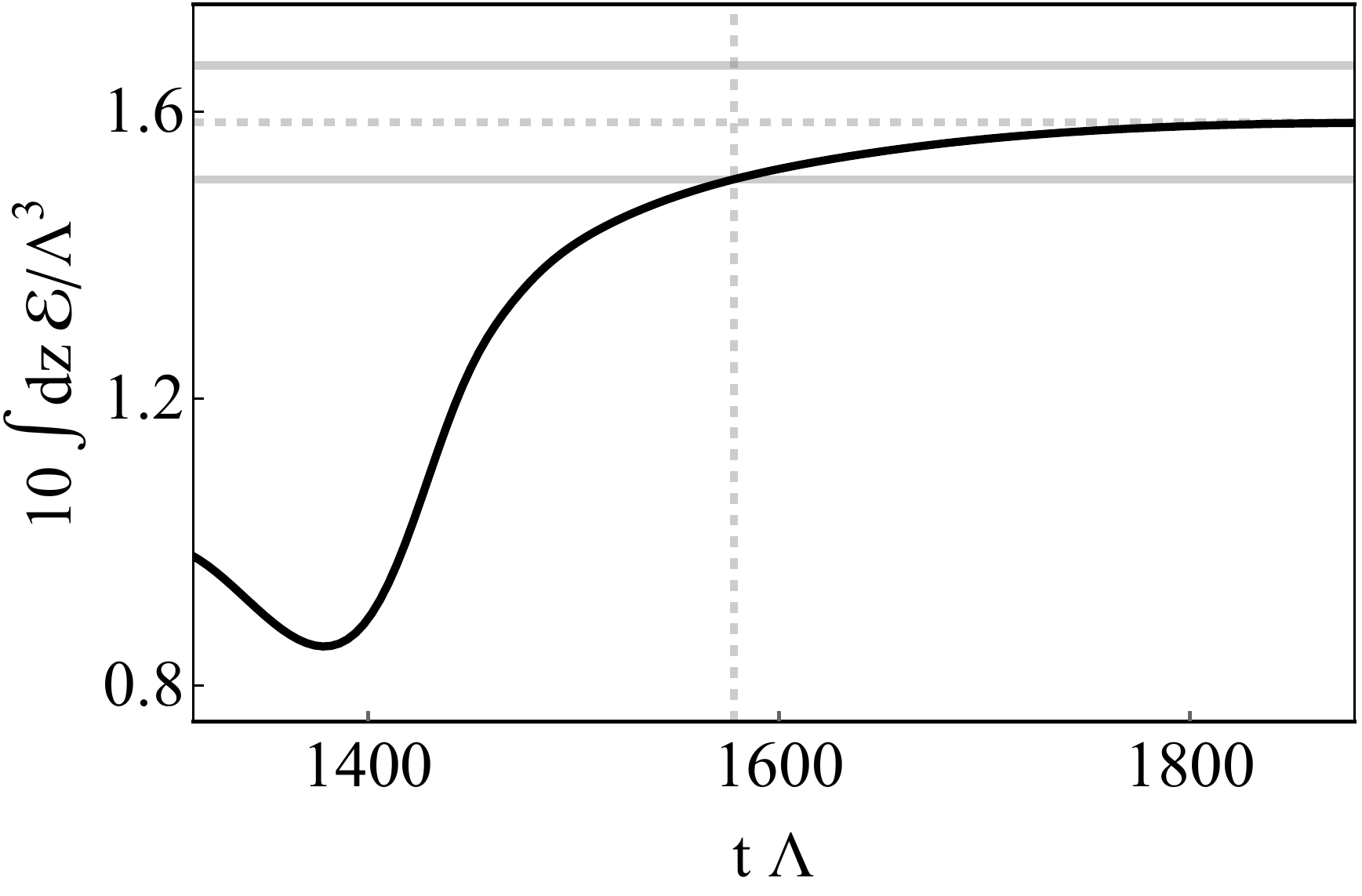} 
			\quad&\quad
			\includegraphics[width=.46\textwidth]{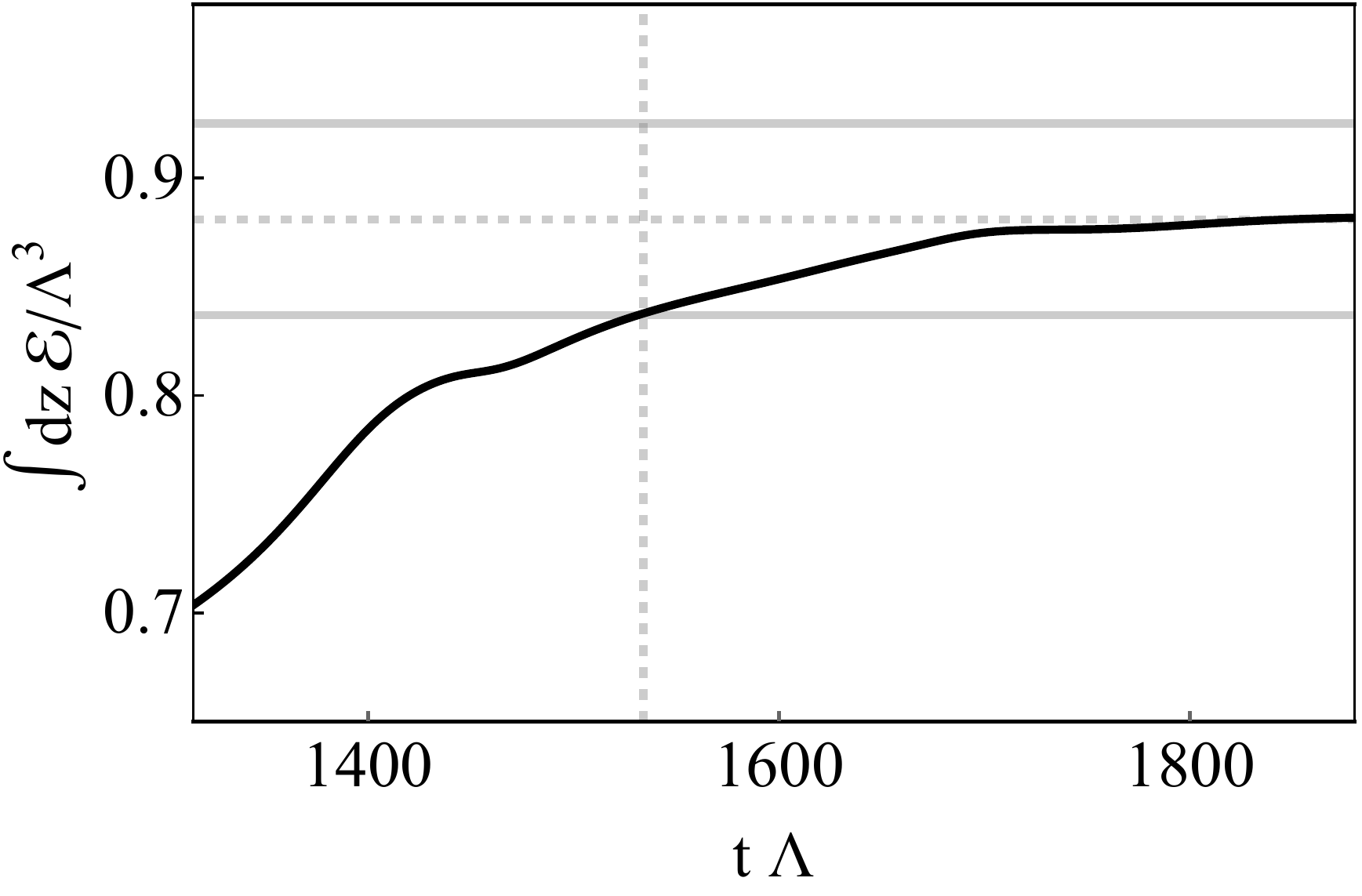} 
		\end{tabular}
	\end{center}
	\vspace{-5mm}
	\caption{
		\label{illus}
		\small 
		Time evolution of the energy density contained in the peak (left) and the domain (right) of \fig{second} starting at the end of the linear regime. The dashed horizontal lines are the values that these integrated energy densities tend to, and the continuous horizontal lines indicate a 5\% band around them. The vertical lines at $t\Lambda=1578$ (left) and $t\Lambda= 1534$ (right) show the times at which the corresponding curves enter these bands. The end of the reshaping period is given by the largest of these times.
	}
	\vspace{7mm}
\end{figure*}
The distinction between a peak and a  domain is of course not a sharp one, since the size of the domain can be reduced continuously until it turns into a peak. 

We define the end of the reshaping period as the time beyond which the energy contained in each peak and domain, defined as the integral between the inflection points in their profiles, does no longer change by more than 5\%. This is illustrated in \fig{illus}, where we plot these integrated energies 

\begin{figure*}[t]
	\begin{center}
		\begin{tabular}{cc}
			\hspace{-5mm}
			\includegraphics[width=.52\textwidth]{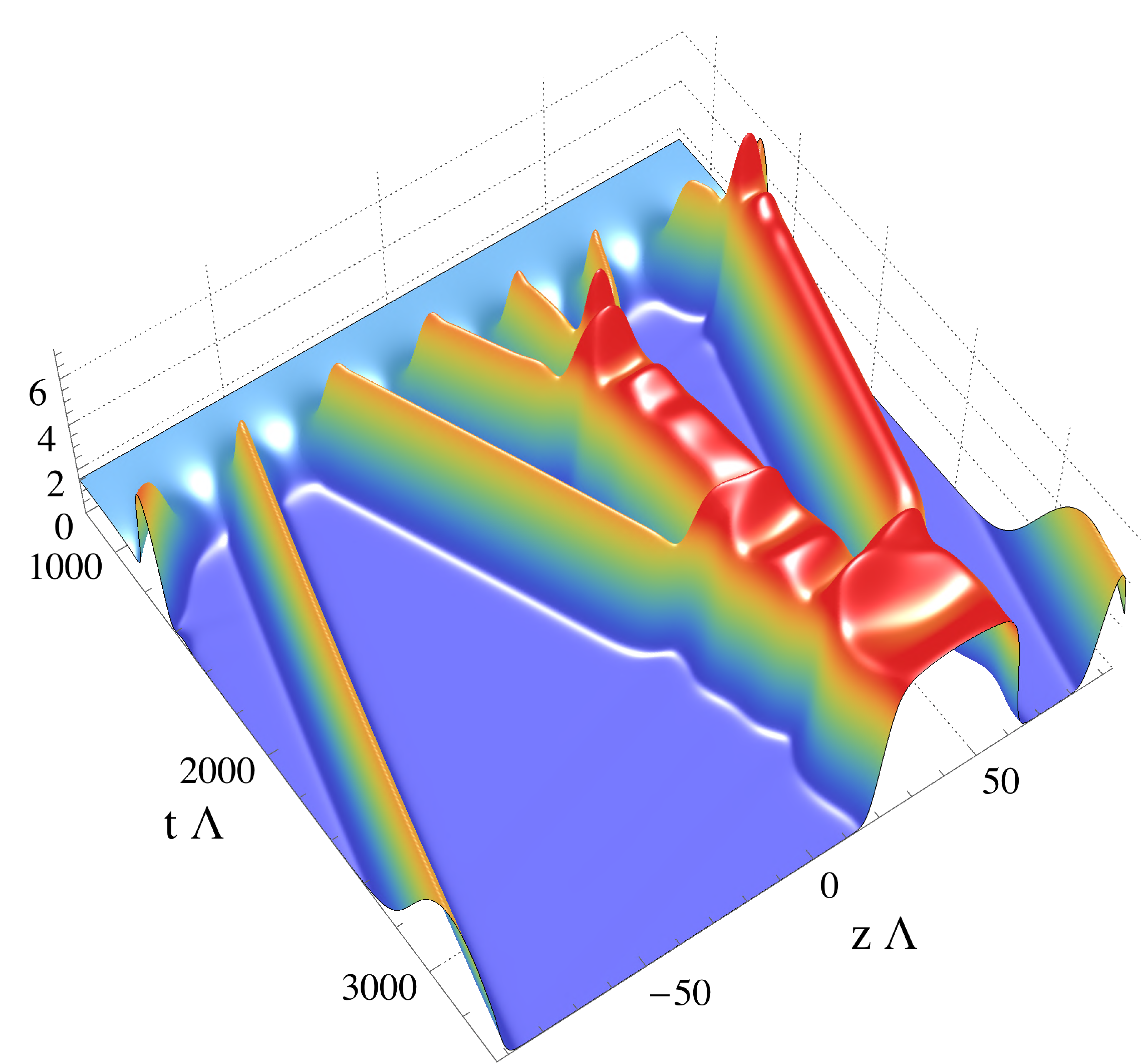} 
			\put(-235,150){\mbox{\small{$10^2$ $\mathcal{E}/\Lambda^4$}}}
			\put(-184,133){\mbox{\small{$\downarrow$}}}
			\quad
			\includegraphics[width=.52\textwidth]{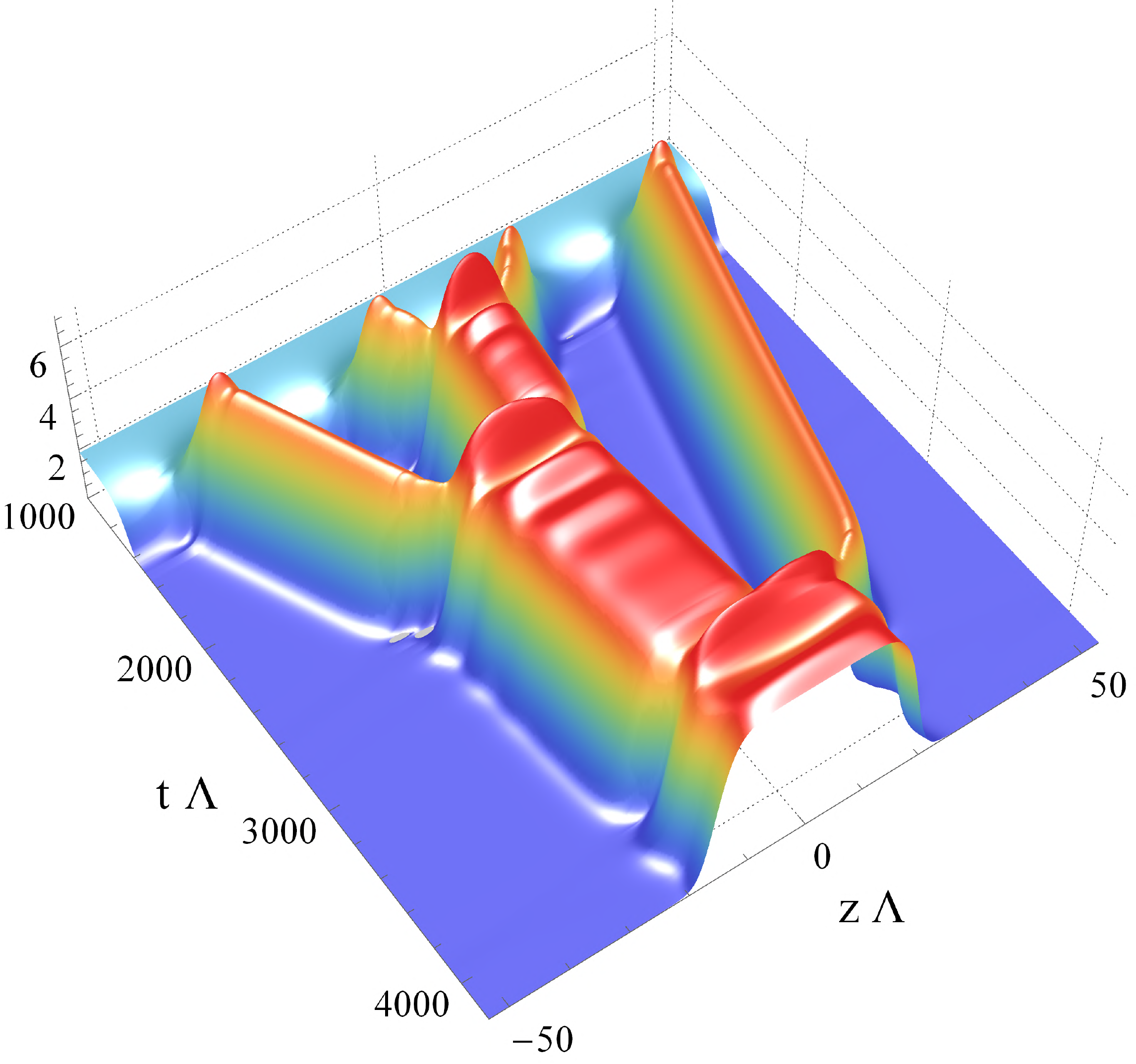} 
			\put(-235,160){\mbox{\small{$10^2$ $\mathcal{E}/\Lambda^4$}}}
		\end{tabular}
	\end{center}
	\vspace{-5mm}
	\caption{\label{ZOOM1} 
		\small Zoom in of \fig{3Denergy}. At the end of the reshaping period  seven peaks form on the left plot and four on the right one. Snapshots at those times are shown in \fig{snap} in dashed green. From that point on, the peaks move rigidly with almost constant shapes at slowly varying velocities. They subsequently collide and merge, forming larger structures. Eventually, all the structures merge into a single  one and the system reaches a phase-separated configuration (see  \fig{3Denergy}). The peak indicated with an arrow on the left plot corresponds to the peak shown in \fig{peakundeformedfig}.
	}
	\vspace{7mm}
\end{figure*}

\begin{figure*}[h!!]
	\begin{center}
		\begin{tabular}{cc}
			\includegraphics[width=.46\textwidth]{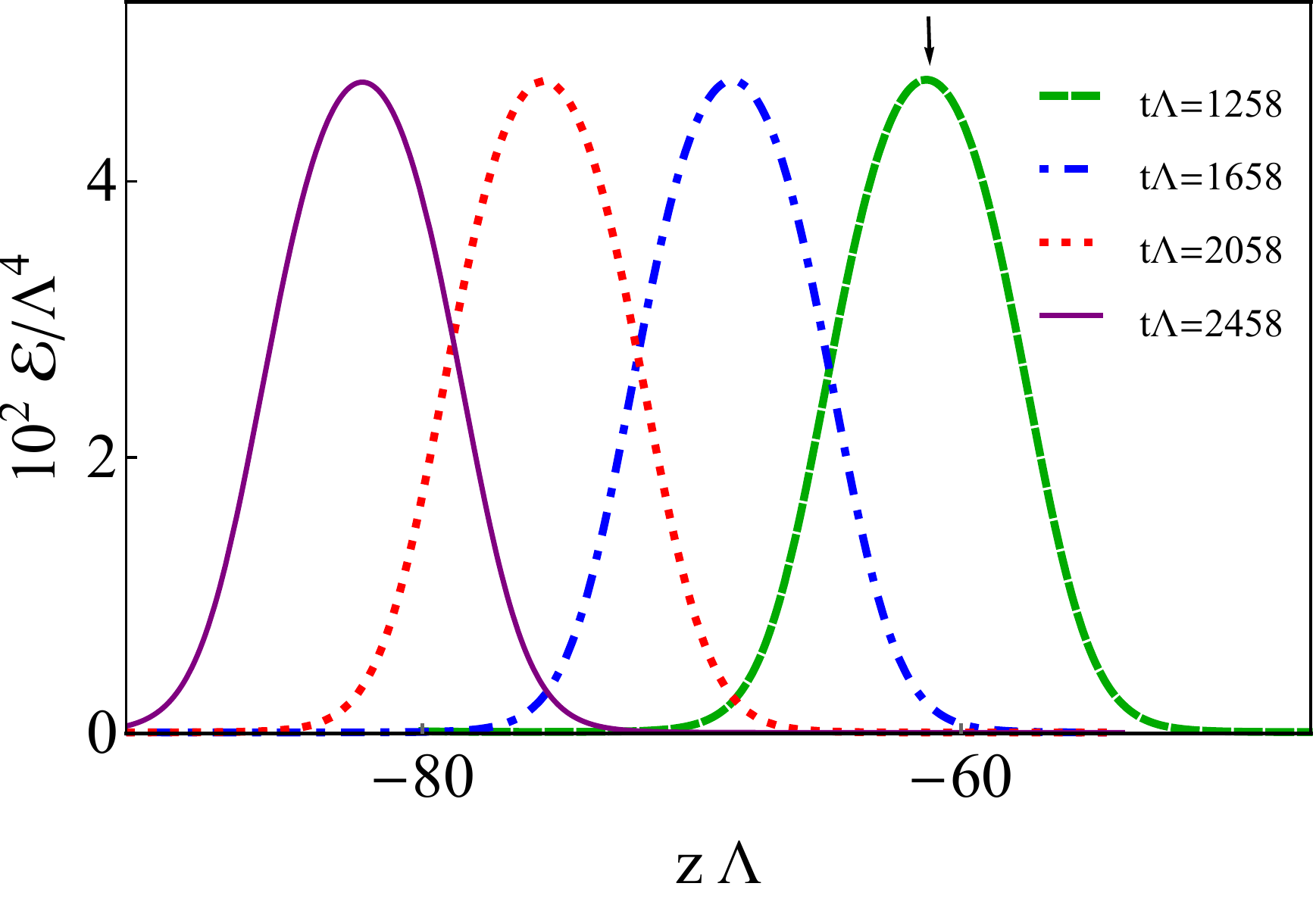} 
			\quad&\quad
			\includegraphics[width=.46\textwidth]{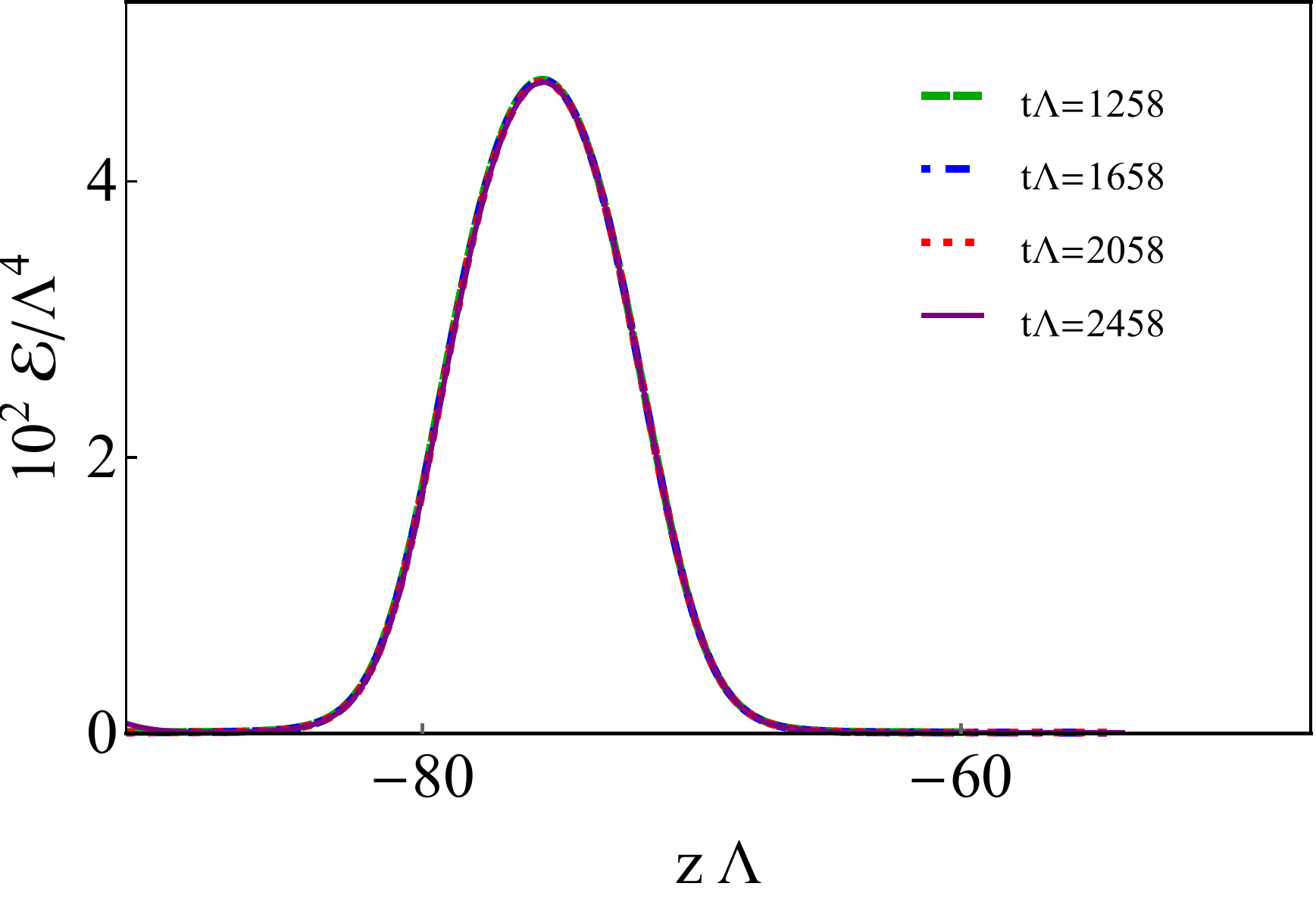} 
		\end{tabular}
	\end{center}
	\vspace{-5mm}
	\caption{\label{peakundeformedfig} 
	  \small (Left) Profile of one of the peaks in \fig{3Denergy} at different times. The peak indicated by an arrow is  the same as the one indicated by an arrow in \fig{snap}(left) and \fig{ZOOM1}(left).
	  (Right) The same profiles as in (Left) shifted by a constant amount to show that the peak moves undeformed. 
	}
	\vspace{7mm}
\end{figure*}

In our model the reshaping period takes a few hundred times $\Lambda^{-1}$, as can be seen in \figs{snap} and \ref{second}. After this time, the peaks and/or domains move rigidly with slowly varying velocity with almost no distortion of their shape. This can be seen in \fig{ZOOM1}, which is a zoom on early times of \fig{3Denergy}.
The 7 initial peaks in \fig{ZOOM1}(left) correspond to the 7 peaks shown in dashed green in \fig{snap}(left). Similarly, the 4 peaks formed at early times in \fig{ZOOM1}(right) correspond to the 4 peaks in  dashed green of \fig{snap}(right). To confirm that the profile of the peaks moves almost undeformed, in \fig{peakundeformedfig}(left) we plot the profile of one of these peaks at different times. In \fig{peakundeformedfig}(right) we plot the same profiles shifted by a constant amount to check that they coincide. The first snapshot in \fig{peakundeformedfig} corresponds exactly to the peak indicated with an arrow in \fig{snap}(left) at the end of the reshaping period. We see that, on the scale of this plot, the shape is indeed ``frozen'' after this moment, and that it moves rigidly.
However, a finer analysis reveals that neither the shape nor the velocity of the peaks are strictly constant in time, hence our use of the terms ``almost constant shape'' and ``slowly varying velocity'' above. Specifically, we have verified that the maximum local energy density of a peak stays constant to within 1\% over very long periods of time of order $\Lambda \Delta t \sim 15000$. Over a similar amount of time the velocity of a peak can decrease by a factor of two. Presumably, the almost constancy of the shape and the relatively slow variation of the velocity rely on two features. One is the fact that the initial velocity of the peaks and/or domains is not too high, and the other is the fact that the hierarchy between the stable equilibrium energy densities is large, $\Elow/\Ehigh \ll 1$, which implies that a peak or domain can move  through the cold phase with little ``friction''. Presumably the distortion in the shape of the moving peaks and/or domains and their slowing down will become more pronounced  as the ratio $\Elow/\Ehigh$ increases. We leave further investigation of this point for future work.

\subsection{Mergers}
\label{mergers}

\begin{figure*}[t]
\begin{center}
\includegraphics[width=.57\textwidth]{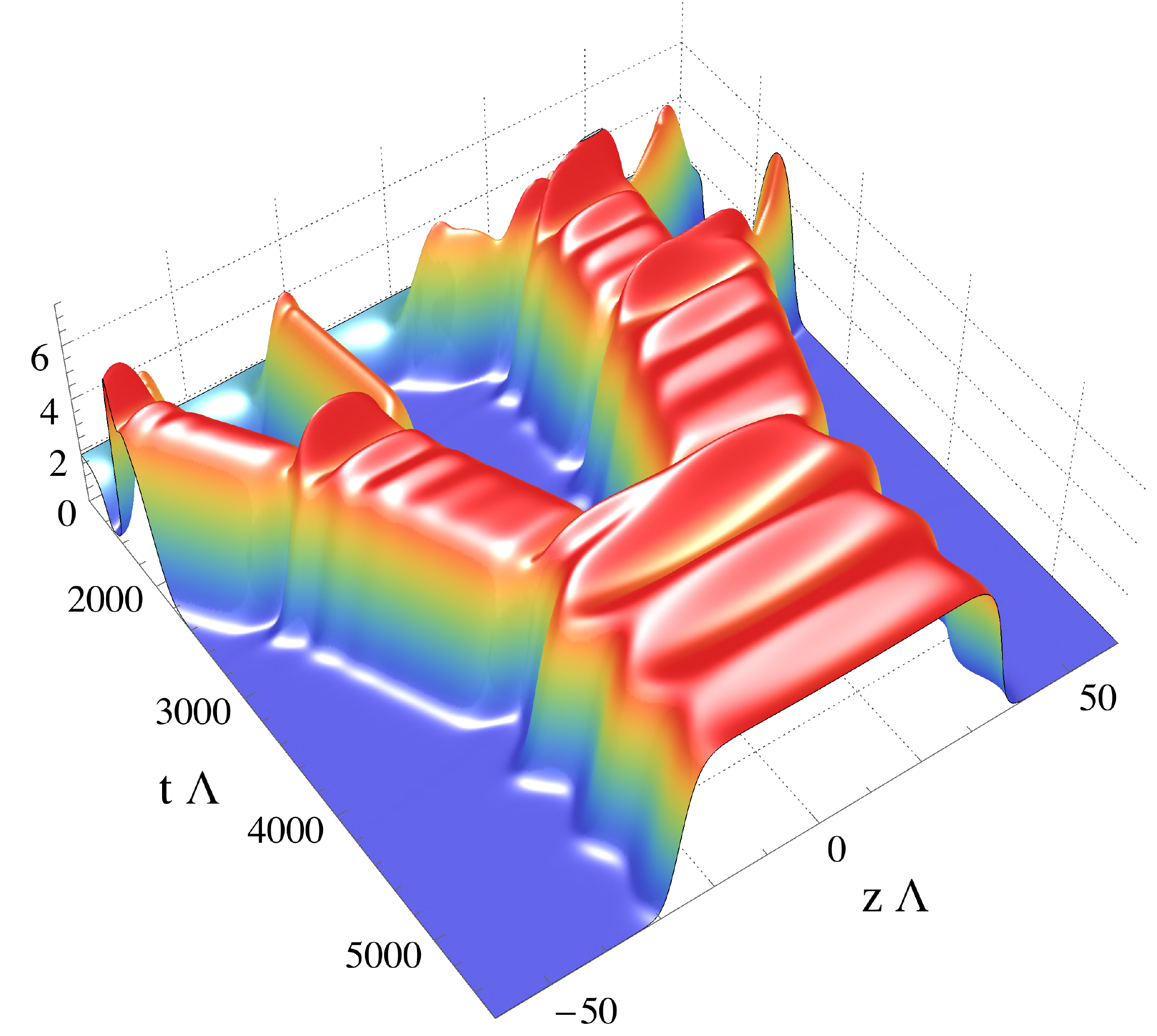} 
\put(-264,168){\mbox{\small{$10^2$ $\mathcal{E}/\Lambda^4$}}}
\end{center}
\vspace{-5mm}
\caption{\label{reflecting1} 
\small Energy density of the evolution of an initial state with a box $L \Lambda=213$ perturbed with an $n=1$ mode. We zoom into the region where the phase mergers take place. 
}
\vspace{7mm}
\end{figure*}

In generic situations the different structures formed at the end of the reshaping period move with different velocities, so they eventually collide with one another. In \fig{ZOOM1} we can see several of these collisions. In all the cases that we have considered in this paper, a collision of any two structures leads to a merger, i.e.~the result of the collision is a single, larger structure.\footnote{However, preliminary investigations indicate that for sufficiently high velocities the result of the collision consists of two objects moving away from each other.} The merger between two peaks may result in a larger peak or a domain. The merger between a peak and a domain or between two domains results in a larger domain. A particularly clear illustration is shown in \fig{reflecting1}, which has the largest  box that we have considered, $L \Lambda \simeq 213$, and an initial $n=1$ mode. The large box allows several peaks to coalesce to form two separate 
domains that move with non-vanishing velocity, and that finally collide forming a larger domain.
\begin{figure*}[t!!!]
\begin{center}
\includegraphics[width=.32\textwidth]{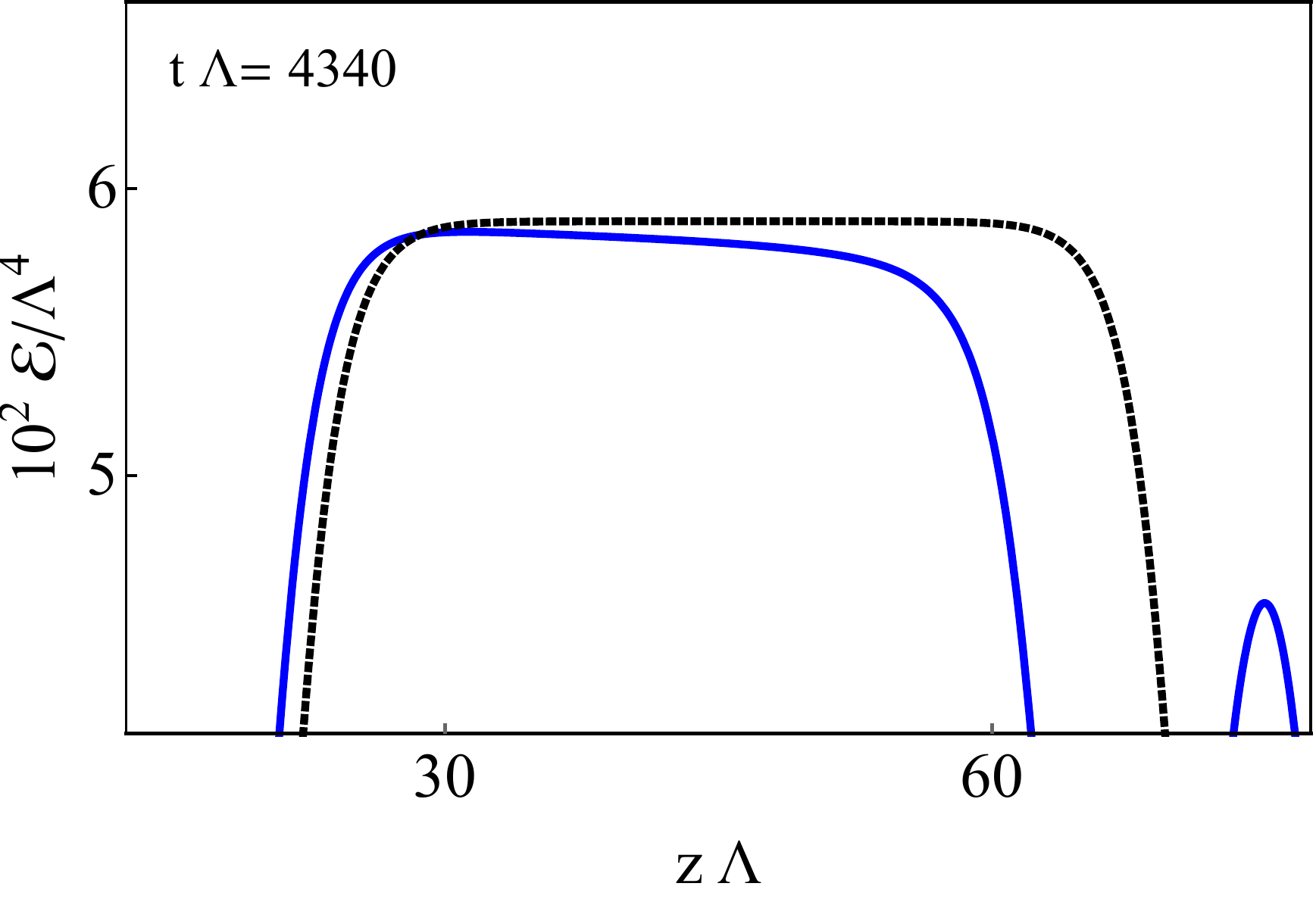} 
\includegraphics[width=.32\textwidth]{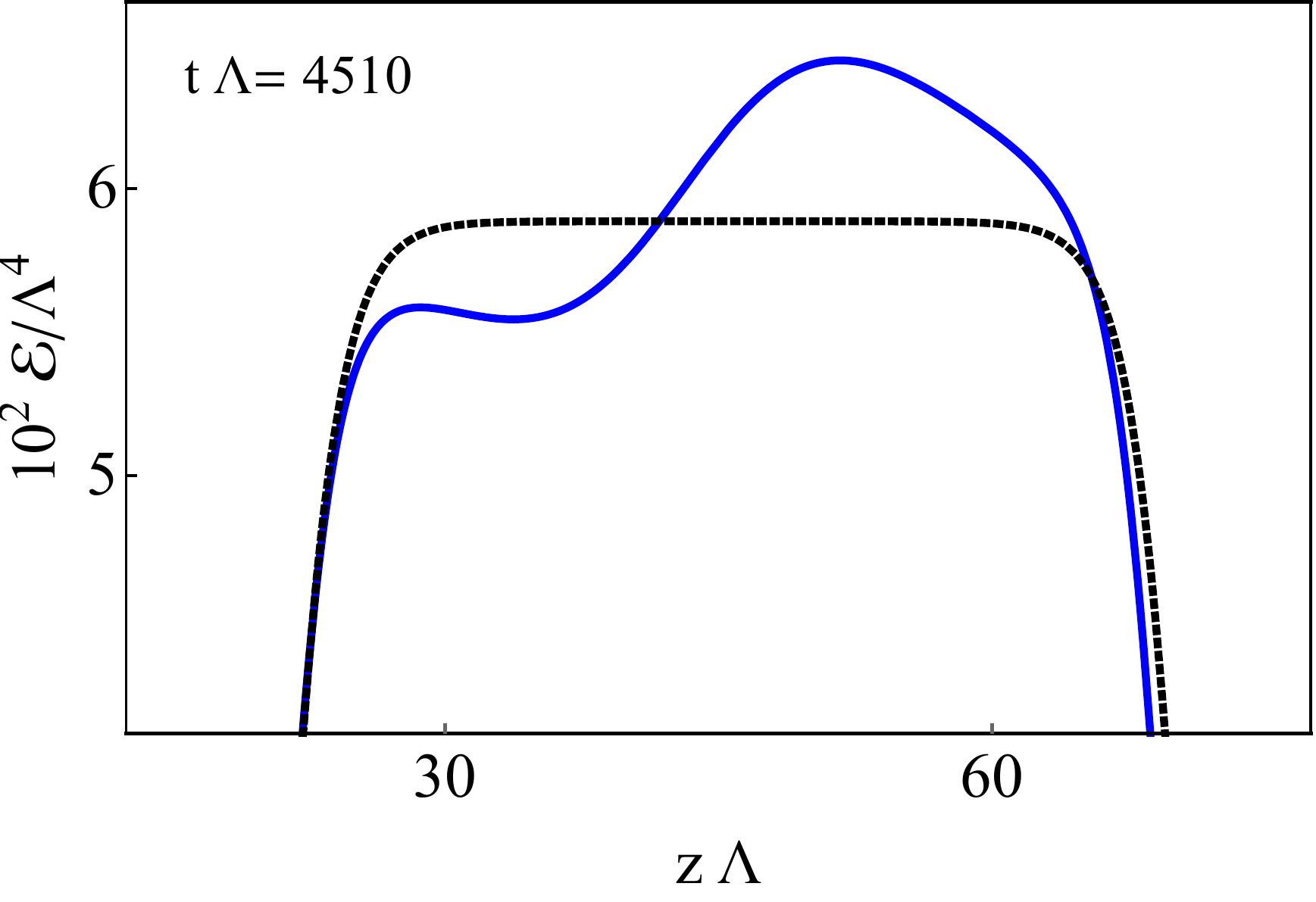} 
\includegraphics[width=.32\textwidth]{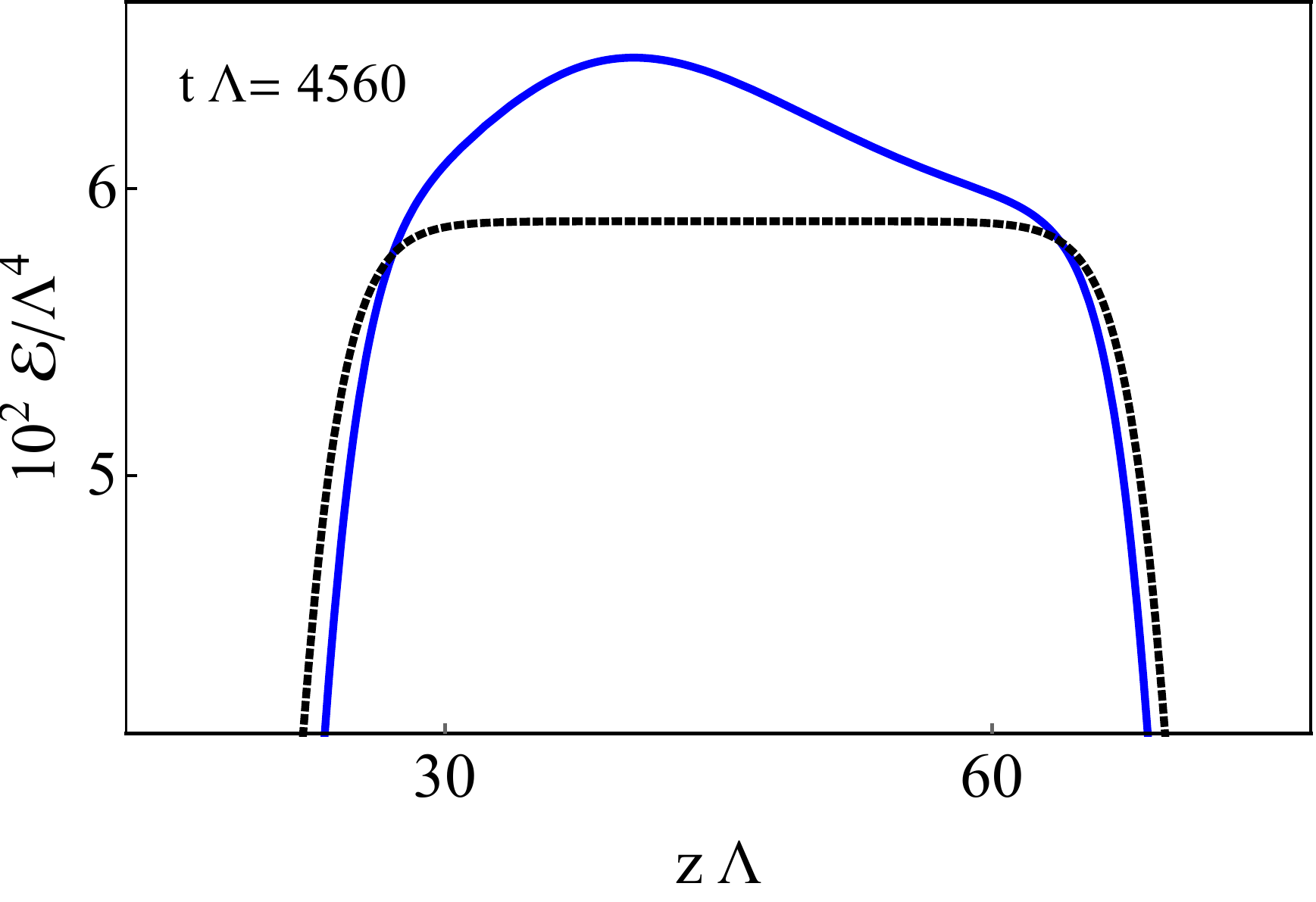} 
\includegraphics[width=.32\textwidth]{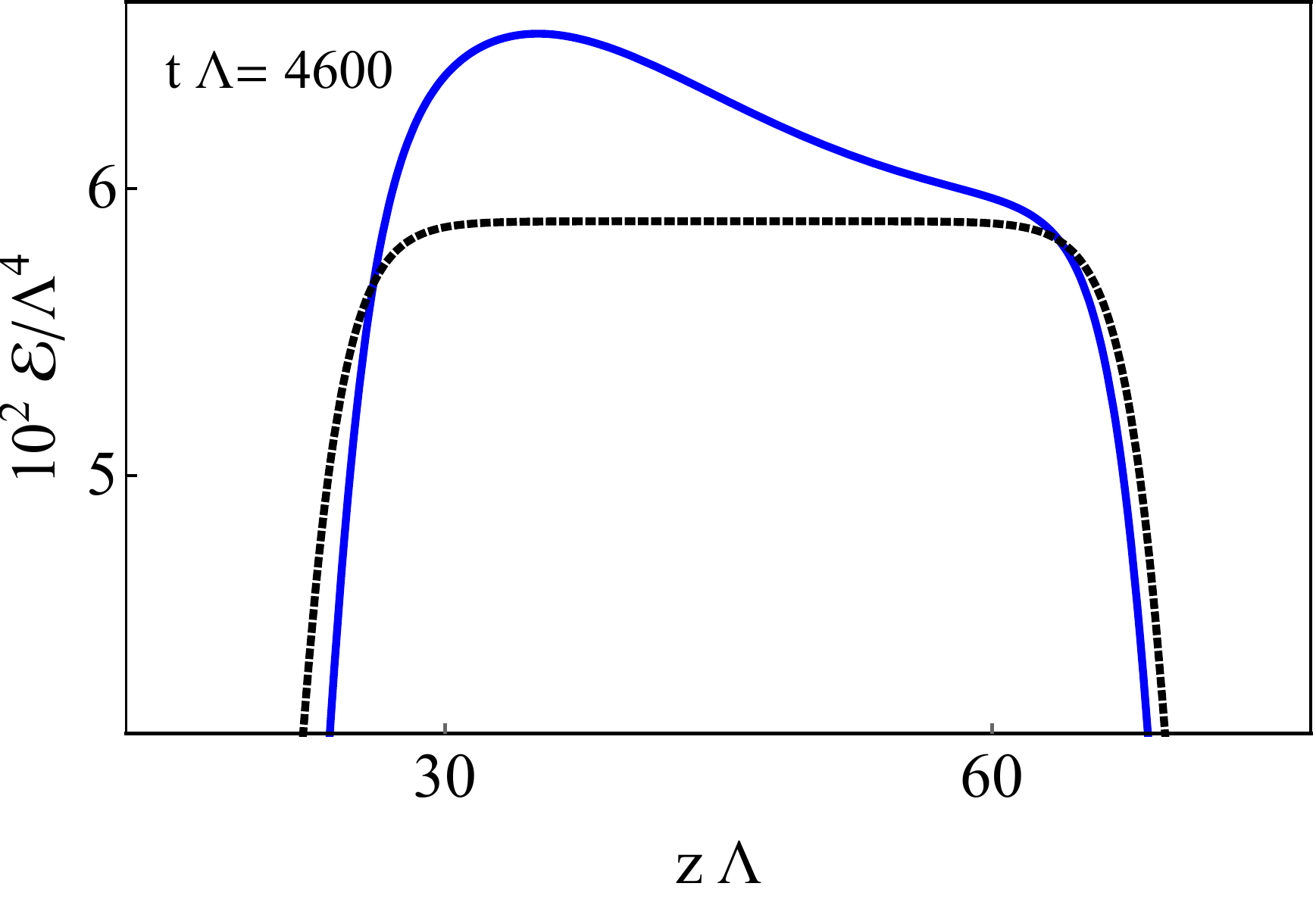} 
\includegraphics[width=.32\textwidth]{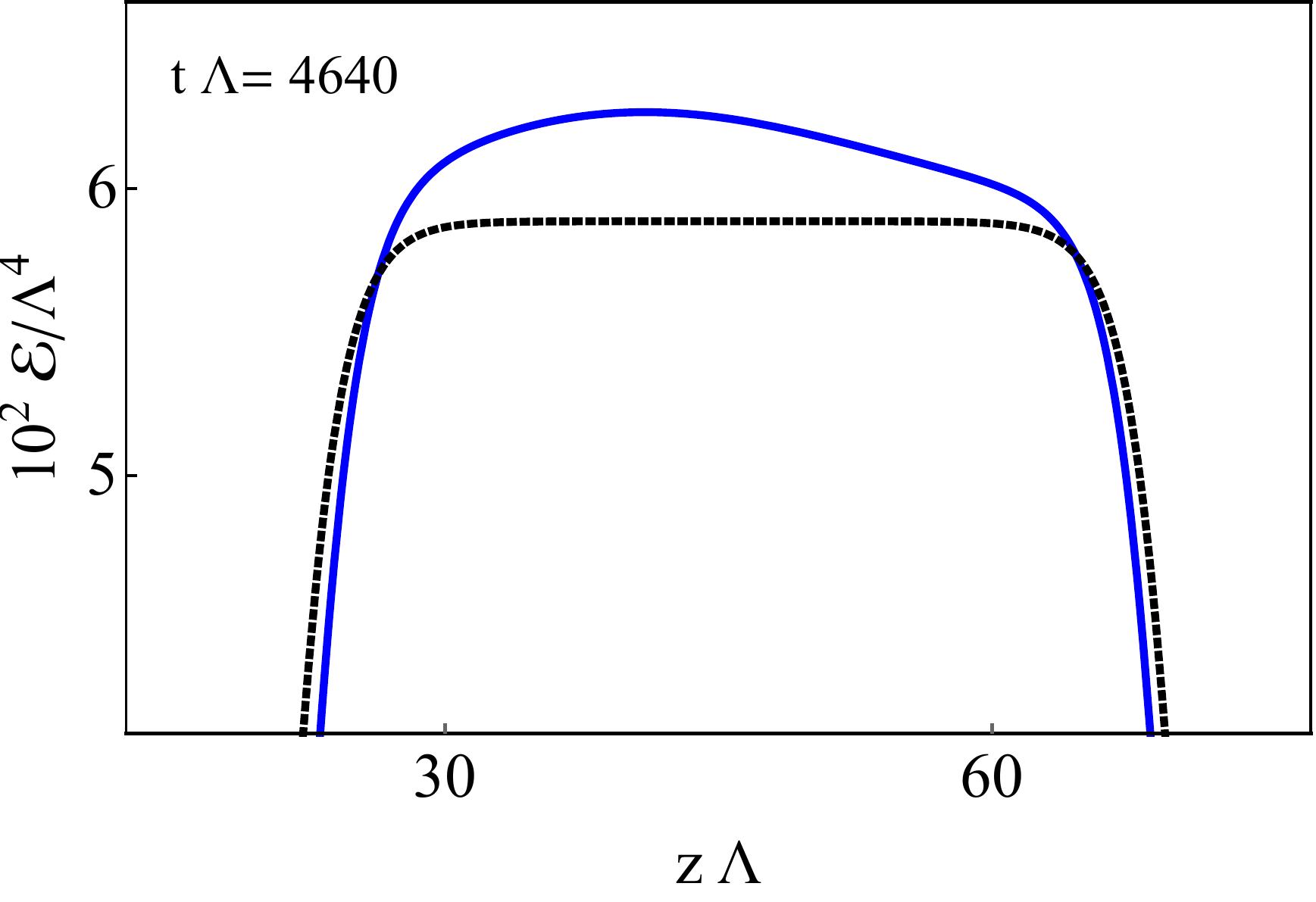} 
\includegraphics[width=.32\textwidth]{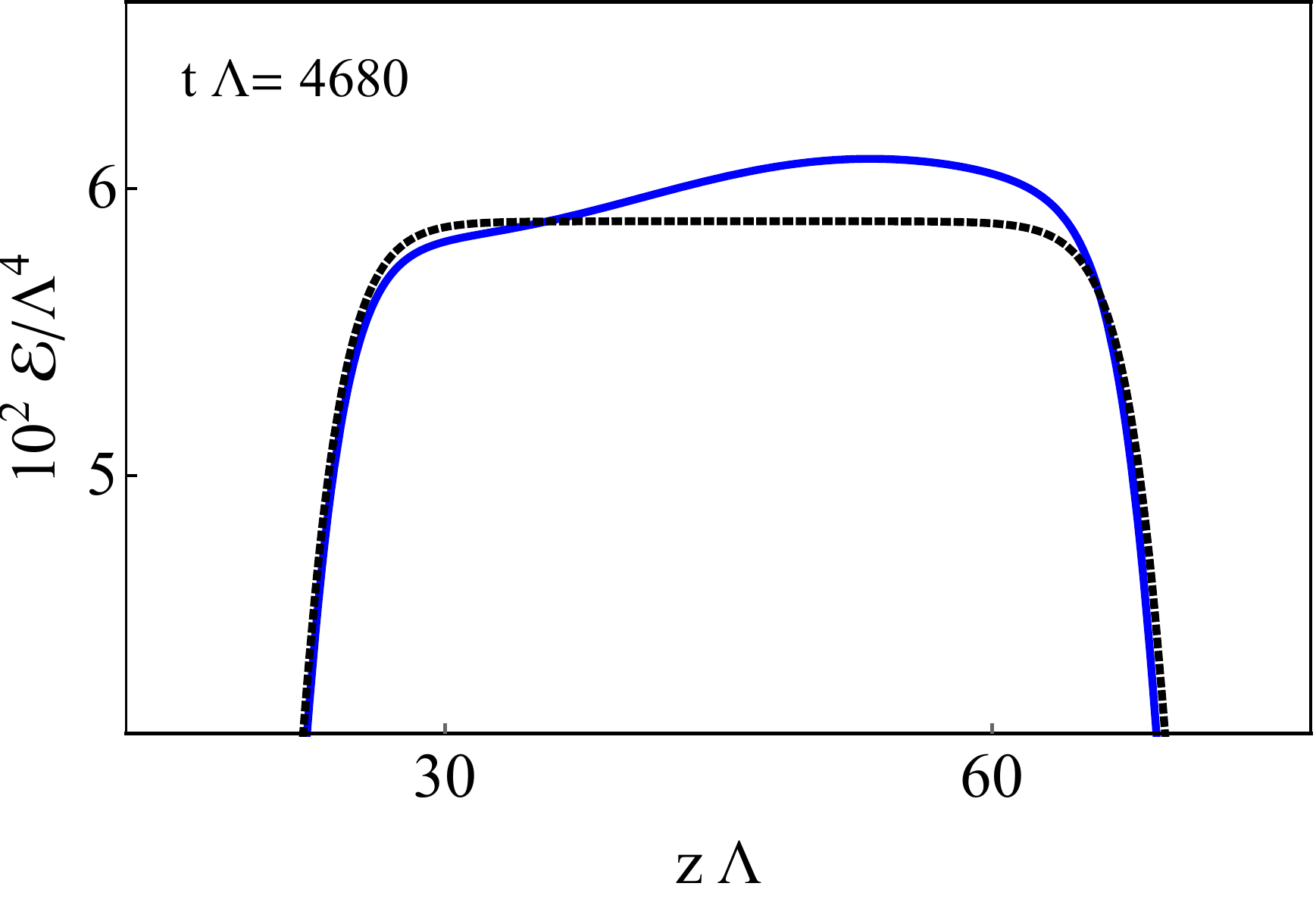}
\end{center}
\vspace{-5mm}
\caption{\label{atte} 
\small The continuous blue curves are snapshots of the energy density for the last merger between a peak and a domain in \fig{3Denergy}(left). The dashed black curve shows the profile  at asymptotically late times. In the first plot the peak is moving leftward and is about to collide with the domain. In the second and third plots the peak has merged with the domain, creating a perturbation that is moving leftward.  After bouncing off the interface on the left side,  the perturbation  is moving rightward in the bottom three plots.  The widening and the attenuation of the perturbation as it travels on top of the domain are clearly seen. 
}
\vspace{7mm}
\end{figure*} 

The final structure formed after a collision relaxes to equilibrium through  damped oscillations, as can seen in e.g.~\fig{ZOOM1}. Although these are oscillations around an inhomogeneous state, we will show in the next subsection that, if the final structure is a phase domain, then at late times these oscillations correspond to  linearised sound mode oscillations around the high energy phase. 



\begin{figure*}[t]
\begin{center}
\includegraphics[width=.46\textwidth]{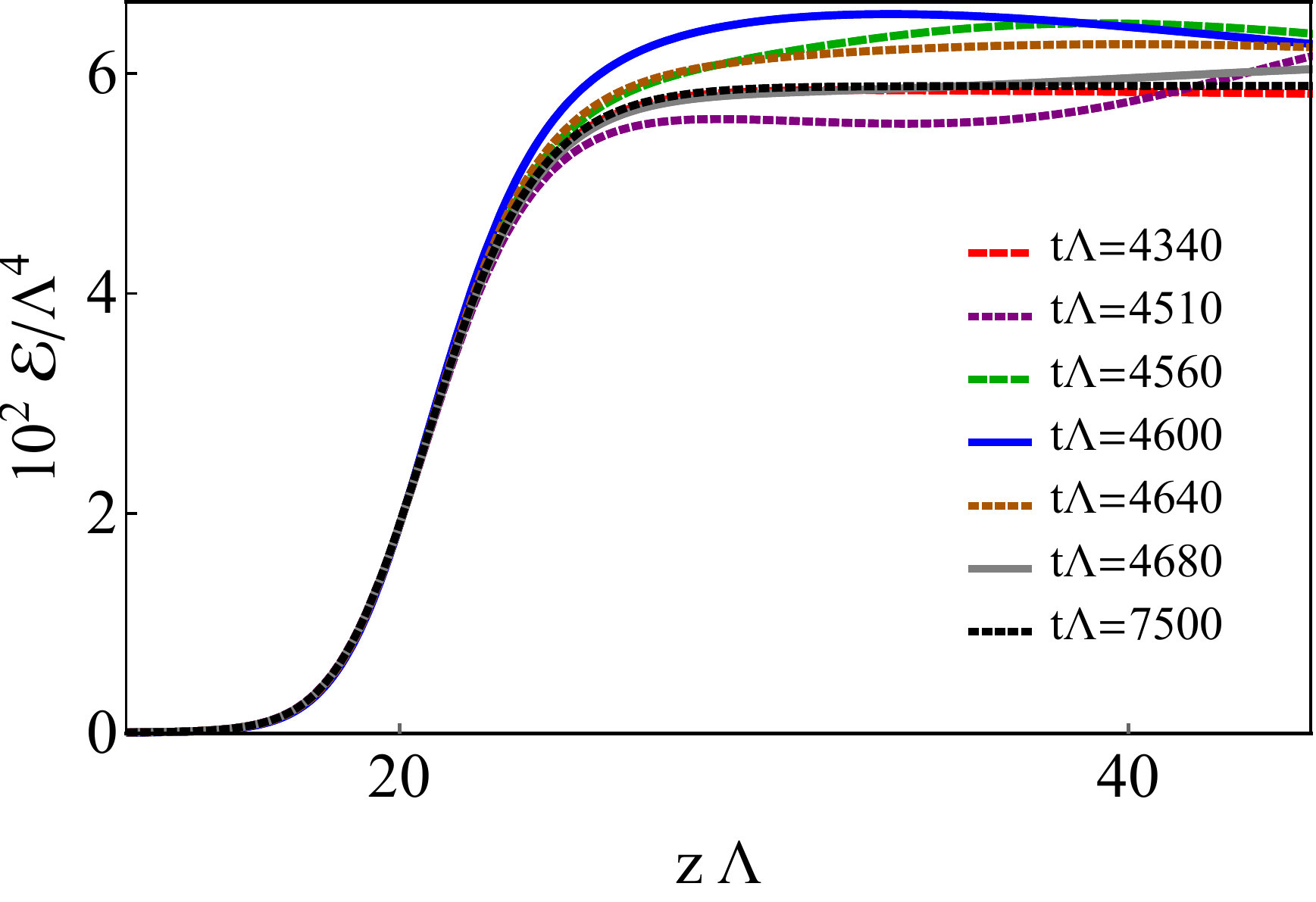} 
\end{center}
\vspace{-5mm}
\caption{\label{reflecting} 
\small Same curves as in \fig{atte} shifted by a constant amount in order to show that the shape of the interface on the left side is hardly modified by the bouncing of the perturbation. 
}
\vspace{7mm}
\end{figure*}

In the case of a collision between a peak and a domain the dynamics can be qualitatively understood even from early times due to the separation of scales provided by the different sizes of the colliding structures. Indeed, in this case the peak creates a perturbation on top of the domain. This perturbation suffers attenuation and widening as it travels from one side of the domain to the other, as illustrated in \fig{atte}.
It would be interesting to verify if the attenuation and the widening are the same as for a perturbation traveling on an infinite plasma with the same energy density, as one would expect for large domains.  When the perturbation reaches the end of the domain it bounces off entirely in the sense that no energy escapes the domain. In other words, the  interface between the domain and the cold phase acts as a rigid wall with negligible transmission coefficient. This is illustrated in \fig{reflecting}, where we plot the same curves as in \fig{atte} shifted by a constant amount in order to show that the shape of the interface on the left side is hardly modified by the bouncing of the perturbation. In fact, the total energy comprised between the midpoints of the two interfaces is practically constant in time (within $0.1\%$) once the peak has merged with the domain. 
After the perturbation has bounced back and forth a few times, the system is well described by the linearised sound mode, as we will verify in \sect{relaxation}.

\subsection{Unstable static configurations}
\label{unstable}
\begin{figure*}[t!!!]
\begin{center}
\begin{tabular}{cc}
\hspace{-6mm}
\includegraphics[width=.50\textwidth]
{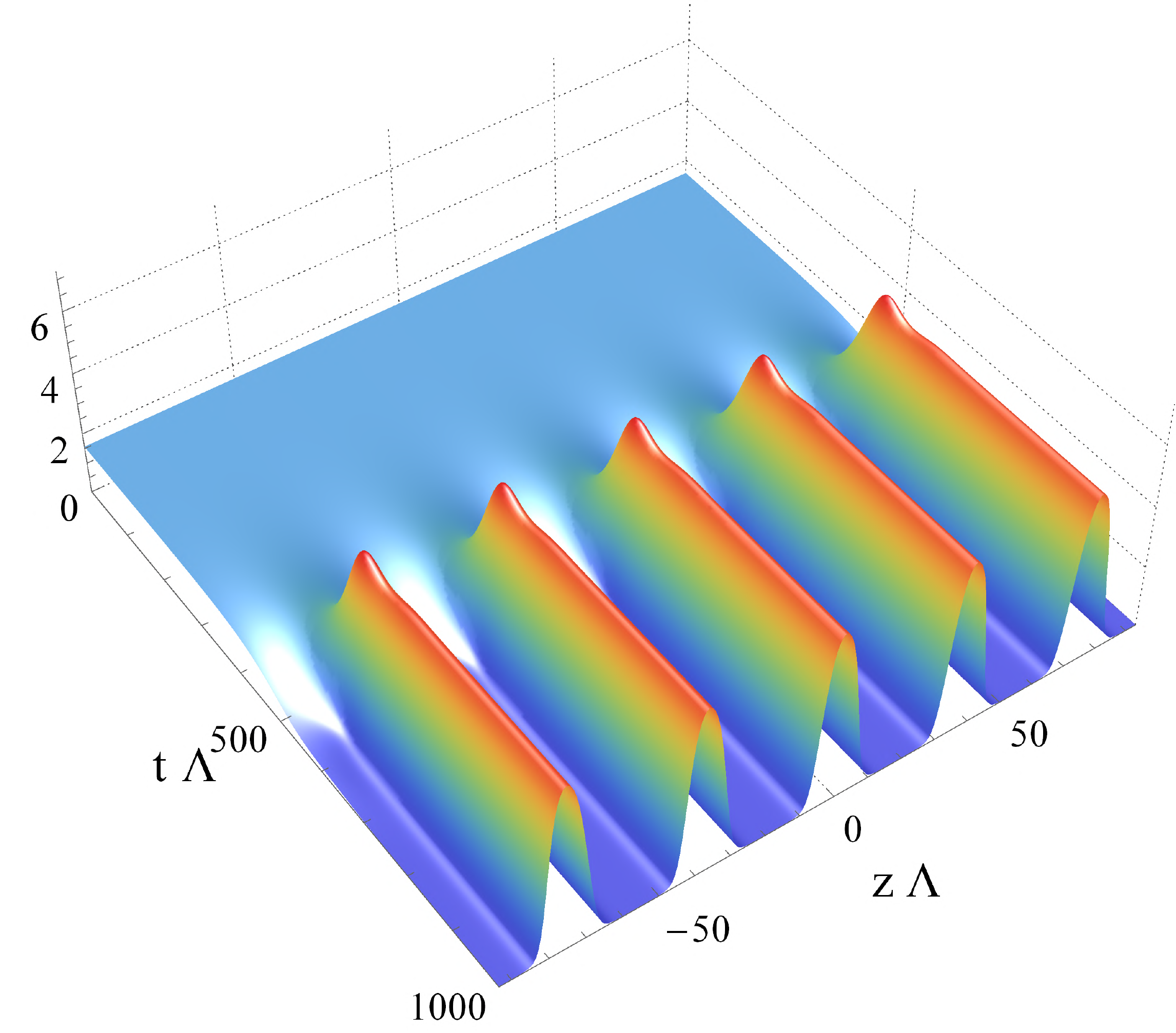} 
\put(-234,148){\mbox{\small{$10^2$ $\mathcal{E}/\Lambda^4$}}}
\includegraphics[width=.50\textwidth]{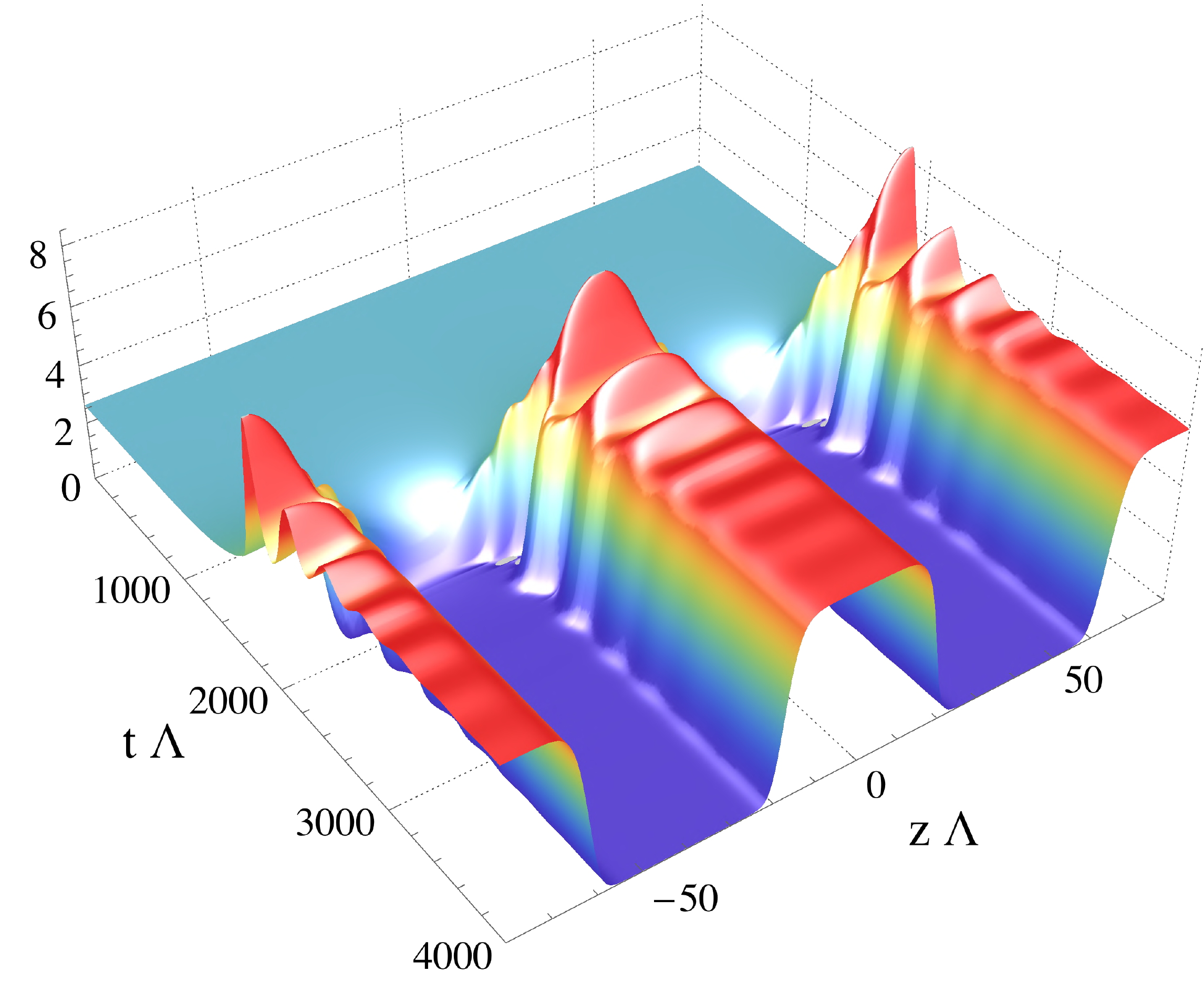} 
\put(-225,148){\mbox{\small{$10^2$ $\mathcal{E}/\Lambda^4$}}}
\end{tabular}
\end{center}
\vspace{-5mm}
\caption{
\label{what} 
\small  (Left) Energy density of the evolution of an initial state with 
\mbox{$\Ezero=\E_3, L\Lambda \simeq 187$} and initial mode $n=5$. 
(Right) Energy density of the evolution of an initial state with 
\mbox{$\Ezero =\E_1, L\Lambda \simeq 160$} and initial mode $n=2$.
}
\vspace{7mm}
\end{figure*}

In non-generic situations the result at the end of the reshaping period may be an almost-static configuration. Typically this happens when a single mode (and multiples thereof) completely dominates the  configuration. The reason for this is that the positions of the maxima and minima of a single Fourier mode are time-independent. An example of this kind of situation with a dominant  $n=3$ mode is illustrated by Fig.~2 of \cite{Attems:2017ezz}, and two further examples with $n=5$ and $n=2$ are shown in \fig{what}. Although these configurations seem static at late times on the scale shown in these plots, they are actually not, as we will now see.

\begin{figure*}[h!!!]
	\begin{center}
		\begin{tabular}{c}
			\hspace{-4mm}
			\includegraphics[width=.9\textwidth]{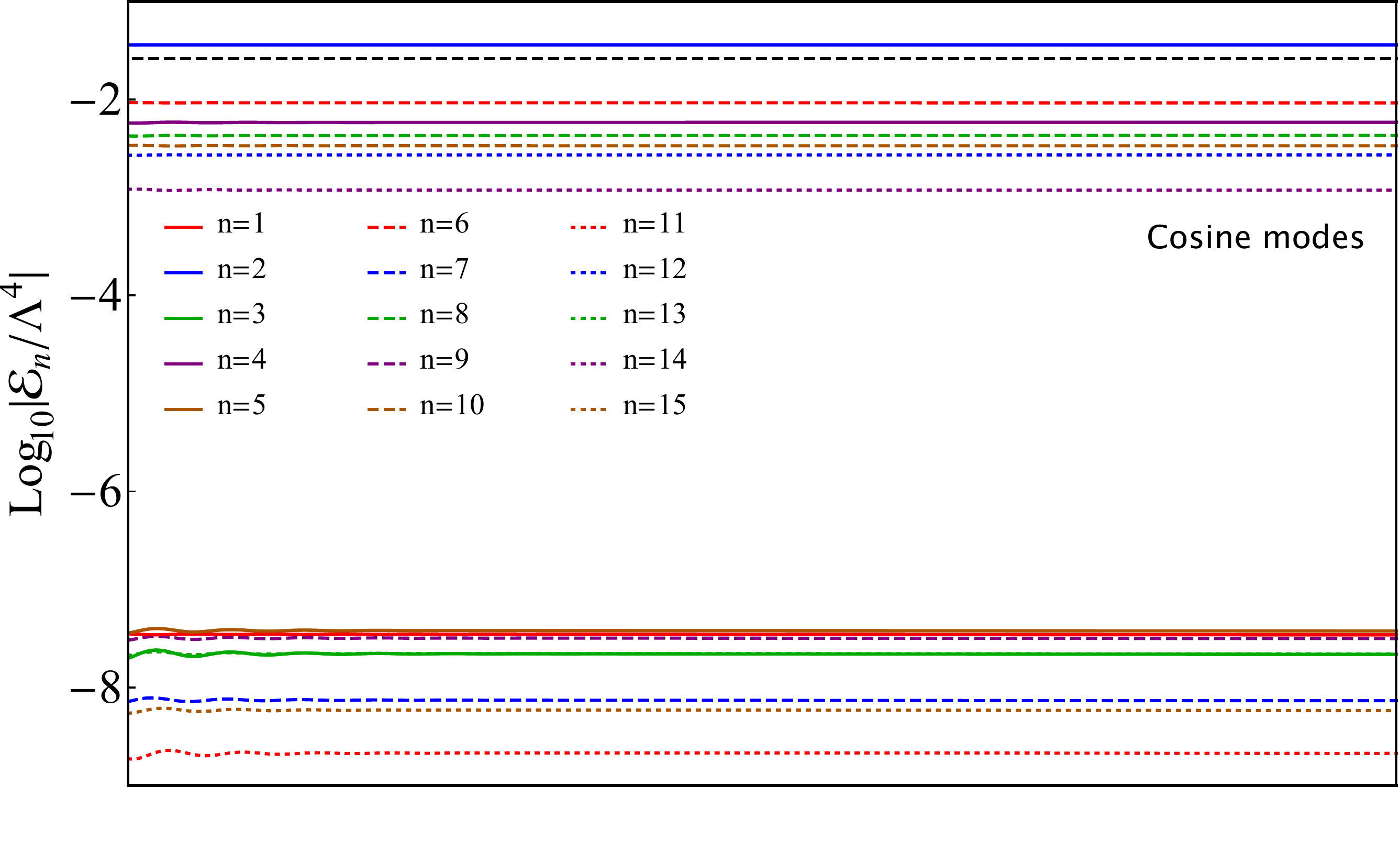} 		
		\\[-0.041\textwidth]
			\includegraphics[width=.9\textwidth]{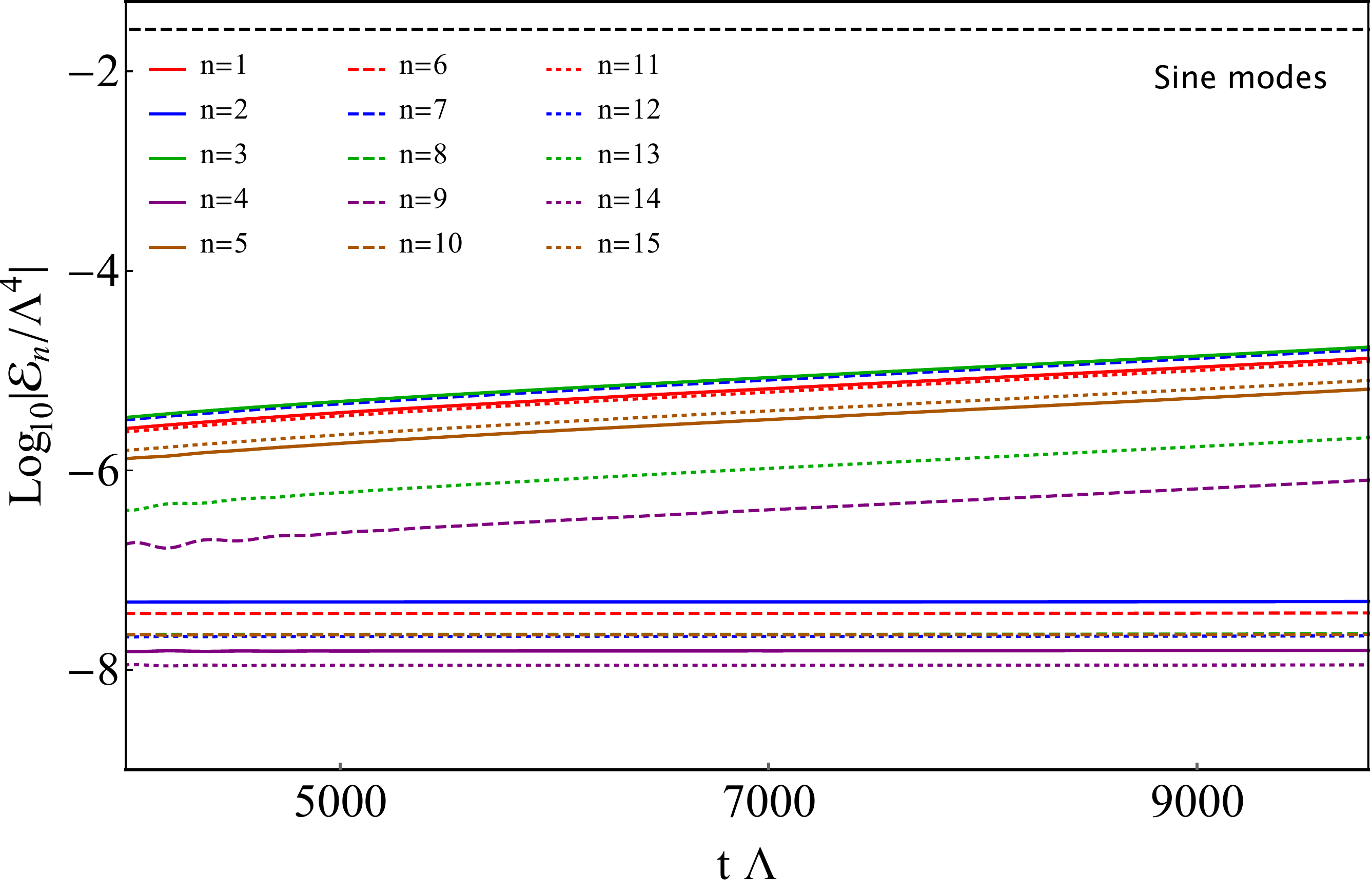} 
			\hspace{0.0093\textwidth}
				\end{tabular}
	\end{center}
	\vspace{-5mm}
	\caption{
		\label{decom1} 
		\small 
		Late time evolution of some Fourier modes of the energy density shown in \fig{what}(right). The black, dashed horizontal lines at the top indicate the average energy density or, equivalently, the $n=0$, constant mode. 
	}
	\vspace{7mm}
\end{figure*}

\begin{figure*}[t!!!]
	\begin{center}
		\begin{tabular}{c}
			\hspace{-4mm}
			\includegraphics[width=.9\textwidth]
			{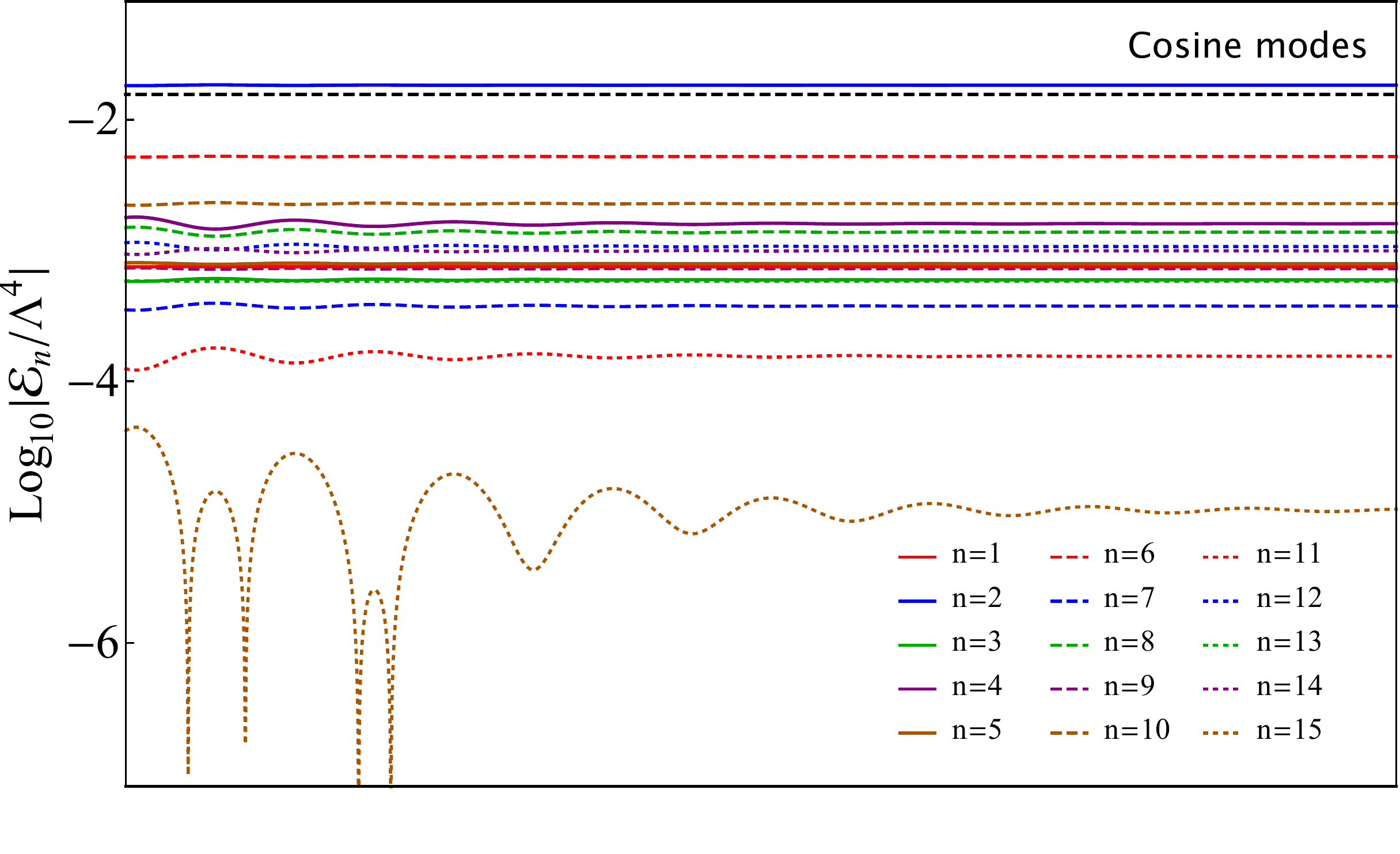} 			
			\\[-0.041\textwidth]
			\includegraphics[width=.9\textwidth]{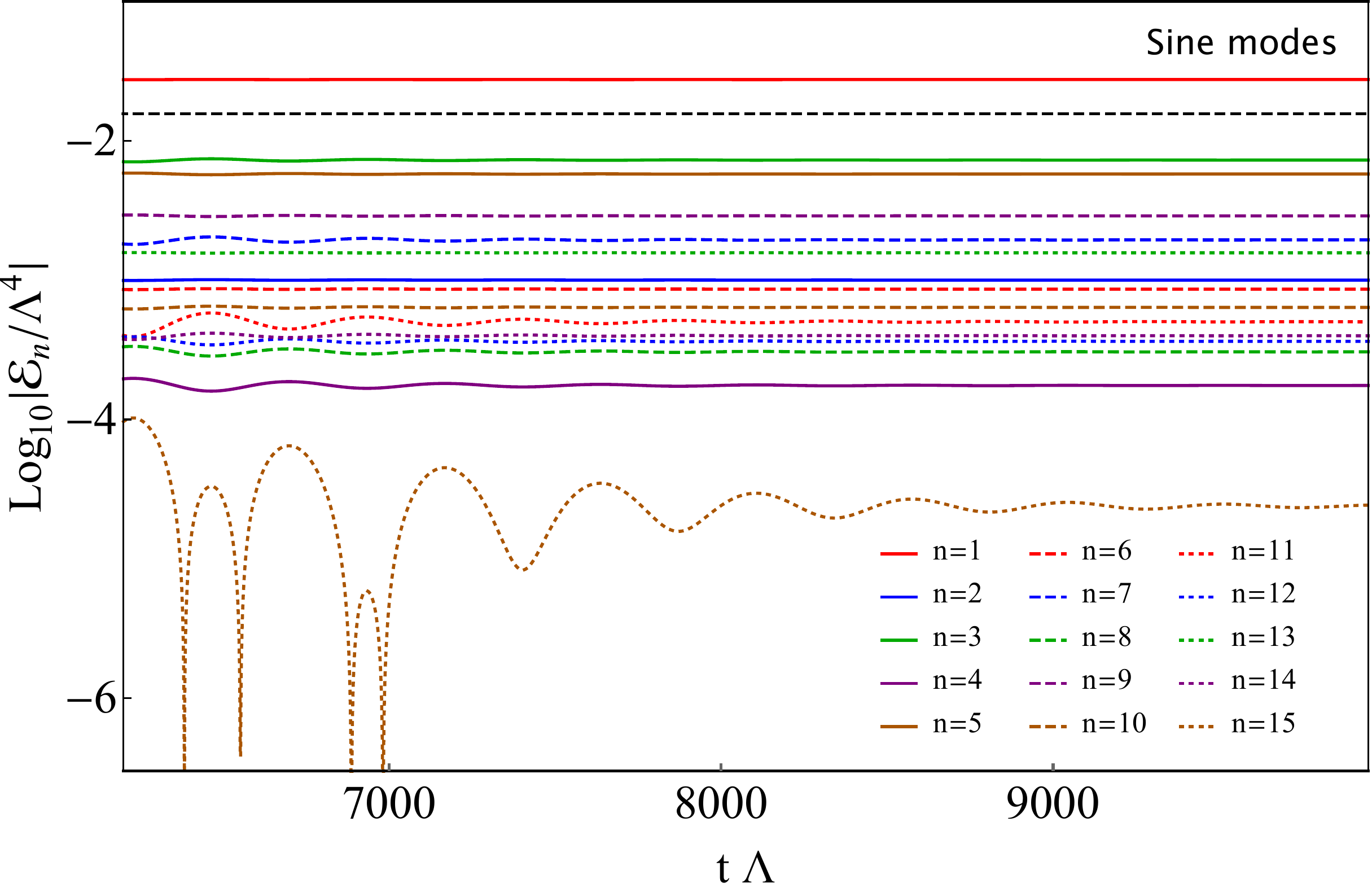} 
			\hspace{0.0093\textwidth}
				\end{tabular}
	\end{center}
	\vspace{-5mm}
	\caption{
		\label{decom2} 
		\small 
		Late time evolution of some Fourier modes of the energy density shown in \fig{3Denergy}(left). The black, dashed horizontal lines at the top indicate the average energy density or, equivalently, the $n=0$, constant mode. 
	}
	\vspace{7mm}
\end{figure*}

Consider first the case of the simulation of \fig{what}(right), whose Fourier modes at very late times are shown in \fig{decom1}.
We see that the cosine modes approach constant values at late times whereas some sine modes, although very small in amplitude, are growing exponentially with growth rate 
\be
\label{raterate}
\gamma \simeq 2.5 \times 10^{-4}\Lambda\,.
\ee
Note that this  is two orders of magnitude smaller than the typical growth rates of the unstable modes around the initial homogeneous state (see \fig{comparison}), which is the reason why the late-time part of \fig{what}(right) looks approximately static.  This situation should be contrasted with that of  phase-separated configurations. For example, the Fourier modes at very late times of the simulation of \fig{3Denergy}(left) are  displayed in \fig{decom2}. In this case  all modes approach constant values, showing that this configuration becomes truly static at late times.\footnote{We have checked on the gravity side that at asymptotically late times the vector field $\partial_t$ becomes both Killing and hypersurface-orthogonal. In other words, the configuration at late times is truly static and not just stationary.}
Since this configuration is also periodic,  by taking multiple copies in a bigger box we conclude that truly static configurations with multiple, equally spaced domains also exist (recall the discussion in \sect{box}), and the same is true for multi-peak configurations. Our analysis then suggests that all multi-domain configurations are actually in unstable equilibrium towards merging into a single domain. Presumably on the gravity side this is due to the gravitational attraction between the high-energy regions of the horizon at different points on the circle. 

The physics of the simulation in \fig{what}(right) can therefore  be understood as follows. The system starts in a homogeneous state perturbed by an unstable $n=2$ mode. The initial amplitude of this mode is very small compared to the homogeneous, $n=0$ mode, but very large compared to any other mode, including of course numerical noise. 
Therefore  time evolution initially takes the system very close to a static configuration with two domains on antipodal points of the $z$-circle. Note that in the truly static configuration all the sine modes would be exactly zero by symmetry. However, although they are very small, they are non-zero in our case because they are generated by numerical noise in the initial, homogeneous state. These modes can thus be viewed as perturbations of the exactly static configuration with two antipodal domains. Since some of these modes are unstable,  they drive the system away from the antipodal configuration. 

\begin{figure*}[t]
\hspace{-6mm}
\includegraphics[width=.5\textwidth]
{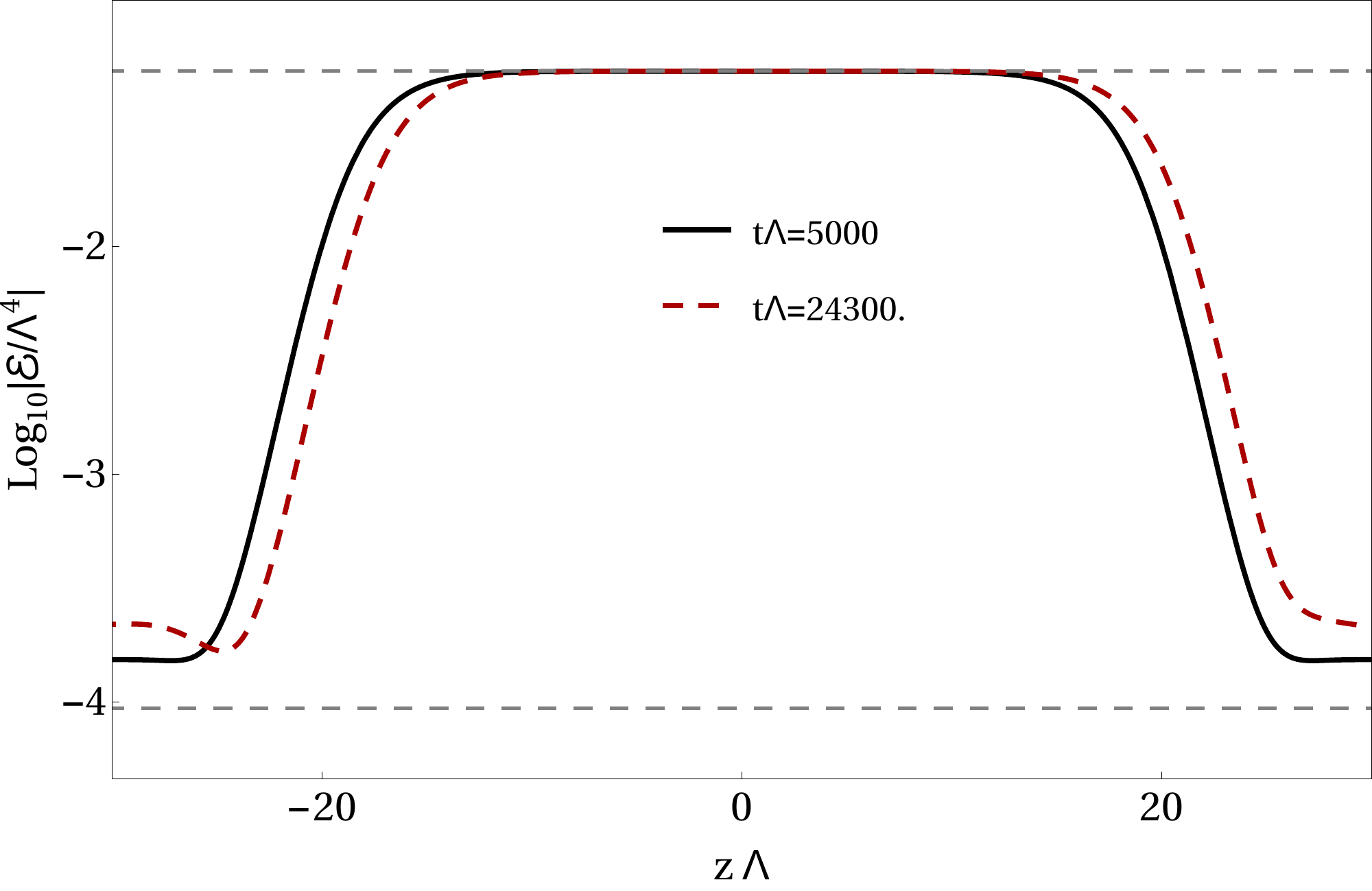}
\quad\quad
\includegraphics[width=.5\textwidth]
	{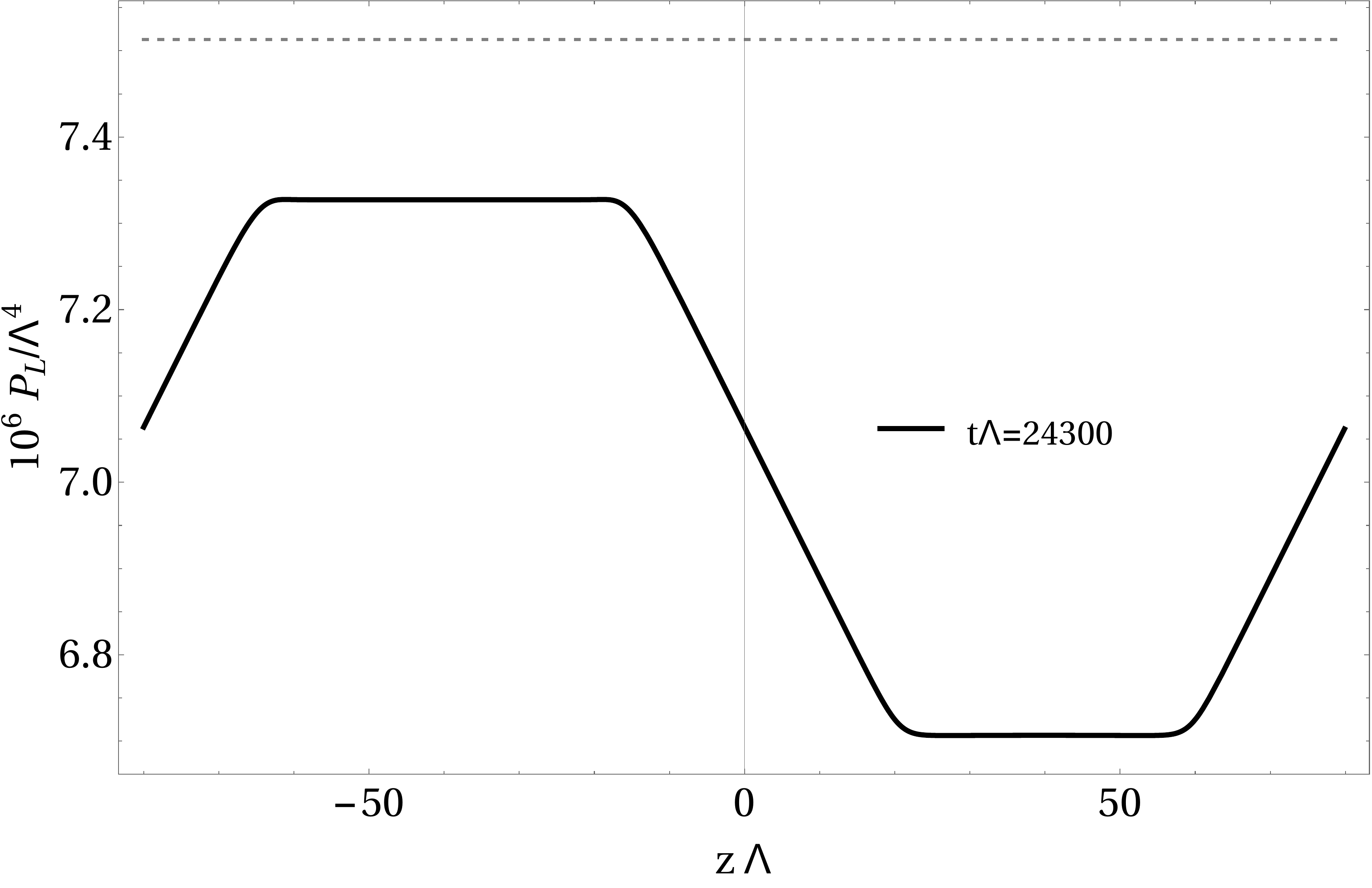}
\vspace{-5mm}
\caption{\label{push}
\small  Snapshots for the simulation of  \fig{what}(right).
	(Left) Plot of the energy density at two very late times
	The top and bottom  dashed, grey, horizontal lines correspond to $\Ehigh$ and $\Elow$, respectively. 
\small (Right) Plot of the longitudinal pressure at a very late time. The  dashed, grey, horizontal line is $P_c$, the equilibrium pressure at $T_c$.  
}
\vspace{7mm}
\end{figure*}

\begin{figure*}[h!!!]
	\begin{center}
			\hspace{-4mm}
			\includegraphics[width=.63\textwidth]
			{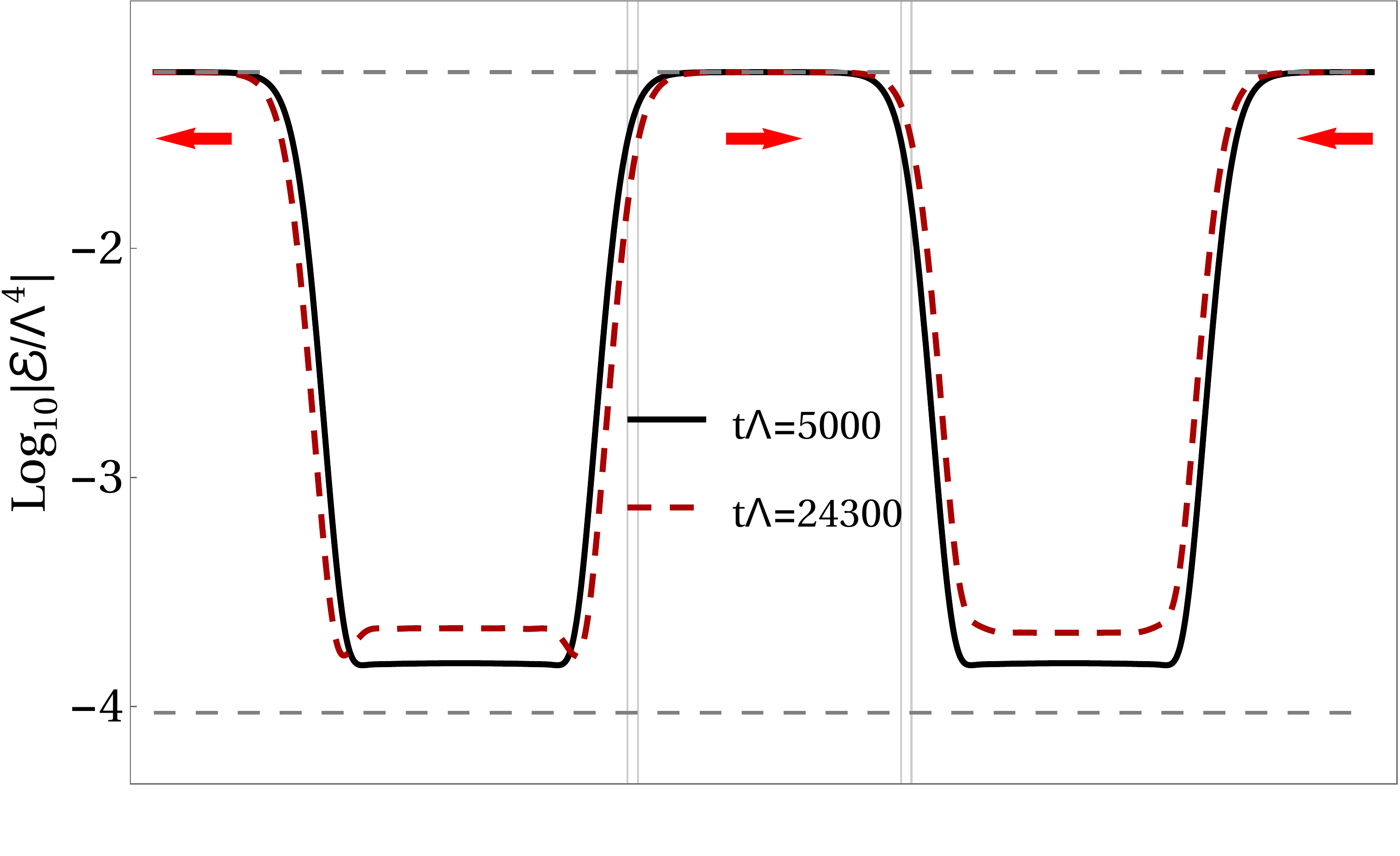}
			\\[-0.027\textwidth]
			\includegraphics[width=.63\textwidth]
			{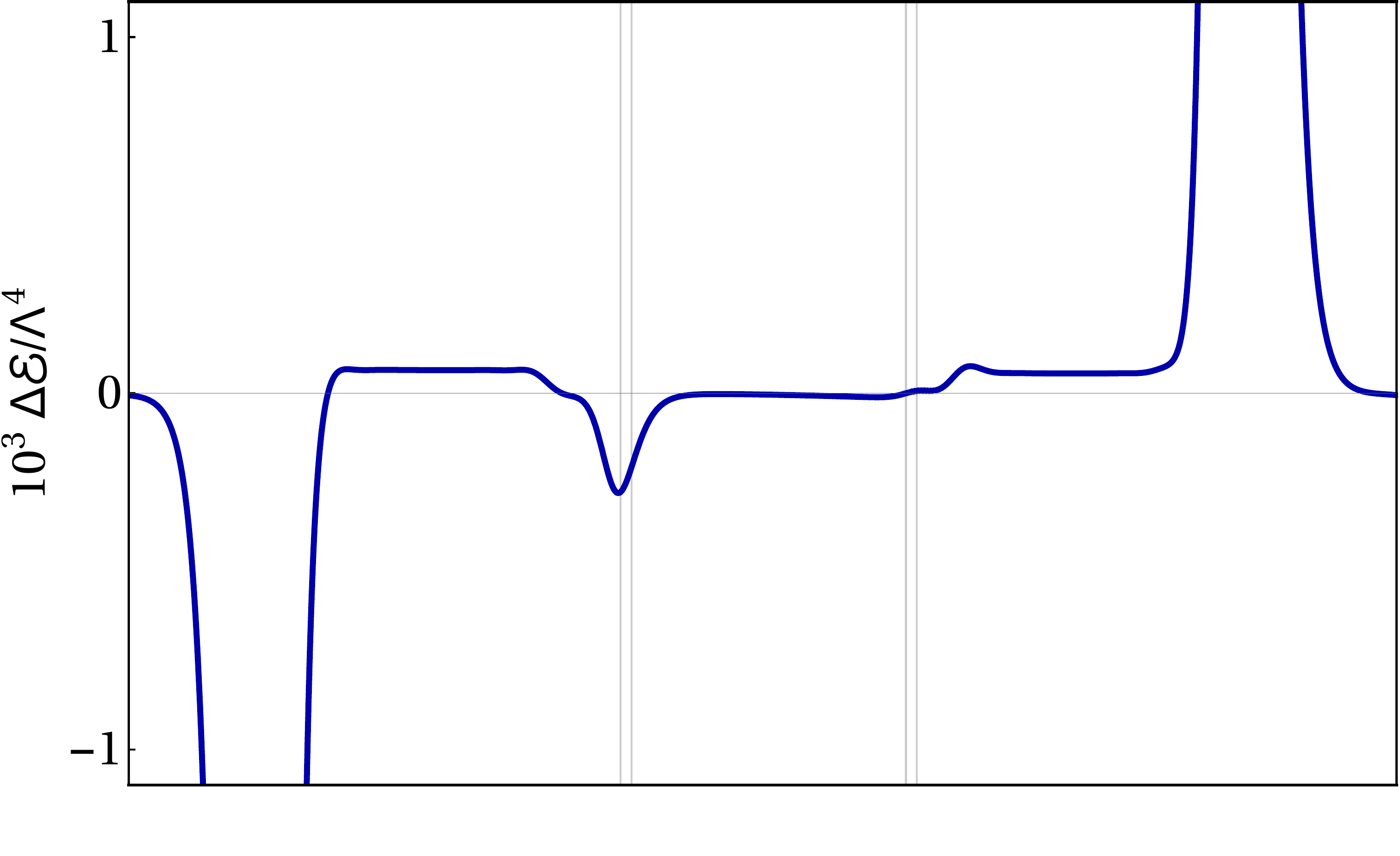}
			\hspace{0.0093\textwidth}
			\\[-0.027\textwidth]
			\includegraphics[width=.63\textwidth]
			{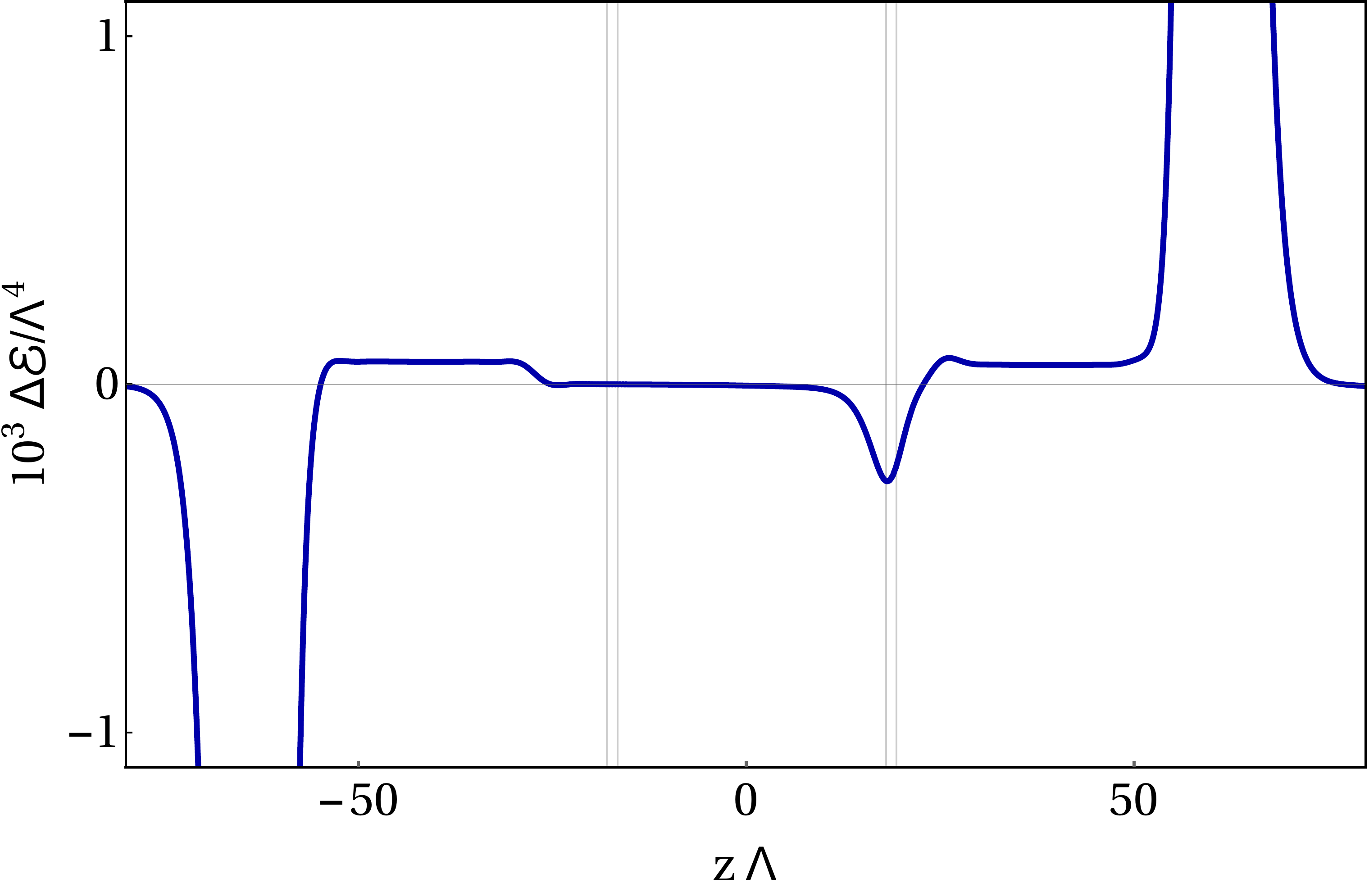}
			\hspace{0.0093\textwidth}
	\end{center}
	\vspace{-5mm}
	\caption{
		\label{arrows} 
		\small 
		(Top) Snapshots of the energy density at two very late times for the simulation of  \fig{what}(right). The top and bottom  dashed, grey, horizontal  lines correspond to $\Ehigh$ and $\Elow$, respectively.  The arrows indicate the direction of motion of each domain. (Middle) Difference between the  energy densities  in (top), $\E(t \Lambda =24300) - \E(t \Lambda =5000)$, having shifted the red, dashed curve to the left by an amount $\Delta z - \Delta \ell/2$ (see main text). (Bottom) Same difference between the two energy densities in (top) but now having shifted the red, dashed  curve to the left by an amount $\Delta z + \Delta \ell/2$ (see main text). In all three rows the grey, vertical lines indicate the inflection points of the interfaces of the central domain. 
	}
	\vspace{5mm}
\end{figure*}

The evolution  proceeds by moving the two domains towards each other, by simultaneously compressing them while keeping the shape of the interfaces fixed, and by increasing the energy density in the low-energy regions. To illustrate this in \fig{push}(left) we plot the energy profiles of the central domain at two late and widely separated times. 
The small but appreciable relative displacement between the two curves shows that the central domain is moving towards the right. A similar plot shows that the other domain is moving towards the left, as indicated by the arrows in \fig{arrows}(top). 
The compression of the domains and the rigidity of the interfaces are illustrated by the middle and bottom rows of \fig{arrows}, which are produced as follows. In the time interval between the two times shown in \figs{push}(left) and \ref{arrows}(top) the central domain moves to the right an amount $\Lambda \Delta z= 1.380$ (defined as the average motion of the two interfaces) and the size of the domain decreases by an amount 
$\Lambda \Delta \ell= 0.0264$ (defined as the relative motion of the two interfaces). Thus to produce \fig{arrows}(middle) we shift the red, dashed  curve to the left by an amount 
$\Delta z - \Delta \ell/2$, so that the inflection points of the  interfaces on the right-hand side of the central domain are on top of one another. Then we plot the difference between the shifted curve and the continuous, black curve. We see that the result vanishes in the centre of the domain and also on the location of the right interface. This shows that the value of the energy density in the centre of the domain has remained constant and that the shape of the right interface has not changed. The fact that the result is negative at the location of the left interface shows that the domain has decreased in size. To produce \fig{arrows}(bottom) we repeat the procedure except that we shift the red, dashed  curve to the left by an amount  $\Delta z + \Delta \ell/2$, so that now it is the left inflection points of the central domain that fall on top of one another. The result shows that the left interface has also moved with constant shape. We obtain analogous results for the other domain. Since both domains decrease in size  the energy that they carry also decreases. \figs{push}(left) and \ref{arrows}(top) show that this excess energy is transferred to the low-energy region between the domains, whose energy density clearly increases as the domains approach each other. Due to the different velocities of the interfaces the cold region shape differ between being squeezed or opened up, where minimas appear. The direction of motion of the domains is consistent with the fact that the longitudinal pressure is lower in the low-energy region towards which the domains are moving than in the region that they leave behind, as shown in \fig{push}(right). The minimum of the longitudinal pressure deviates by more than $10\%$ from the critical pressure, $P_c$. Mechanically speaking, the high
pressure in one region pushes the domains towards the low-pressure region. 

The picture above  is also consistent with several features of the unstable modes in \fig{decom1}. First, all cosine modes are stable, because adding a cosine perturbation to the antipodal configuration does not displace the domains towards each other but instead changes the distribution of energy between the two domains. While a large perturbation of this type could potentially take the system towards a single-domain configuration, this does not seem to lead to an instability at the linear level. Second, sine modes with even mode numbers are also stable. In this case this is due to the fact that a perturbation of this type shifts the position of the antipodal points simultaneously in the same direction and hence does not change the relative distance between them. Third, all the sine modes with odd mode number grow exponentially with the same growth rate \eqq{raterate}. The reason why these modes are unstable is that they do change the relative distance between the domains. The reason why they all grow with the same growth rate is a consequence of the rigid motion of the two domains.

\begin{figure*}[t!!!]
	\begin{center}
		\begin{tabular}{c}
			\hspace{-4mm}
			\includegraphics[width=.9\textwidth]
			{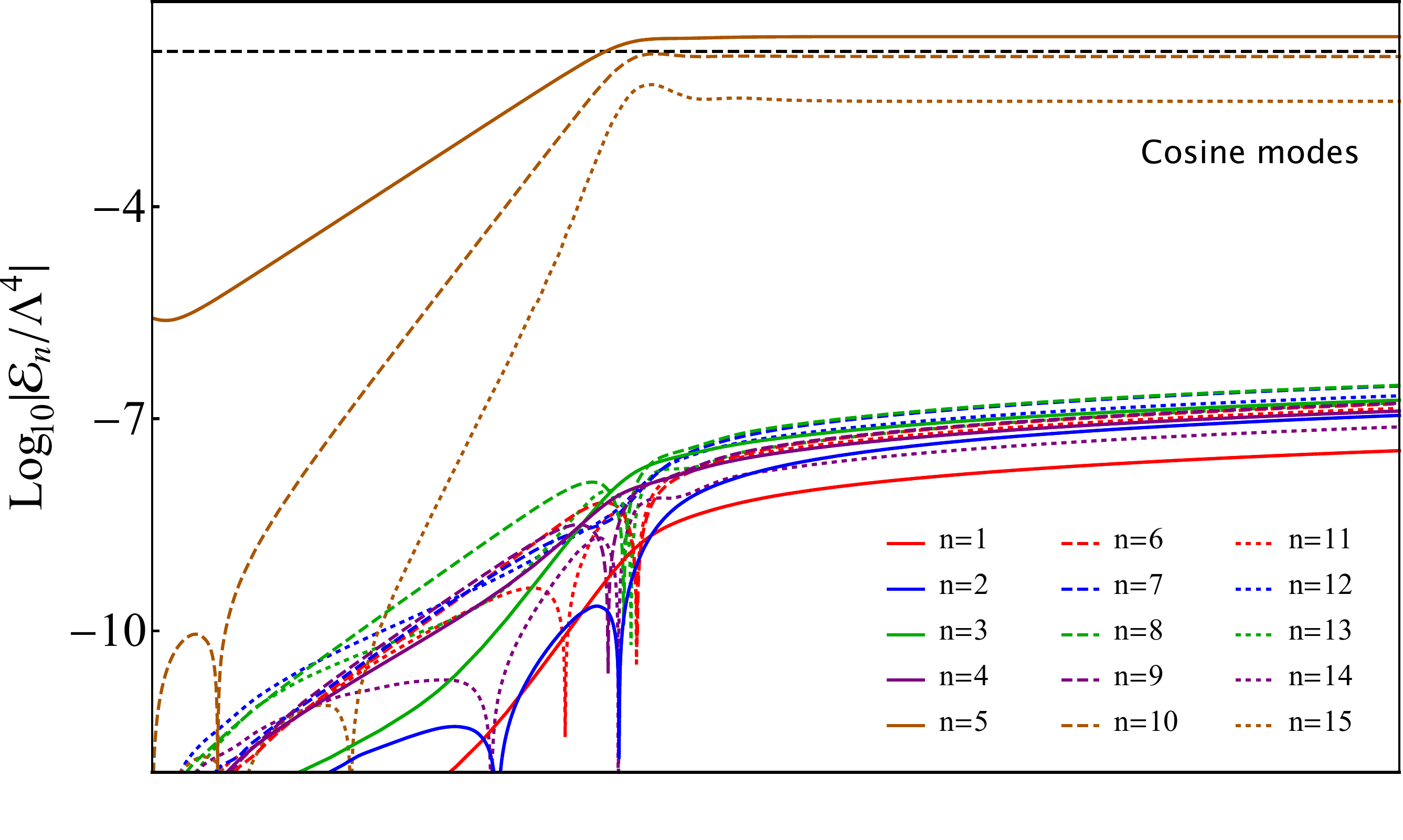} 			
			\\[-0.041\textwidth]
			\includegraphics[width=.9\textwidth]{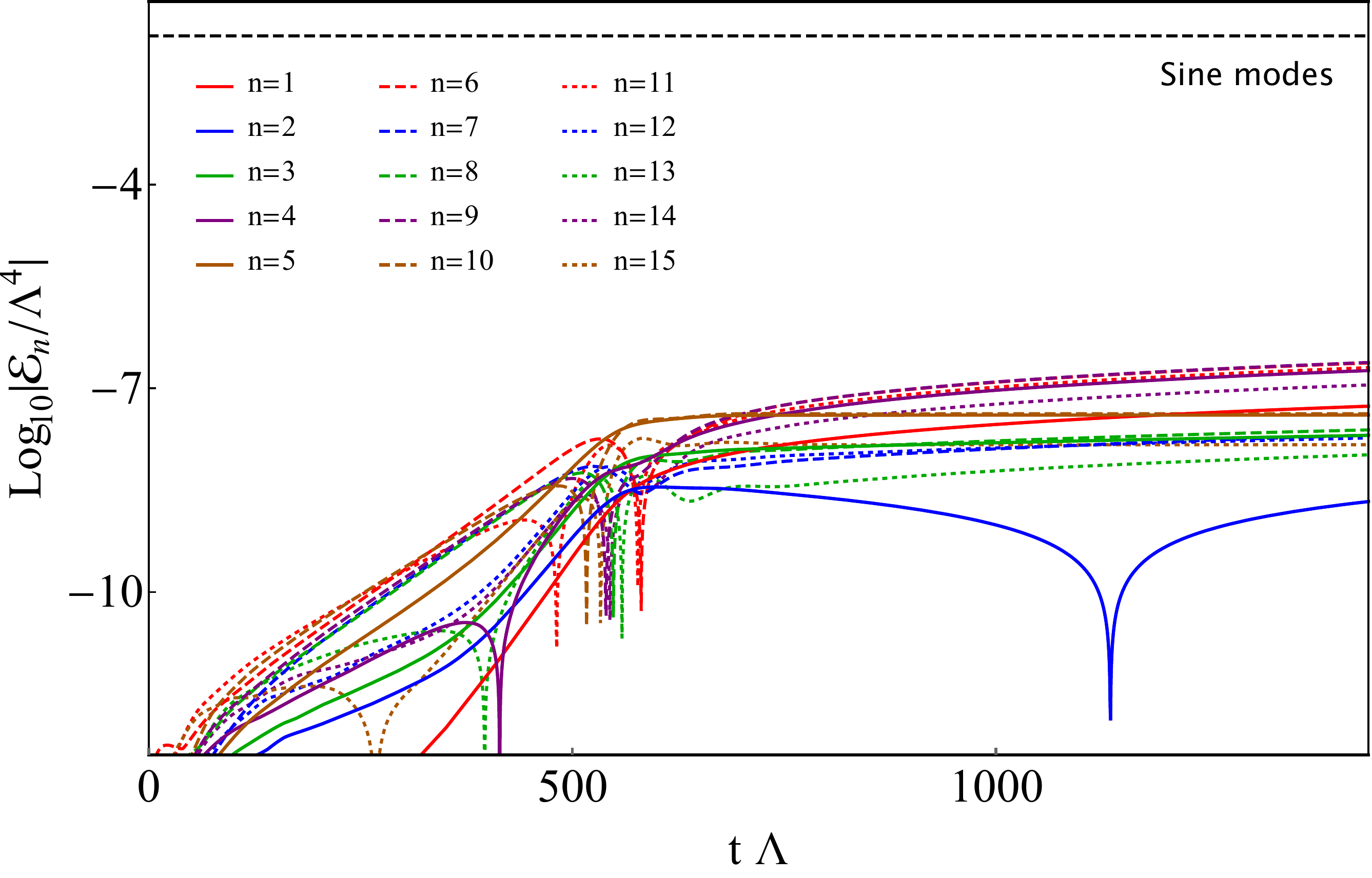} 
			\hspace{0.0093\textwidth}
				\end{tabular}
	\end{center}
	\vspace{-5mm}
	\caption{
		\label{decom3} 
		\small 
		Late time evolution of some Fourier modes of the energy density shown in \fig{what}(left). The black, dashed horizontal lines at the top indicate the average energy density or, equivalently, the $n=0$, constant mode. 
	}
	\vspace{7mm}
\end{figure*}

\begin{figure*}[t]
\begin{center}
\includegraphics[width=.7\textwidth]{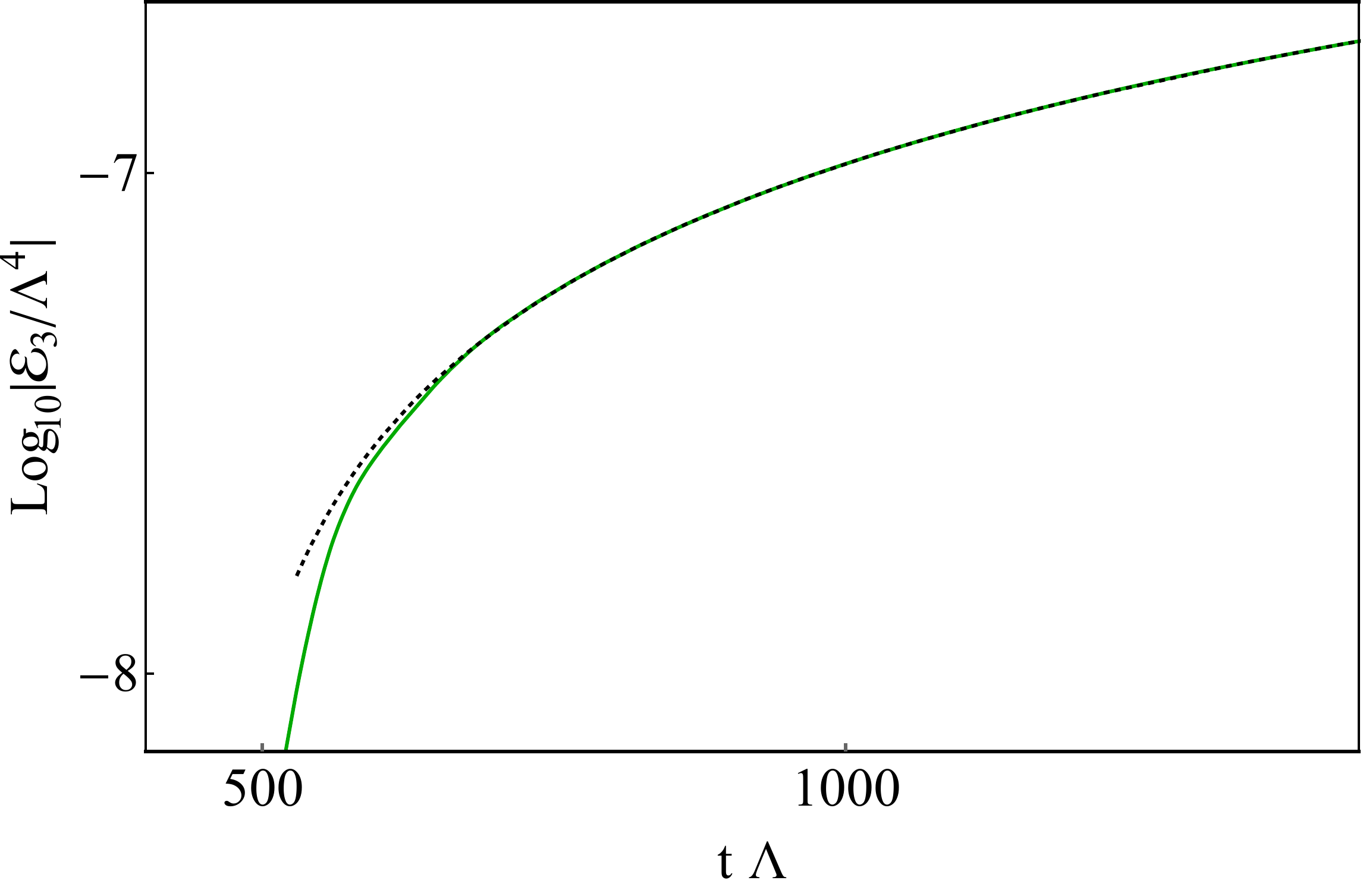} 
\end{center}
\vspace{-5mm}
\caption{\label{fitfit} 
\small Fit \eqq{fiteq} to the time evolution of the $n=3$ cosine mode shown in \fig{decom3}. 
}
\vspace{7mm}
\end{figure*}

We now turn to \fig{what}(left), whose Fourier modes are shown in \fig{decom3}. 
Since the modes are a combination of one growing and one decaying exponential and the times are not  long enough, the unstable modes  do not yet appear in the figure as straight lines. For example, \fig{fitfit} shows a fit to the unstable, $n=3$ cosine mode of the form
\be
\label{fiteq}
10^7 \times \frac{{\E}_3}{\Lambda^4} \, \simeq \,
a_- e^{\gamma_-  t}  +
a_+ e^{\gamma_+  t}
\ee 
with
\be
a_-= -4.17  \sac a_+= 3.26  \sac \gamma_-= -3.5 \times 10^{-4} \Lambda
\sac \gamma_+ = 2.0 \times 10^{-4} \Lambda \,.
\ee
Note that the growth rates are of the same order of magnitude as \eqq{raterate}. 

\begin{figure*}[t]
\begin{center}
\includegraphics[width=.7\textwidth]
	{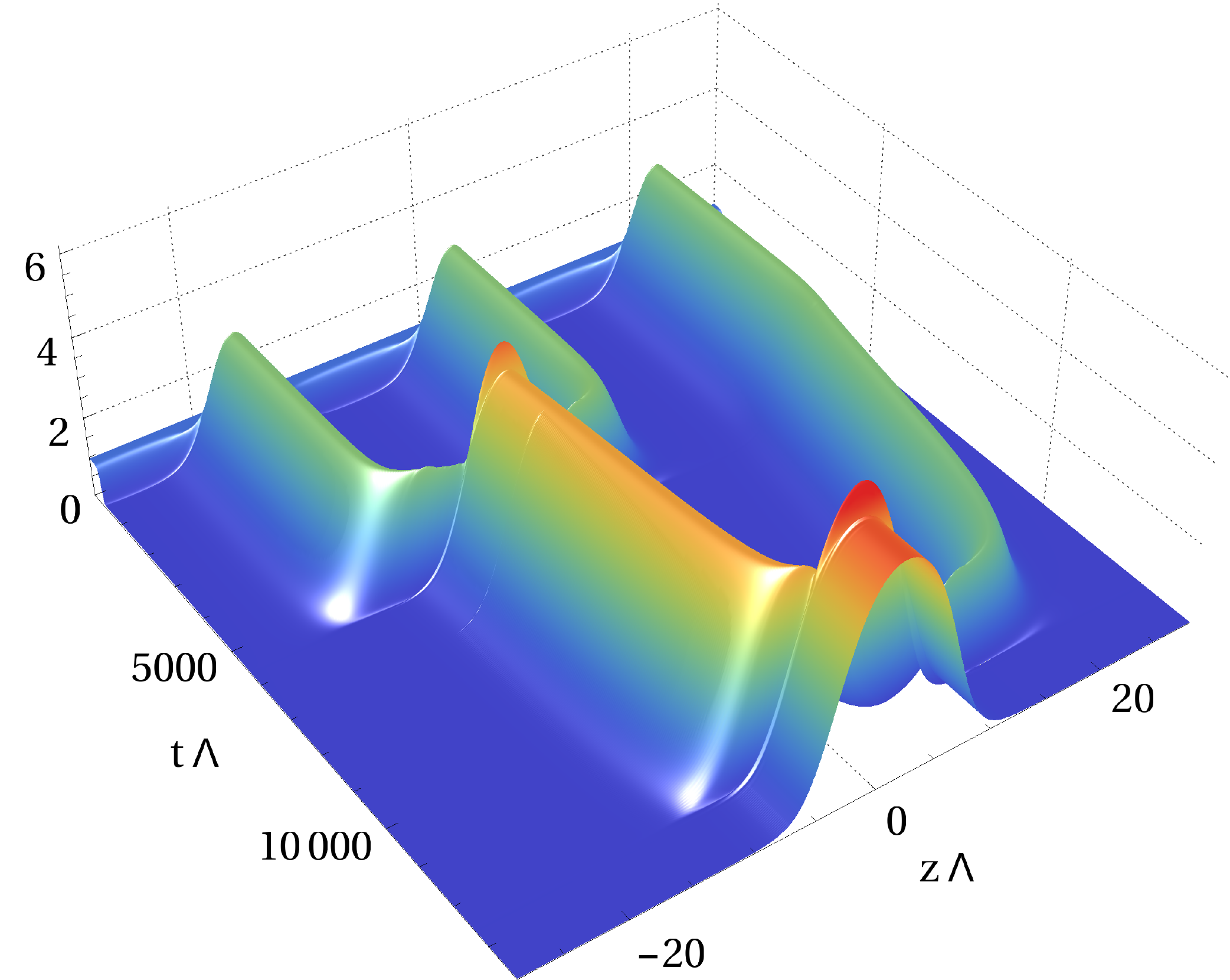}
	\put(-315,190){\mbox{\small{$10^2$ $\mathcal{E}/\Lambda^4$}}}
\end{center}
\vspace{-5mm}
\caption{\label{triplepeak} 
	\small Evolution up to $t \Lambda = 13790$ of the energy density of the initial state considered in \cite{Attems:2017ezz}, which has \mbox{$\Ezero =\E_5, L\Lambda \simeq 57$} and initial mode $n=3$.
}
\vspace{7mm}
\end{figure*}

Fig.~2 of ref.~\cite{Attems:2017ezz} shows an $n=3$ analog of the $n=5$ and $n=2$ runs of Fig.~\ref{what}, namely a seemingly static, triple-peak configuration. The exactly static configuration has been constructed by directly solving a manifestly static problem in \cite{private}. \fig{triplepeak} shows that, contrary to what was stated in \cite{Attems:2017ezz}, this static configuration is unstable, as expected from the  discussion in this section. Indeed, on the scale shown in the figure, the triple-peak configuration appears static only in the range of times 
$300 \lesssim t \Lambda \lesssim 4500$. At $t \Lambda \sim 5500$ the first and second peaks merge, while the third peak is slowly moving left. At  $t \Lambda \sim 12500$ all three peaks have merged into a single peak. The instability of the triple-peak configuration was missed in \cite{Attems:2017ezz} because that reference only explored times of order $t\Lambda \lesssim 500$.

In summary, we conclude that the simulations shown in \fig{what} are  slowly evolving but not static because the corresponding static configurations are unstable. Therefore for sufficiently long times the different structures will presumably merge. The configurations in \fig{what} are dominated by an $n=5$ and $n=2$ mode, respectively. We have performed an analogous analysis for configurations with $n=3$, which is the case of  Fig.~2 of 
\cite{Attems:2017ezz}, and with $n=4, 6, 7$. In all cases the conclusion is the same. This leads us to  conjecture that, given a fixed-size box (recall the discussion in \sect{box}), the only stable configurations are those with a single structure. In particular, all static, non-phase separated configurations in large enough boxes should be dynamically unstable. For large enough boxes the only stable states should be phase-separated configurations, which we will study in detail in \sect{phaseseparationsection}, whereas for smaller boxes they would correspond to configurations with a single peak. 

We close this section by pointing out that the conclusion about the instability of multi-peak configurations is outside the scope of the Gubser-Mitra conjecture \cite{Gubser:2000ec,Gubser:2000mm}. Recall that this states that a black brane with a non-compact translational symmetry is classically stable if, and only if, it is locally thermodynamically stable. The non-compactness assumption is violated since the $z$-direction is periodically identified, and the translational symmetry assumption is violated because the multi-peak configurations are inhomogeneous. The system is still translationally invariant along the transverse directions, but these are simply spectator directions and one could periodically identify them, thus making them also compact.

\subsection{Domain relaxation}
\label{relaxation}

After two structures merge and form a phase domain, or after the latter  forms directly at the end of the reshaping period, the domain oscillates and relaxes to equilibrium. Although the merger is a non-linear process, we will show that the subsequent relaxation can be very well described by linear theory, in particular by the linear sound mode perturbations around the high-energy phase. This may seem surprising given that the full configuration is inhomogeneous. 
However, as we have already seen in \fig{atte} and as we will further confirm below, the interfaces at the end of the domain behave as rigid walls, thus effectively confining the oscillations to the interior of the domain. 

\begin{figure*}[t]
\begin{center}
\begin{tabular}{cc}
\includegraphics[width=.46\textwidth]{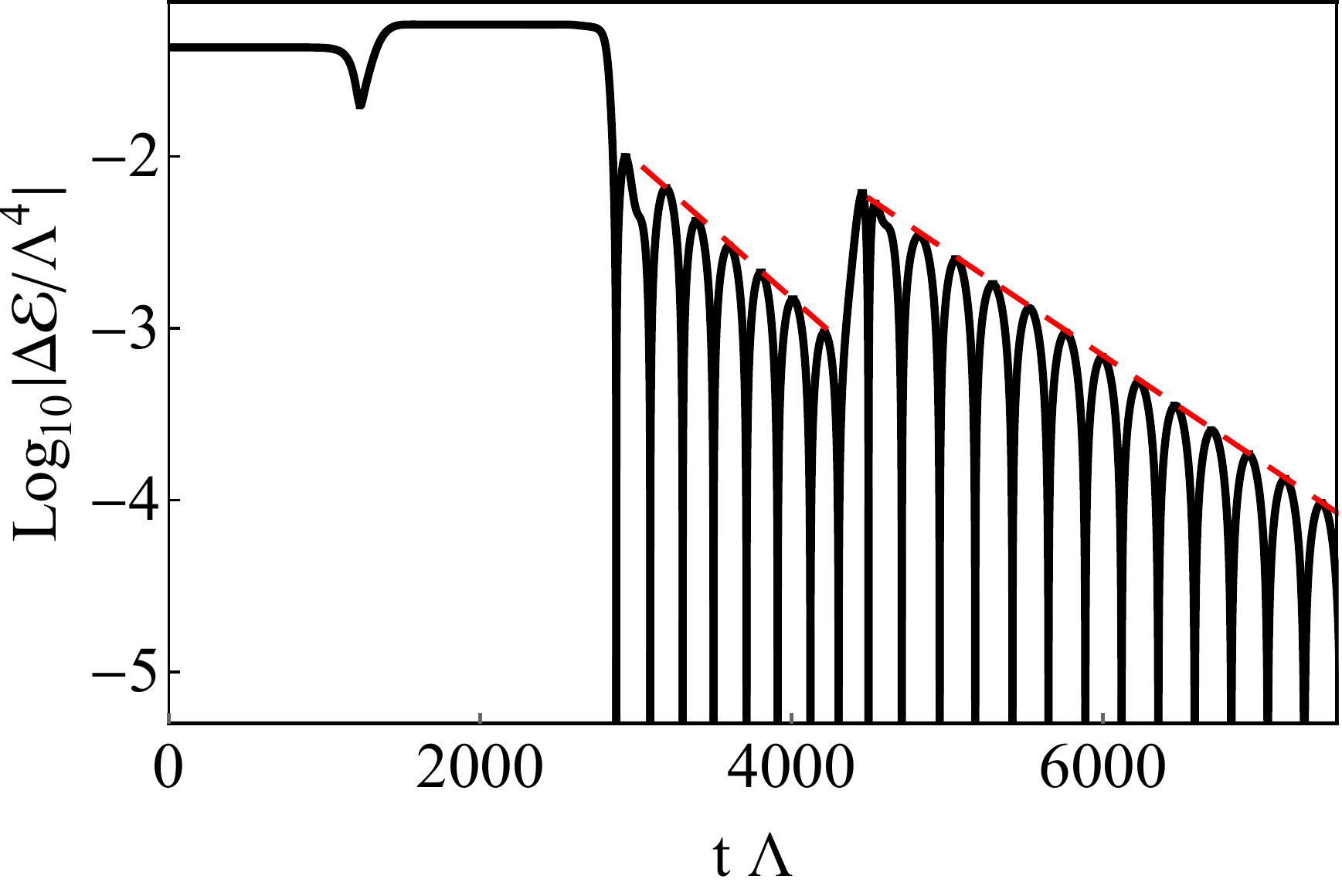} 
\quad&\quad
\includegraphics[width=.46\textwidth]{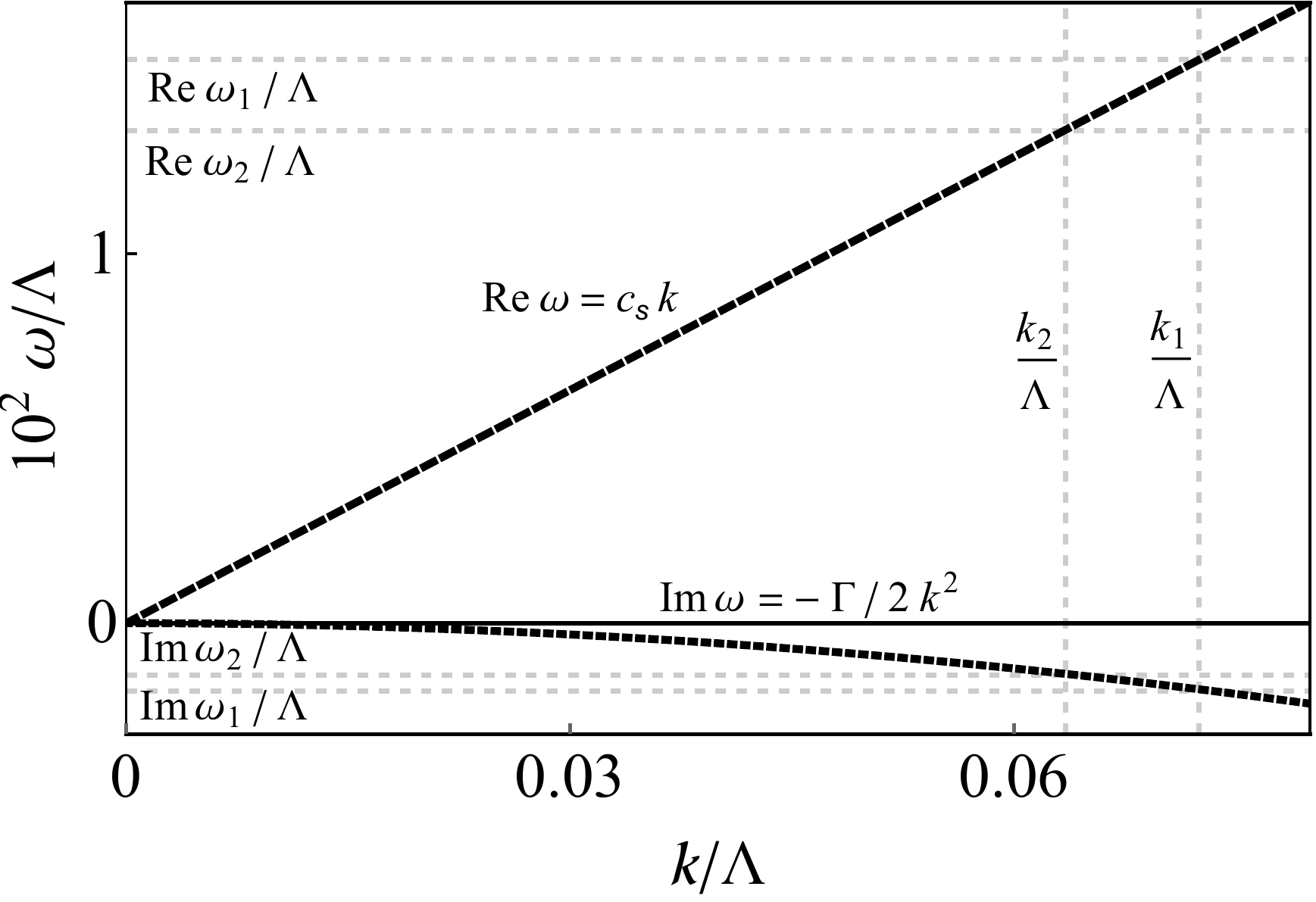} 
\end{tabular}
\end{center}
\vspace{-5mm}
\caption{\label{dampe} 
\small (Left) The black curve shows the difference  
$\Delta \mathcal{E}=\mathcal{E}-\Ehigh$  on a slice at constant $z\Lambda=46$ in \fig{3Denergy}, which corresponds to the center of the final domain. The slopes of the fitted straight lines (in dashed red) determine $\rm{Im} \, \omega_1$ and  
$\rm{Im} \, \omega_2$, whereas the separation between maxima determine  
$\rm{Re} \, \omega_1$ and $\rm{Re} \, \omega_2$.   
(Right) The two independent values of $k$ obtained from comparing 
$\rm{Re} \, \omega_{1,2}$ and $\rm{Im} \, \omega_{1,2}$ with the real and imaginary parts of \eqq{comp} agree excellently with one another.  
}
\vspace{7mm}
\end{figure*}

\begin{figure*}[t]
\begin{center}
\begin{tabular}{cc}
\includegraphics[width=.46\textwidth]{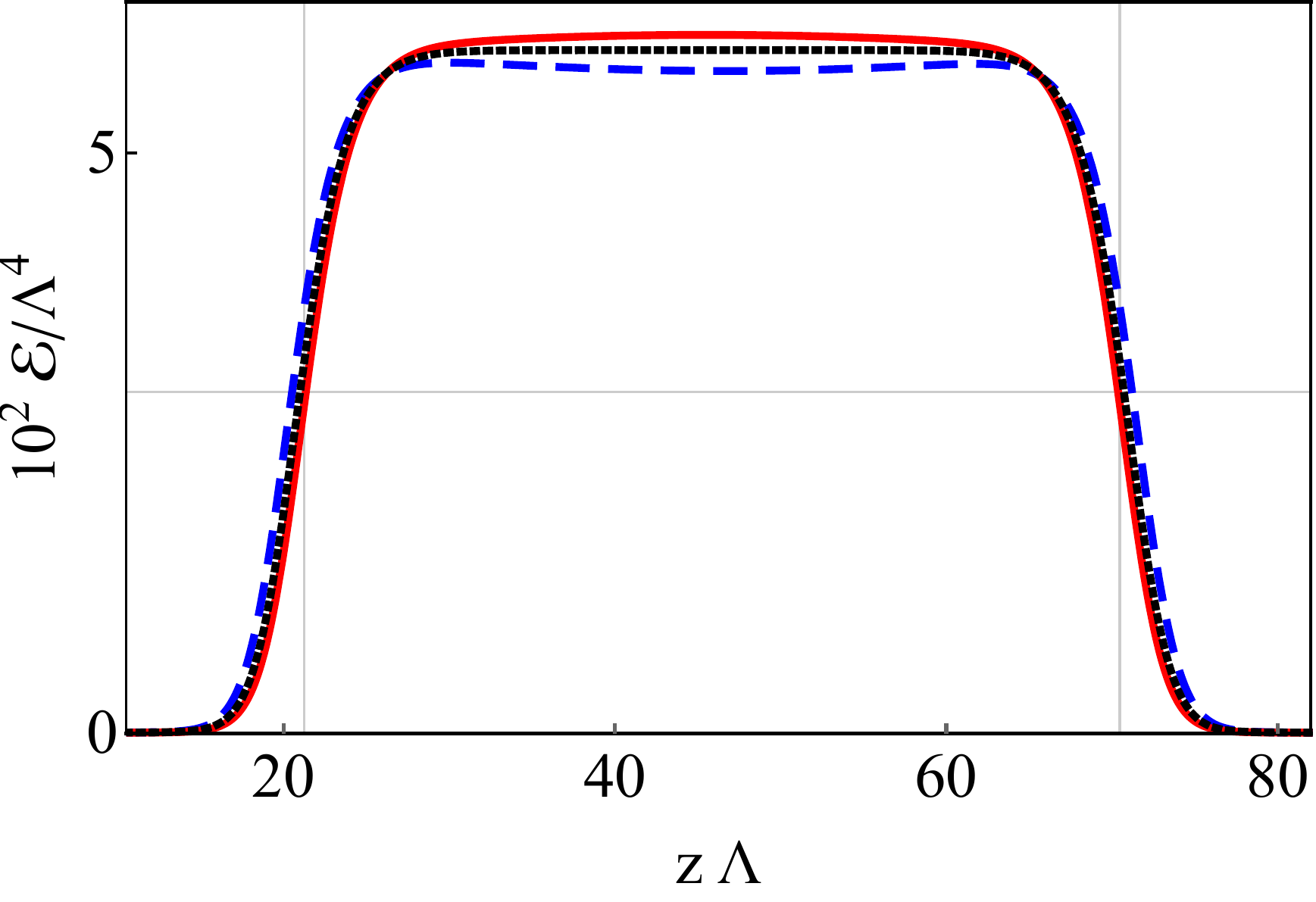} 
\quad&\quad
\includegraphics[width=.46\textwidth]{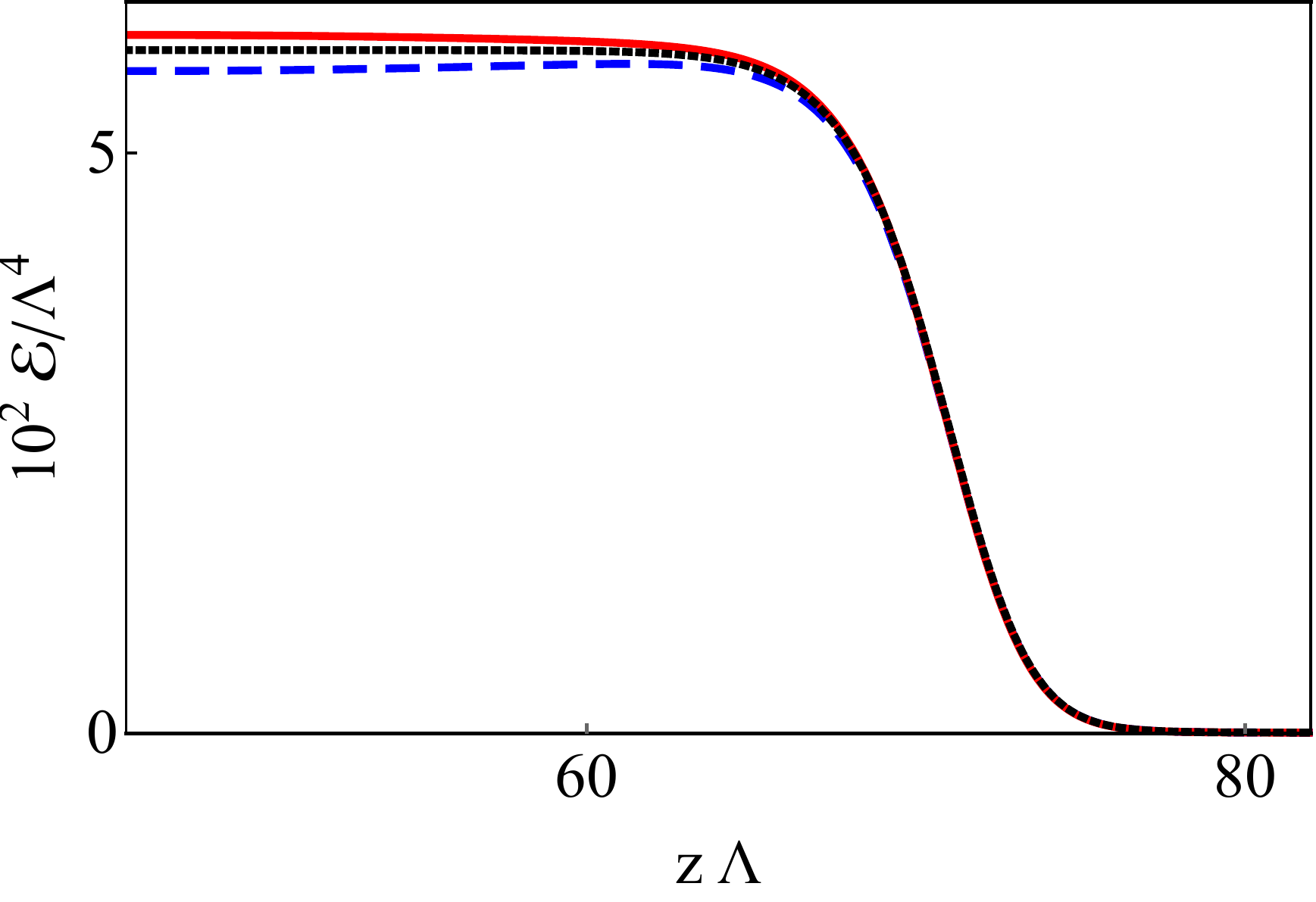} 
\end{tabular}
\end{center}
\vspace{-5mm}
\caption{\label{size} 
\small (Left) Late-time oscillations of the final domain in \fig{3Denergy} once the size of the perturbation created by the last merger has become comparable to the size of the domain itself. The distance between vertical lines is half a wavelength, $\lambda_2/2$. The horizontal line is the half of the energy of the hot stable phase. (Right) Same curves as on the left plot shifted by a constant amount in order to illustrate that the interface oscillates rigidly.  
}
\vspace{7mm}
\end{figure*}

Soon after its formation, the relaxation of the domain is controlled by the largest sound mode that fits within it. To see this, in \fig{dampe} we examine the two mergers that take place in \fig{3Denergy}(left) between the central domain and the two peaks that hit  it from the right (i.e.~from larger values of $z$). 
Specifically, in \fig{dampe}(left) we show the time evolution of the energy density at a constant position $z \Lambda=46$, which corresponds to the center of the final domain. We see two regions of exponentially damped oscillations corresponding to the relaxation of the domain after each of the two hits. In each of them we extract the imaginary part of the frequency from the dampening coefficient, i.e.~from the slope of the straight lines in the figure, and the real part of the frequency from the period of the oscillations. In other words, in each region we perform a fit to an amplitude of the form
\be
\mathcal{E}-\Ehigh \propto \exp [-i ( \omega_R + i \omega_I ) t] \, . 
 \ee
We then compare the result to that predicted by the dispersion relation \eqq{disp} expanded to quadratic order:
\be
\omega(k) \simeq \left |c_s \right | k - \frac{i}{2} \Gamma k^2  \,,
\label{comp}
\ee
with $c_s$ and $\Gamma$ evaluated at $\mathcal{E}=\Ehigh$. Once \eqq{comp} is assumed, the comparison of the real and imaginary parts of 
$\omega$ with the values  extracted from the fit determines two independent values of $k$. These agree excellently with one another (within 0.7\%), as illustrated in \fig{dampe}(right), so we will not distinguish between them. Each value of $k$ has an associated wavelength given by 
$\lambda=2\pi/k$. Remarkably, in each case the size of the domain measured between midpoints of the interface agrees almost exactly with 1/2 of the  corresponding $\lambda$. For the second merger this is illustrated in \fig{size}(left), where we see that the vertical lines, which we have drawn at a distance $\lambda_2 /2=50 /\Lambda$ from one another, intersect the interfaces at their midpoints. In \fig{size}(right) we illustrate once more the rigidity of the interfaces by shifting by a constant amount the curves of \fig{size}(left).  This rigidity implies that, effectively, the oscillations obey Dirichlet boundary conditions at the ends of the domain, which is the reason why the size of the domain equals $\lambda/2$ as opposed to $\lambda$.

In the previous analysis we have shown that the oscillations are associated with the longest wave-length sound mode of the high energy phase that fits within the phase domain. This suggests the possibility of describing not just the time dependence of the energy density at the center of the domain but the full spacetime evolution of the oscillations by using the analytical expression for the sound mode. Mathematically, this means that the form of the energy density in the domain should be of the following form:
\be
\label{analyticaloscillations1}
\mathcal{E}(t,z) \simeq \tilde{\mathcal{E}}_\mt{final}(z, \Delta z(t))+ A_0 \, e^{{\rm{Im}} \, \omega(t-t_0)} \, \cos\big[ 
{\rm{Re}} \, \omega (t-t_0)+\psi_0\big] \, \cos\big[k (z-z_0)\big] \,.
\ee
The second term on the right-hand side describes the oscillation in time and space of the sound mode, where $t_0$ is an arbitrarily chosen time, $\psi_0$ is the phase at $t_0$, $z_0$ is the center of the  domain and $A_0$ is the amplitude at $t_0$. We now explain the meaning of the first term on the right-hand side. 

We have observed that the interfaces move rigidly left and right as the phase domain oscillates, see  \fig{size}(right). Physically, this motion is a consequence of the fact that the total energy inside the phase domain remains constant during the oscillations. Thus, when the cosine of the sound mode oscillates downwards, the two interfaces must move outwards to keep the energy constant, and vice versa, as shown in \fig{size}(left). If we call 
$\Delta z(t)$ the displacement of each interface, then mathematically this means that the oscillations happen on top of a domain with size $\lambda/2+2 \Delta z(t)$ whose energy density can be written approximately as 
\be
\tilde{\mathcal{E}}_\mt{final}(z, \Delta z(t)) \equiv \left\{ \begin{array}{lcc}
	\mathcal{E}_\mt{final}(z+ \Delta z(t)) &  \mbox{if}  & z\leq z_0 \,, \\
	\mathcal{E}_\mt{final}(z - \Delta z(t)) &  \mbox{if}  & z\geq z_0 \,,
\end{array}
\right.
\label{definitionEpsilonDeltaz}
\ee
with $\mathcal{E}_\mt{final}(z)$ the static domain profile at asymptotically late times. This formula is just a simple way of ``stretching'' (for positive $\Delta z$) or ``compressing'' (for negative $\Delta z$) the domain profile by ``gluing in''  or ``cutting out'' a small piece at the centre of the domain, taking advantage of the fact that the energy density is almost exactly constant there. In order to determine $\Delta z$ at each  time, we simply impose conservation of energy, namely that the energy change associated to the rigid shift of the interfaces is exactly compensated by the energy change associated to the oscillations of the sound mode: 
\be
\label{analyticaloscillations2}
2 \Ehigh \Delta z (t)  + \int_{z_0-\lambda/4}^{z_0+\lambda/4} dz A_0 \, e^{{\rm{Im}} \, \omega (t-t_0)} \, 
\cos\big[ {\rm{Re}} \, \omega (t-t_0)+\psi_0\big] \, \cos\big[k (z-z_0)\big] =0\,,
\ee
 where we recall that $\lambda=2\pi/k$. The integral is trivial, and solving for $\Delta z$ we find
\be
\label{analyticaloscillations3}
\Delta z (t)= -\frac{1}{k \Ehigh} A_0 \, e^{{\rm{Im}} \, \omega (t-t_0)} \, 
\cos\big[ {\rm{Re}} \, \omega (t-t_0)+\psi_0\big] \,.
\ee
Note that, strictly speaking, the change in time of the size of the domain implies that  the value of $k$ in \eqq{analyticaloscillations1}, and through the dispersion relation also the value of $\omega$, depend on time. However, correcting these values would result in second-order effects since the second term in \eqq{analyticaloscillations1} is itself small to begin with.

\begin{figure*}[t]
	\begin{center}
		\begin{tabular}{cc}
			\includegraphics[width=.46\textwidth]{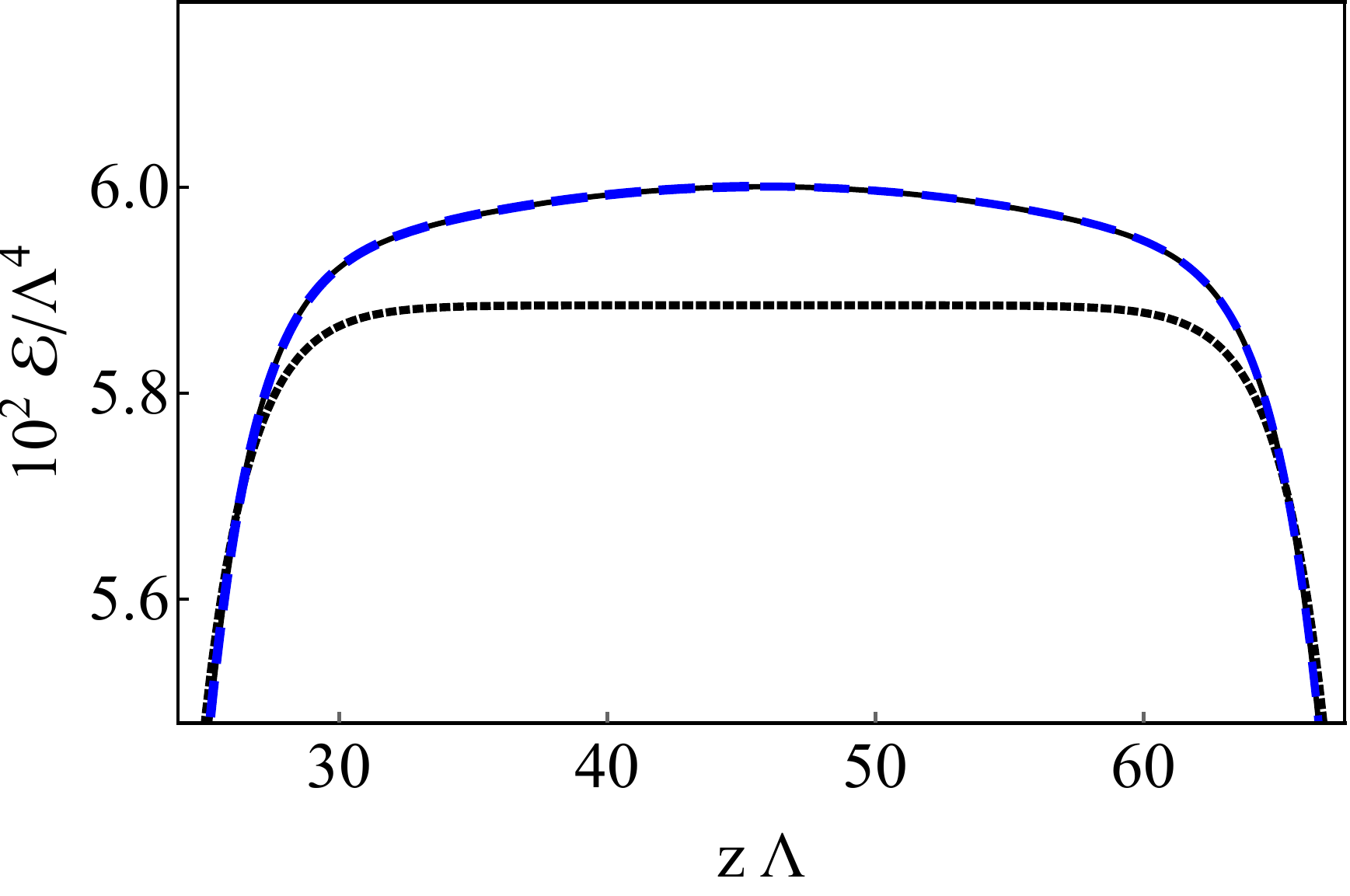} 
			\quad&\quad
			\includegraphics[width=.46\textwidth]{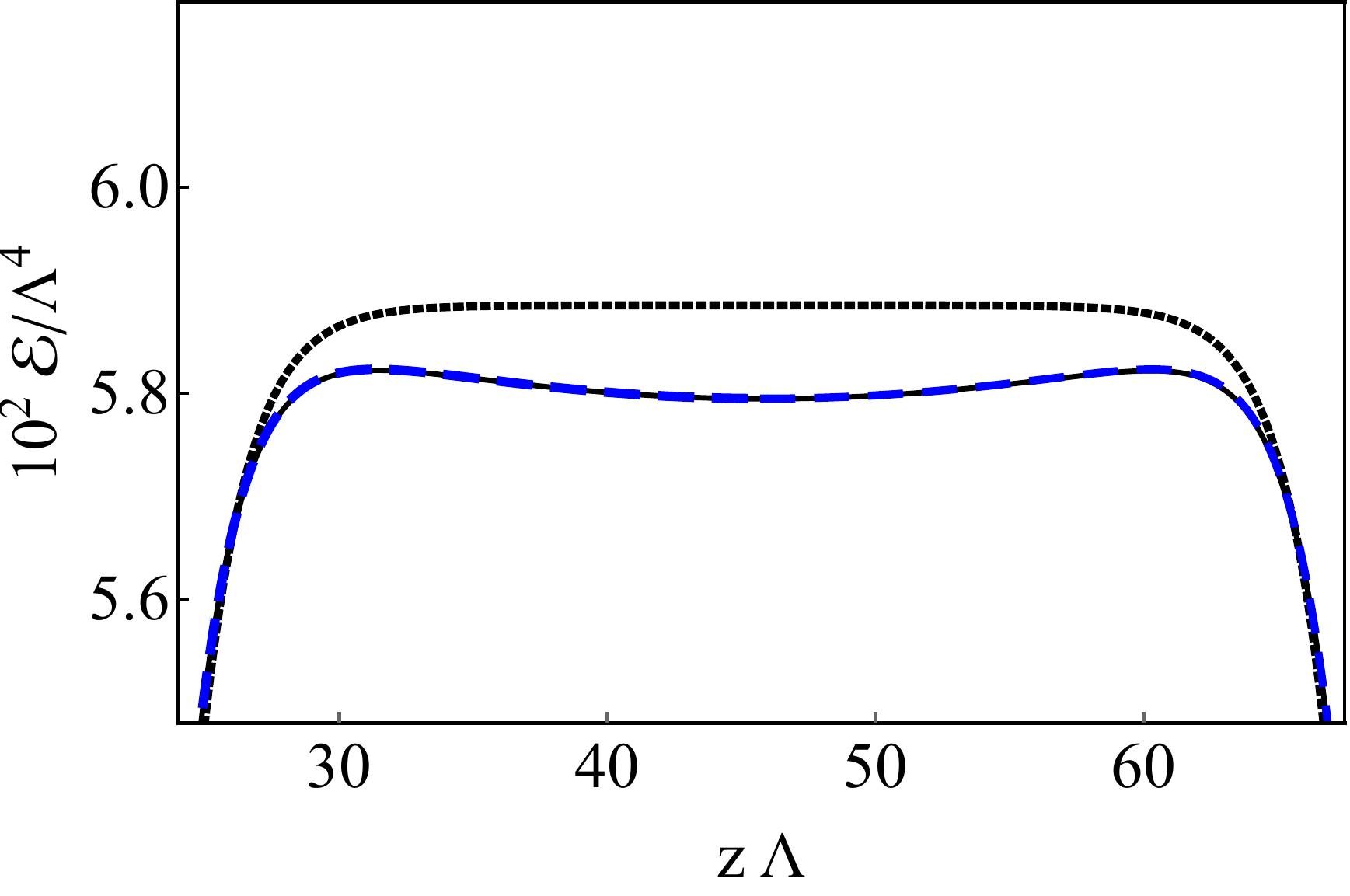} 
		\end{tabular}
	\end{center}
	\vspace{-5mm}
	\caption{\label{OscillationsSoundMode1} 
		\small The solid black curves show snapshots of \fig{3Denergy}(left) at times  $t \Lambda=5490$ (left) and  $t \Lambda=5740$ (right). In each case the dashed blue curve is the result of using expression (\ref{analyticaloscillations1}). For comparison, the black dotted line shows the final static profile at asymptotically late times.
	}
	\vspace{7mm}
\end{figure*}

For concreteness, let us consider applying (\ref{analyticaloscillations1}),  with 
$\Delta z (t)$ given by (\ref{analyticaloscillations3}), to the case of the final phase domain of \fig{3Denergy}(left). The parameters $\rm{Re} \, \omega_2$, 
$\rm{Im} \, \omega_2$ and $k_2$ were already computed.
We fit the different parameters at late times $t_0 \Lambda=6000$ and obtain 
\be
\label{below}
A_0/\Lambda^4=0.0006775 \sac \psi_0=-0.031 \sac z_0 \Lambda=45.86 \,.
\ee
With these values we find a very good description of the full profile at earlier times, specifically within $1\%$ from $t \Lambda \simeq 4860$ (or $0.1\%$ from $t \Lambda \simeq 5430 $) to the end of the evolution.  We illustrate these results in \fig{OscillationsSoundMode1} for two specific times.

\begin{figure*}[h!!!]
\begin{center}
\begin{tabular}{cc}
\includegraphics[width=.46\textwidth]{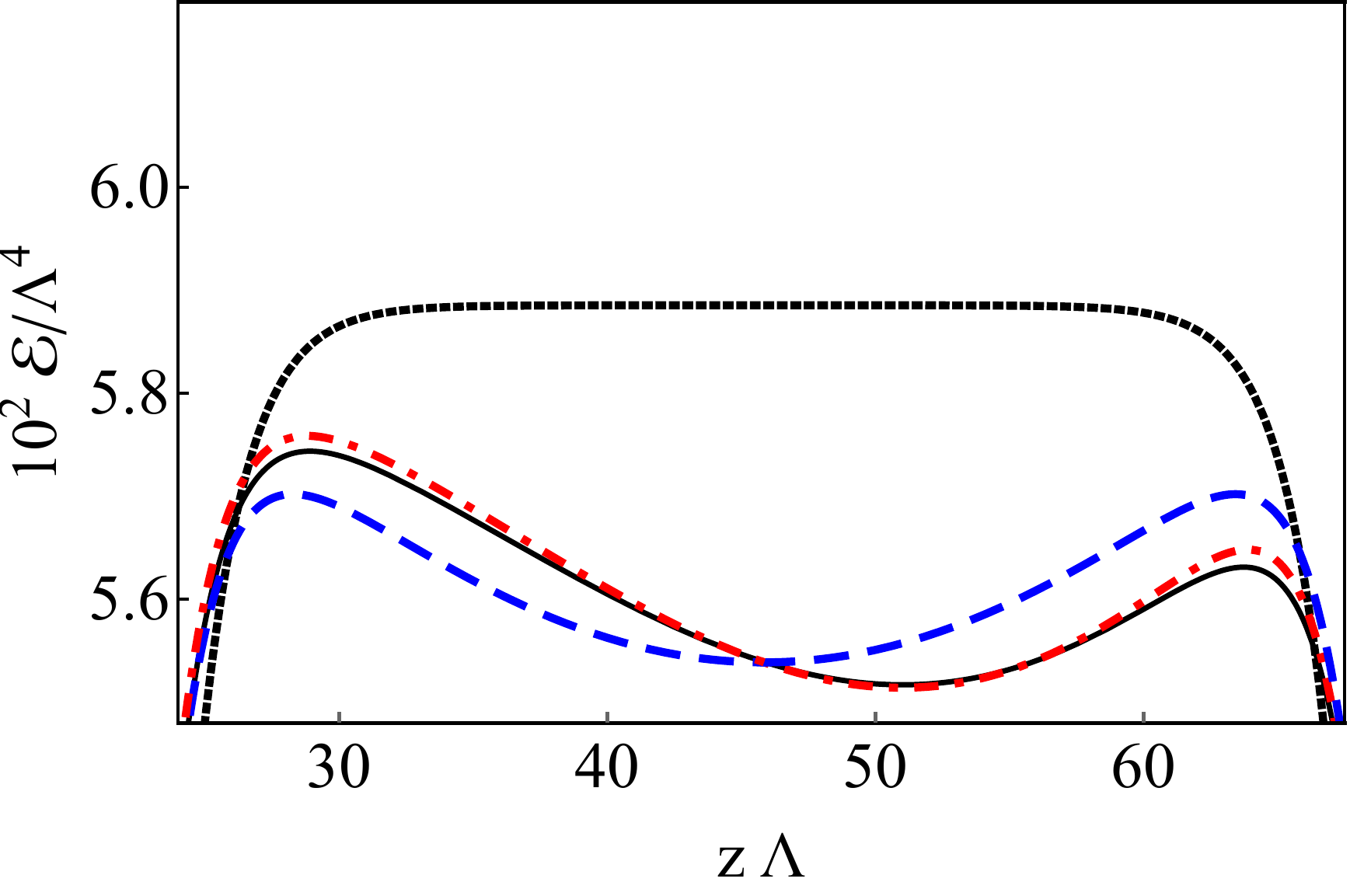} 
\quad&\quad
\includegraphics[width=.46\textwidth]{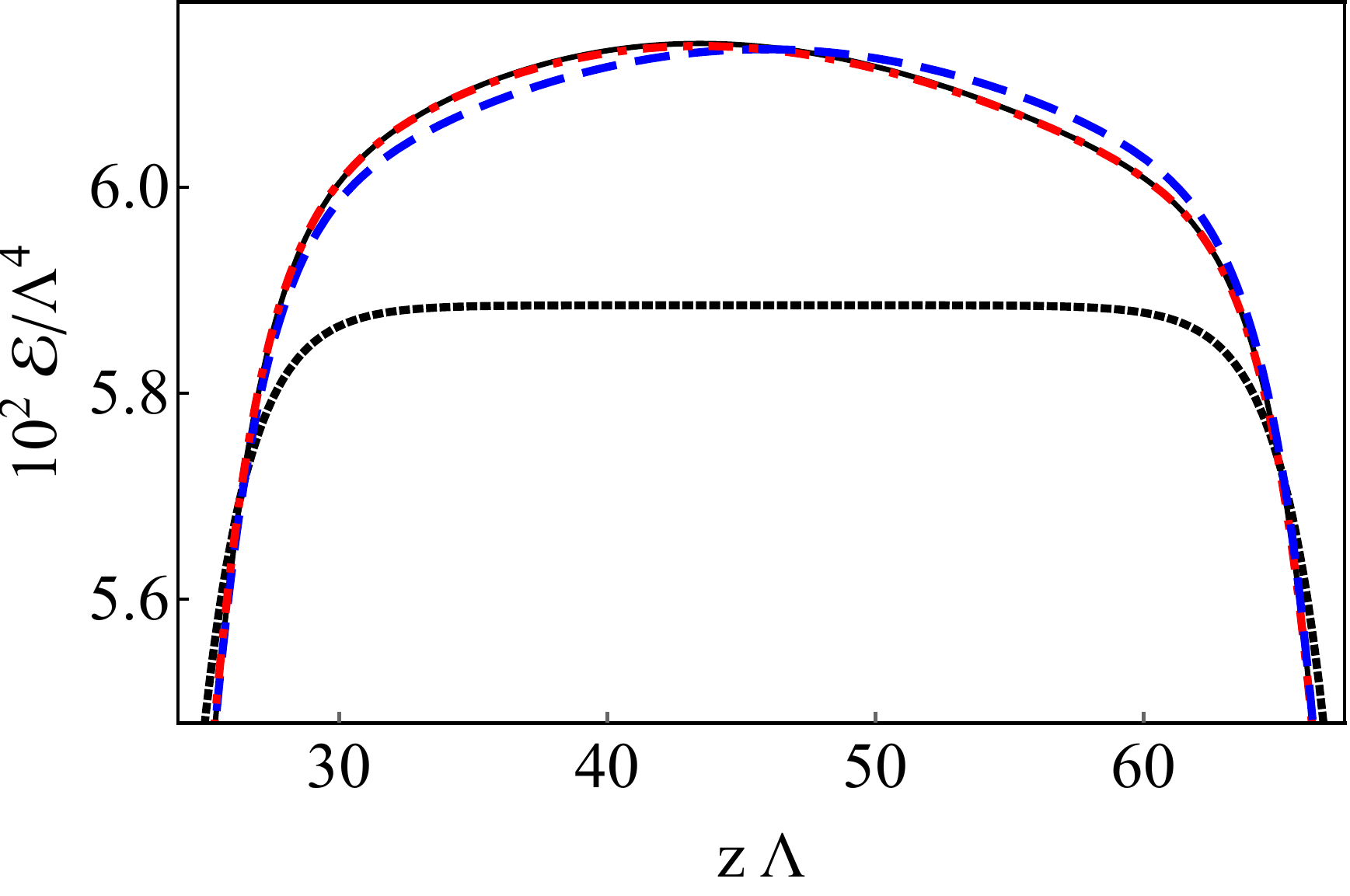} 
\end{tabular}
\end{center}
\vspace{-5mm}
\caption{\label{OscillationsSoundMode2} 
	\small The solid black curve shows snapshots of \fig{3Denergy} at times $t \Lambda=4820$ (left) and  $t \Lambda=5065$ (right). The black dotted curve is the final profile at asymptotically late times. The dashed blue and the 
dot-dashed red curves are the result of using expressions (\ref{analyticaloscillations1}) and (\ref{analyticaloscillations4}), respectively.
}
\vspace{7mm}
\end{figure*}

Above we have only included the mode with the longest wavelength  that fits in the domain. This mode is also the slowest mode to decay, so it is the dominant one at late times. At earlier times the  description can be improved by including higher modes. Let us illustrate this by including the second mode. In this case \eqq{analyticaloscillations1} is replaced by 
\bea
\label{analyticaloscillations4}
\mathcal{E}(t,z) &\simeq& \tilde{\mathcal{E}}_\mt{final}(z, \Delta z(t))+ 
A_0 \, e^{{\rm{Im}} \, \omega(t-t_0)} \, \cos\big[ 
{\rm{Re}} \, \omega (t-t_0)+\psi_0\big] \, \cos\big[k (z-z_0)\big] 
\,\,\,\,\,\,\,\,\,\,\,\,\,\,\,
\\[3mm]
&+& A_0' \, e^{{4 \rm{Im}} \, \omega(t-t_0)} \, \cos\big[ 
{2 \rm{Re}} \, \omega (t-t_0)+\psi_0' \big] \, \sin \big[2 k (z-z_0)\big]
\,.
\eea
The second mode has wavelength $\lambda_2/2$ and, by virtue of (\ref{comp}), double $\rm{Re}\, \omega_2$ and quadruple  $ \rm{Im}\, \omega_2$. It also oscillates in $z$ as a sine as opposed to a cosine, as expected on general grounds. 
We fit the new parameters with the result 
\be
A'_0/\Lambda^4=1.125 \times 10^{-6} \sac \psi'_0=2.85 \,.
\ee
With this  we obtain a good description of the system, specifically within $1\%$ from $t \Lambda=4690$ (or within $0.1\%$ from $t \Lambda=5210$). Note that these times precede those below \eqq{below}, meaning that adding the second mode improves the description at early times. This improvement is also visible in  \fig{OscillationsSoundMode2}, where we can see that at early times the second mode captures the odd (under $z\to-z$) component of the oscillation.  
Notice that the second mode does not contribute to 
(\ref{analyticaloscillations2}) because the integral of the sine over $z$ vanishes.


In summary, we  conclude that soon after the merger the system is well described by linearized hydrodynamics. In Sec.~\ref{sectionhydro} we will verify that, in fact, the full evolution, from the initial spinodal instability to the final state,  is well described by non-linear hydrodynamics.

\section{Phase separation}
\label{phaseseparationsection}
Provided that the box is large enough and that the initial conditions are generic, the  end state of the spinodal instability is a fully phase-separated configuration consisting of one high-energy domain, one low-energy domain and the interfaces between them, as in Figs.~\ref{3Denergy} and \ref{second}(left). This configuration is expected to maximise the entropy given the available total energy and the box size. The entire system is at rest since the net momentum in the initial configuration was zero.  

\begin{figure*}[t]
\begin{center}
\begin{tabular}{cc}
\includegraphics[width=.46\textwidth]{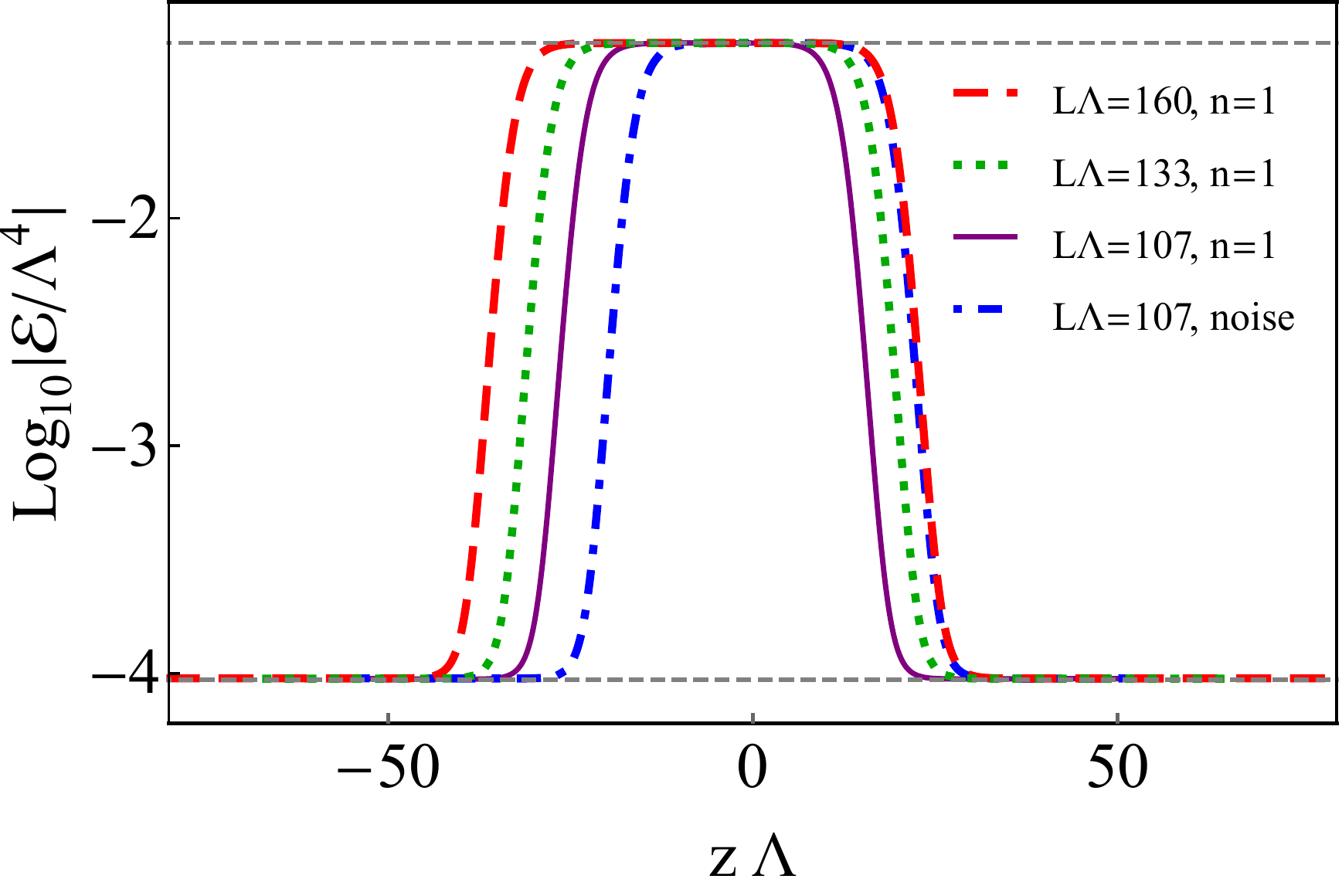} 
\quad \quad
\includegraphics[width=.46\textwidth]{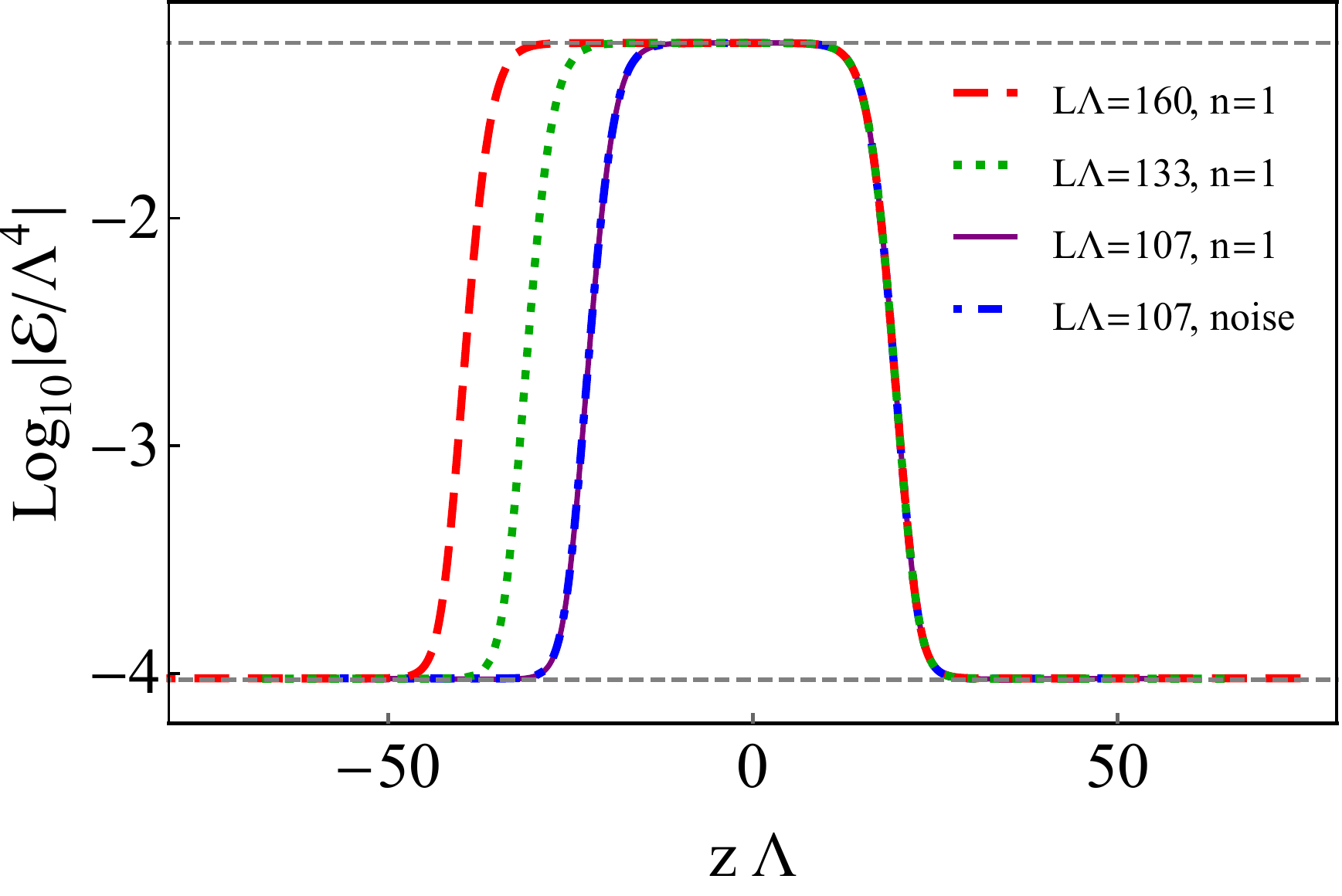} 
\end{tabular}
\end{center}
\vspace{-5mm}
\caption{\label{inter} 
\small (Left) Energy densities in the final, phase-separated configurations for several simulations with the same initial energy density $\E_2$ and different box sizes  $L \Lambda=107$ (triggered either by an $n=1$ mode or by pure numerical noise) and $L\Lambda=133, 160$ (triggered in both cases by an $n=1$ mode).  The horizontal dashed lines indicate the high energy and low energy stable phases obtained from thermodynamics. (Right) Same curves as in (Left) shifted to the right by a constant amount to exhibit the fact that the shape of the interface is the same in all cases.  
}
\vspace{7mm}
\end{figure*}

\begin{figure}[t]
  \begin{center}
    \includegraphics[width=0.55\textwidth]{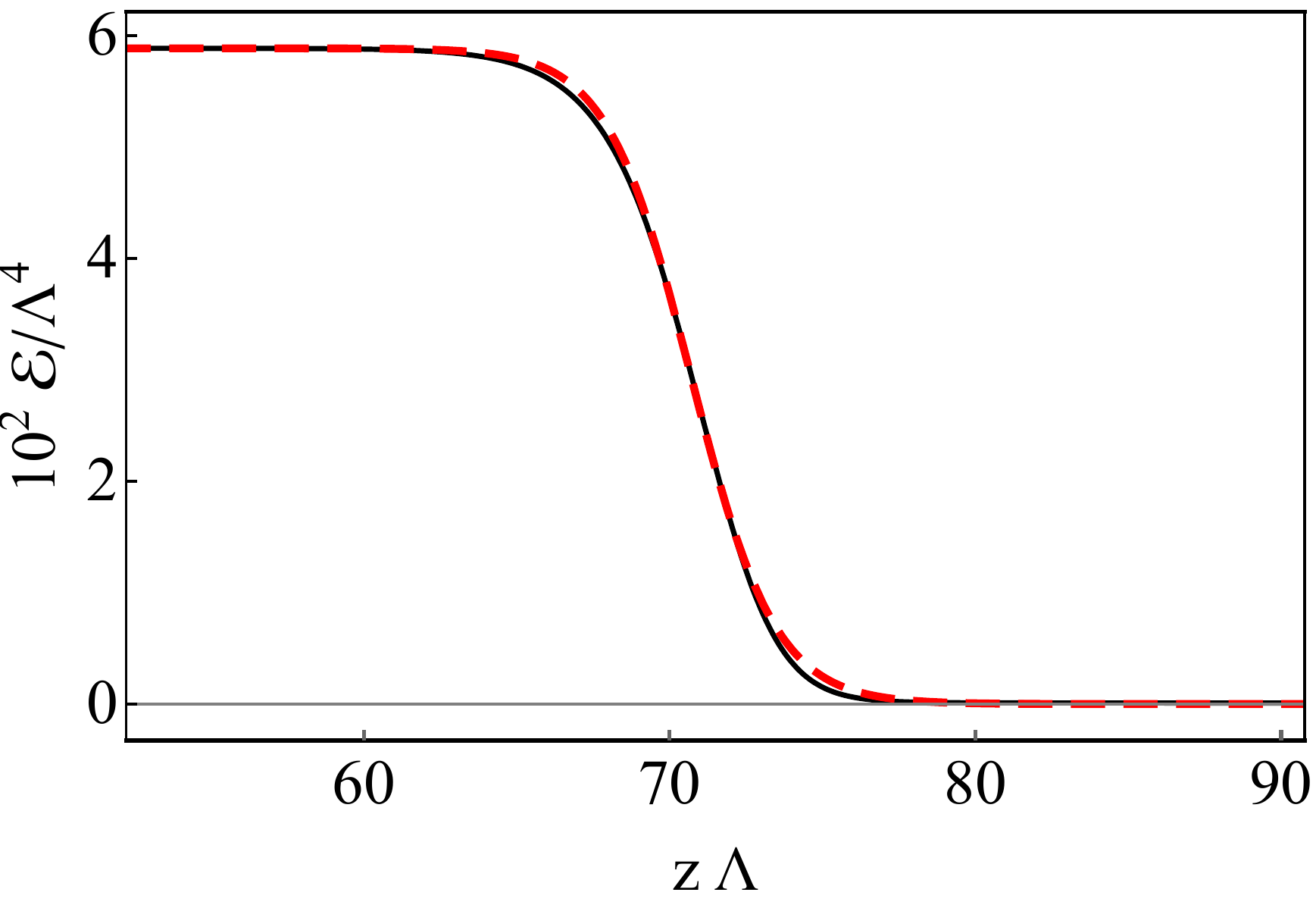}
    \\[4mm]
    \includegraphics[width=0.59\textwidth]{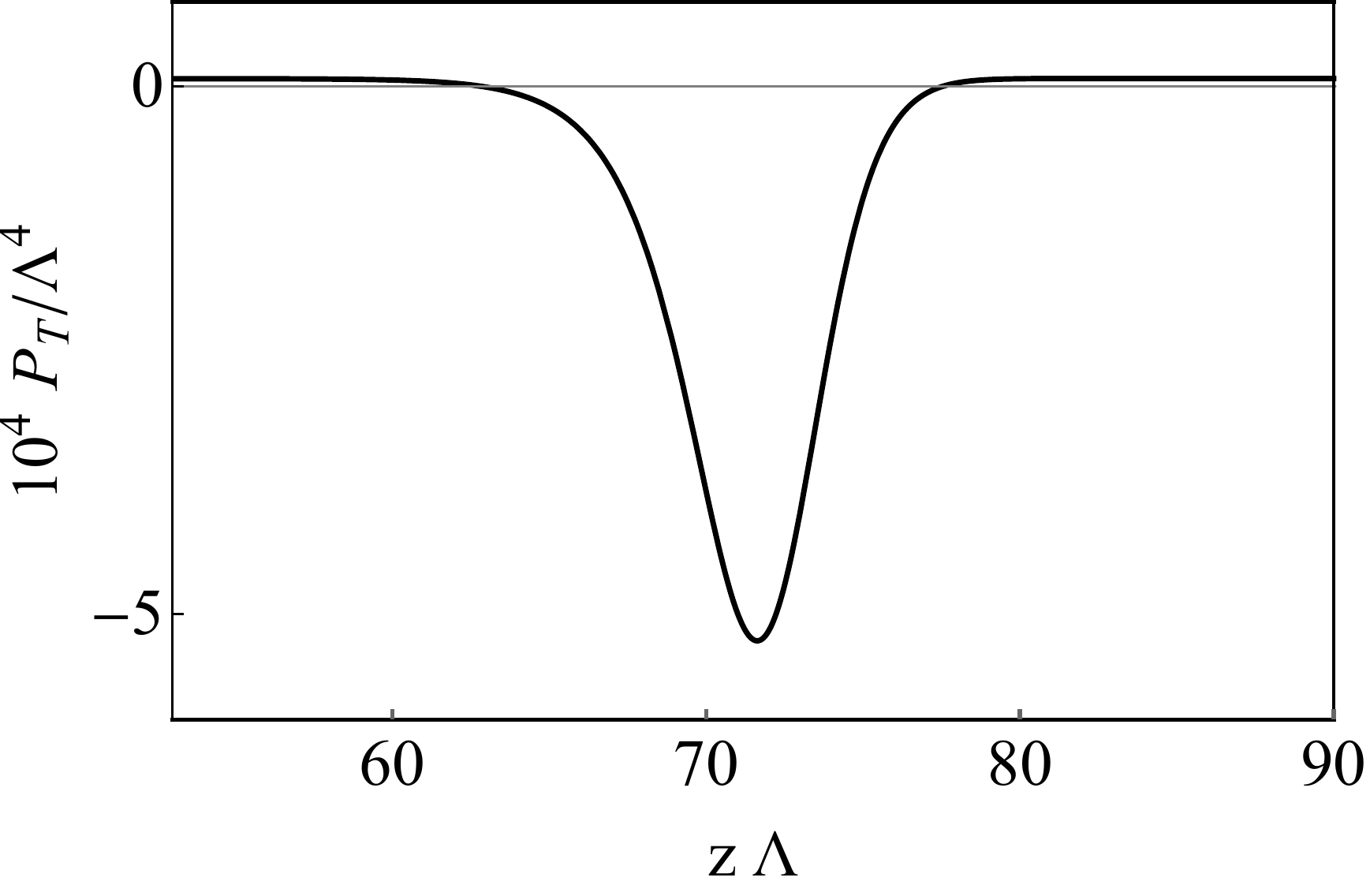}\,\,
  \end{center}
  \vspace{-5mm}
\caption{\label{AnalyticalApproxPlot} 
\small Energy density (top) and transverse pressure (bottom) in the final phase-separated solution of \fig{3Denergy}(left). In (top) the continuous black curve  is the exact energy density, whereas the dashed red curve shows the analytical approximation (\ref{AnalyticalApproximation}). 
}
\end{figure}

In \fig{inter}(left) we plot the energy profiles at late times of several simulations, together with the high- and 
low-energy phases obtained from the thermodynamics of homogeneous configurations. The good agreement between the latter and the energy densities of the domains confirms that this is a phase-separated configuration. 
Moreover, from the surface gravity of the horizon on the gravity side we obtain a temperature that is constant and equal to $T_c$ (within $0.01\%$) across the entire configuration, as expected from phase coexistence. Also as expected, 
we find that the interface that separates one phase from the other is universal, meaning that it is a property of the theory, independent of the initial conditions and of the size of the box. This is clearly demonstrated in \fig{inter}(right), where we show that shifting each curve by a constant
amount all the interfaces agree with one another.
As shown in \fig{AnalyticalApproxPlot}(top), the shape of this universal interface is very well approximated by the function 
\be
\E (z) \simeq
\frac{\Delta \E}{2} \left[ 1 + {\rm Tanh} \left( \frac{z-z_0}{b} \right) \right] \,,
\label{AnalyticalApproximation}
\ee
where $\Delta \E=\Ehigh -\Elow$, $z_0$ is the point at which the energy density is exactly half way between $\Ehigh$ and $\Elow$, and $b\Lambda \simeq 2.75$ can be taken as a definition of the size of the interface. 
The universality of the interface implies that, in a phase-separated configuration, the size of each domain is fixed by the size of the box and by  the total energy in it. 

The surface tension of the interface is defined as the excess free energy in the system, per unit area in the transverse directions $x_\perp$, due to the presence of the interface. In a homogeneous system the free energy density per unit volume is constant and equal to minus the pressure, 
$F=-P$. Our system is only homogeneous   along $x_\perp$, so it is the transverse pressure that appears in this relation 
(see e.g.~\cite{Donos:2013cka}), i.e.~we have $F(z)=-P_T(z)$, and moreover both densities are  \mbox{$z$-dependent}. The transverse pressure in the final phase-separated configuration is shown in \fig{AnalyticalApproxPlot}(bottom).  By definition, at $T=T_c$ the homogeneous, stable, high-energy and low-energy phases have the same free energy density $F_c$, and hence the same transverse pressure, $P_c /\Lambda^4 \simeq 7.5 \times 10^{-6}$. This is the value to which the transverse pressure $P_T(z)$ asymptotes in \fig{AnalyticalApproxPlot}(bottom) away from the interface. 
In the absence of the interface the free energy density per unit transverse area in the box would be $L F_c=-L P_c$. The excess free energy per unit transverse area due to the presence of the interface is therefore 
\be
\sigma = \frac{1}{2} \int_0^L dz \, \Big[ F(z) - F_c \Big] = -
\frac{1}{2} \int_0^L dz \, \Big[ P_T(z) - P_c \Big] 
\simeq 2.9  \times  10^{-3}  \Lambda^3\,,
\label{SurfaceTension}
\ee
where the factor of 1/2 is due to the fact that there are two interfaces in the box. Note that this surface tension is positive because $P_T(z)<P_c$ or, equivalently, because \mbox{$F(z)>F_c$}. This is consistent with the fact that the presence of gradients associated to the interface increases the free energy of the system, as expected on general grounds.

If the box is not large enough then the final state will certainly not be a phase-separated configuration. To be precise, the box must be able to fit at least two interfaces plus two domains of sizes at least as large as the interfaces so that they can be distinguished from the interfaces themselves.  Fine-tuned initial conditions may also prevent the final state from being a phase-separated configuration, as in the examples shown in \fig{what}.

\section{Hydrodynamics}
\label{sectionhydro}
We saw in Sec.~\ref{relaxation} that the relaxation of a domain can be described by linearised hydrodynamics. We will now show that non-linear, second-order hydrodynamics describes the entire spacetime evolution of the system from the beginning  of the spinodal instability to the final, phase-separated configuration. For concreteness we will illustrate this for the evolution shown in 
\fig{3Denergy}(left). Our discussion parallels that in Sec.~4 of \cite{Attems:2018gou} with additional details included.

In modern language we define hydrodynamics as a gradient expansion around local equilibrium that, at any given order, includes all possible gradients of the hydrodynamic variables that are purely spatial in the local rest frame. Let us refer to this as the purely spatial formulation. 
To second order the hydrodynamic stress tensor takes the form 
\be
\label{stresstensor}
T^{\mu \nu} = T^{\mu \nu}_\mt{ideal} + \Pi^{\mu\nu}  \, ,
\ee
with 
\bea
\label{nono}
T^{\mu \nu}_\mt{ideal} &=& \E u^\mu u^\nu + 
P_{\mt{eq}} \Delta^{\mu\nu} \,, \\[2mm]
\Pi^{\mu\nu} &=& \Pi_{(1)}^{\mu\nu} + \Pi_{(2)}^{\mu\nu} \,,
\eea
and 
\bse
\label{omitted0}
\bal
\label{linear}
\Pi_{(1)}^{\mu\nu} &= -\eta\sigma^{\mu\nu} -\zeta \left( \nabla\cdot u \right) \Delta^{\mu\nu} \,, \\[2mm]
\Pi_{(2)}^{\mu\nu} &= \pi_{(2)}^{\mu\nu} + \Delta^{\mu\nu} \Pi_{(2)} \,. 
\end{align}
\ese
In these expressions $P_{\mt{eq}}$ is the equilibrium pressure, $u^\mu$ is the fluid four-velocity, 
\mbox{$\Delta^{\mu\nu}= \eta^{\mu\nu} + u^\mu u^\nu$} is the projector onto spatial directions in the local rest frame, and 
$\Pi_{(1)}^{\mu\nu}$ contains the first-order corrections, with 
$\eta$ and $\zeta$  the shear and bulk viscosities, respectively. The shear tensor is  $\sigma^{\mu\nu} = \nabla^{<\mu} u^{\nu>} $, where $\nabla^\mu\equiv \Delta^{\mu\nu} \del_\nu$, and $A^{<\mu \nu >}$ denotes the symmetric, transverse and traceless part of any rank-two tensor.   As in other holographic models  as e.g.~\cite{DeWolfe:2011ts}, the bulk viscosity remains finite at the points where the speed of sound vanishes  as a consequence of the large-$N_c$ approximation implicit in the holographic set-up \cite{Natsuume:2010bs}.  

\begin{figure*}[t]
	\begin{center}
		\begin{tabular}{cc}
			\includegraphics[width=.43\textwidth]{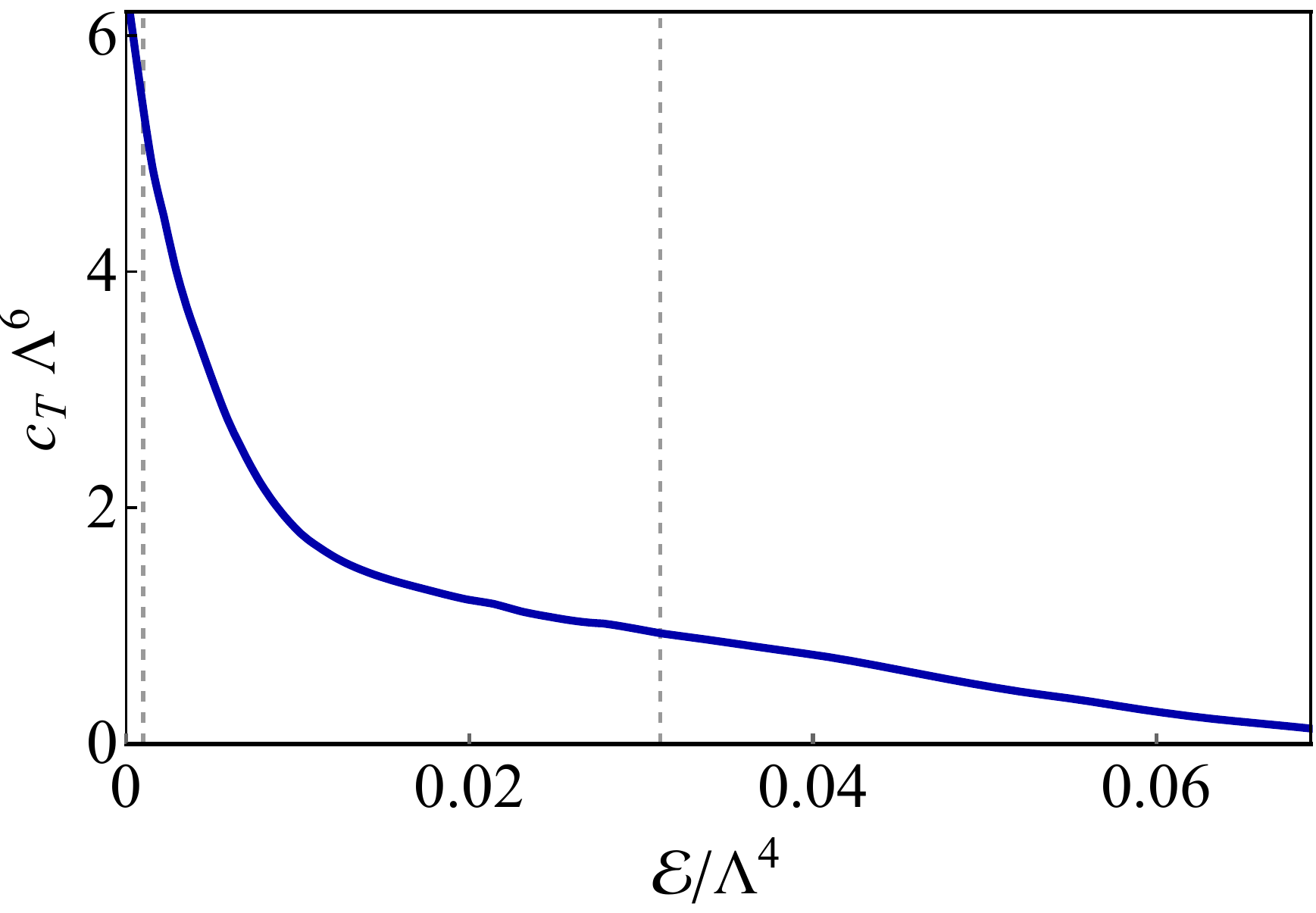} 
			\quad&\quad
			\includegraphics[width=.43\textwidth]{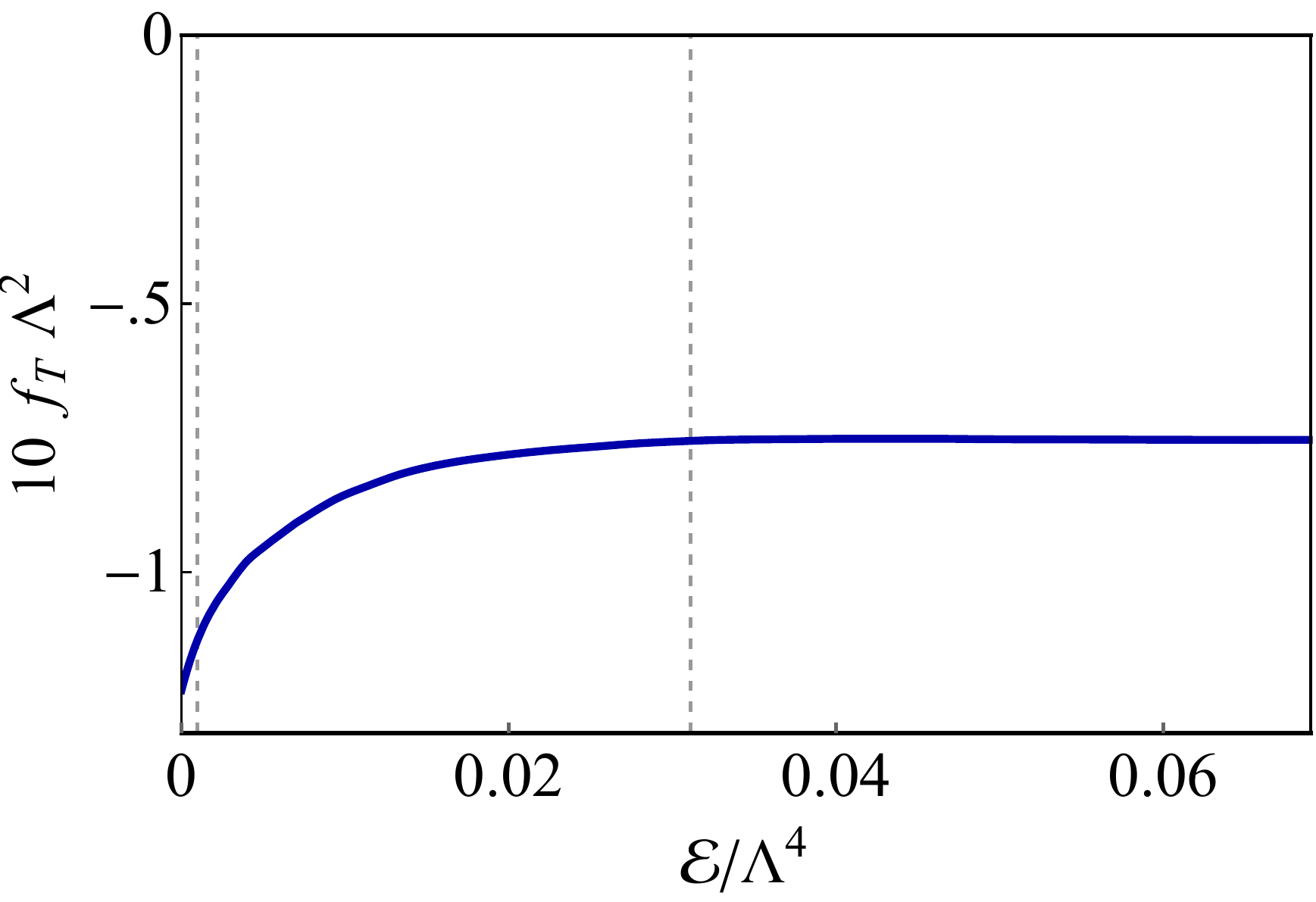} 
			\\[4mm]
			\includegraphics[width=.43\textwidth]{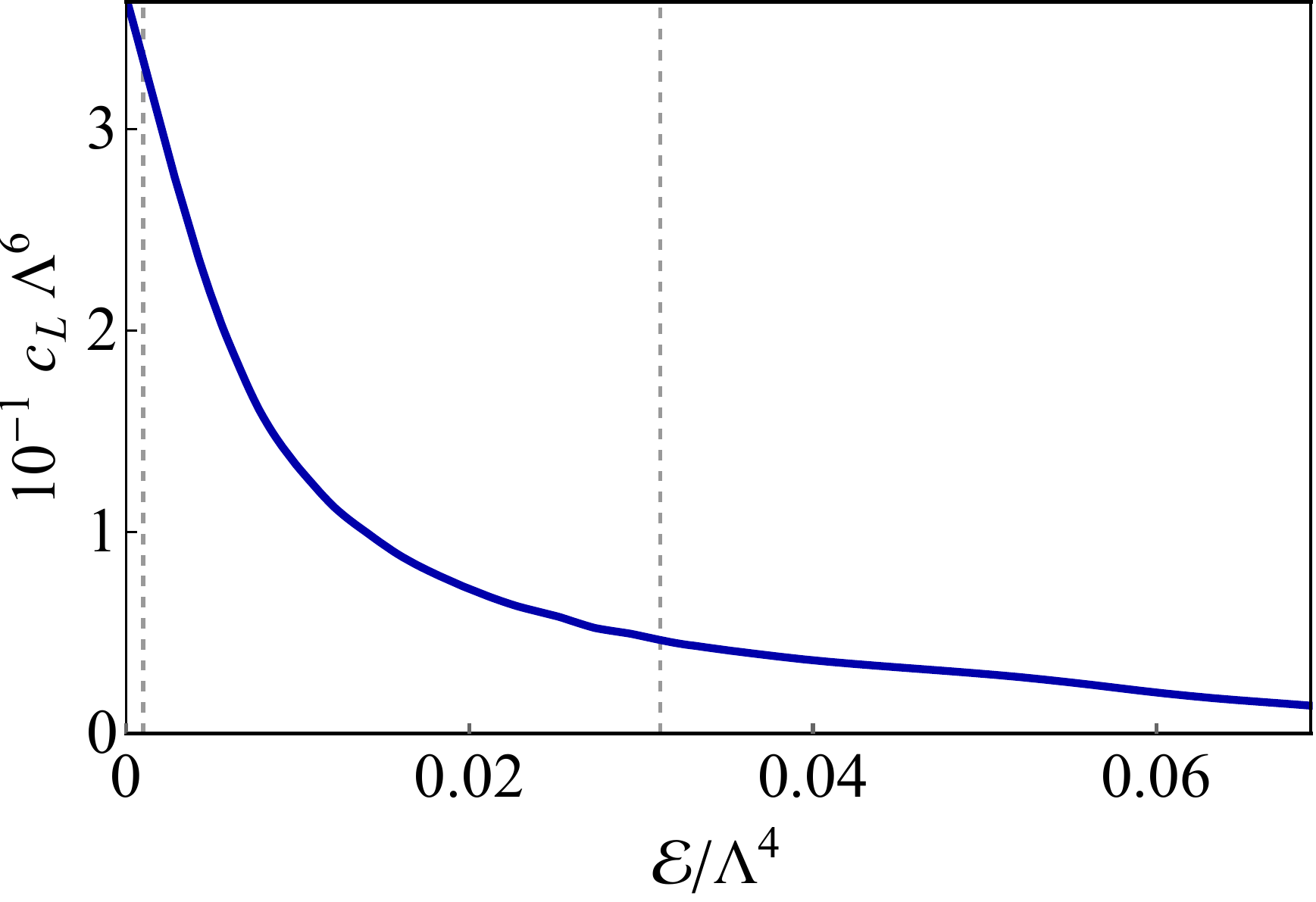} 
			\quad&\quad
			\includegraphics[width=.43\textwidth]{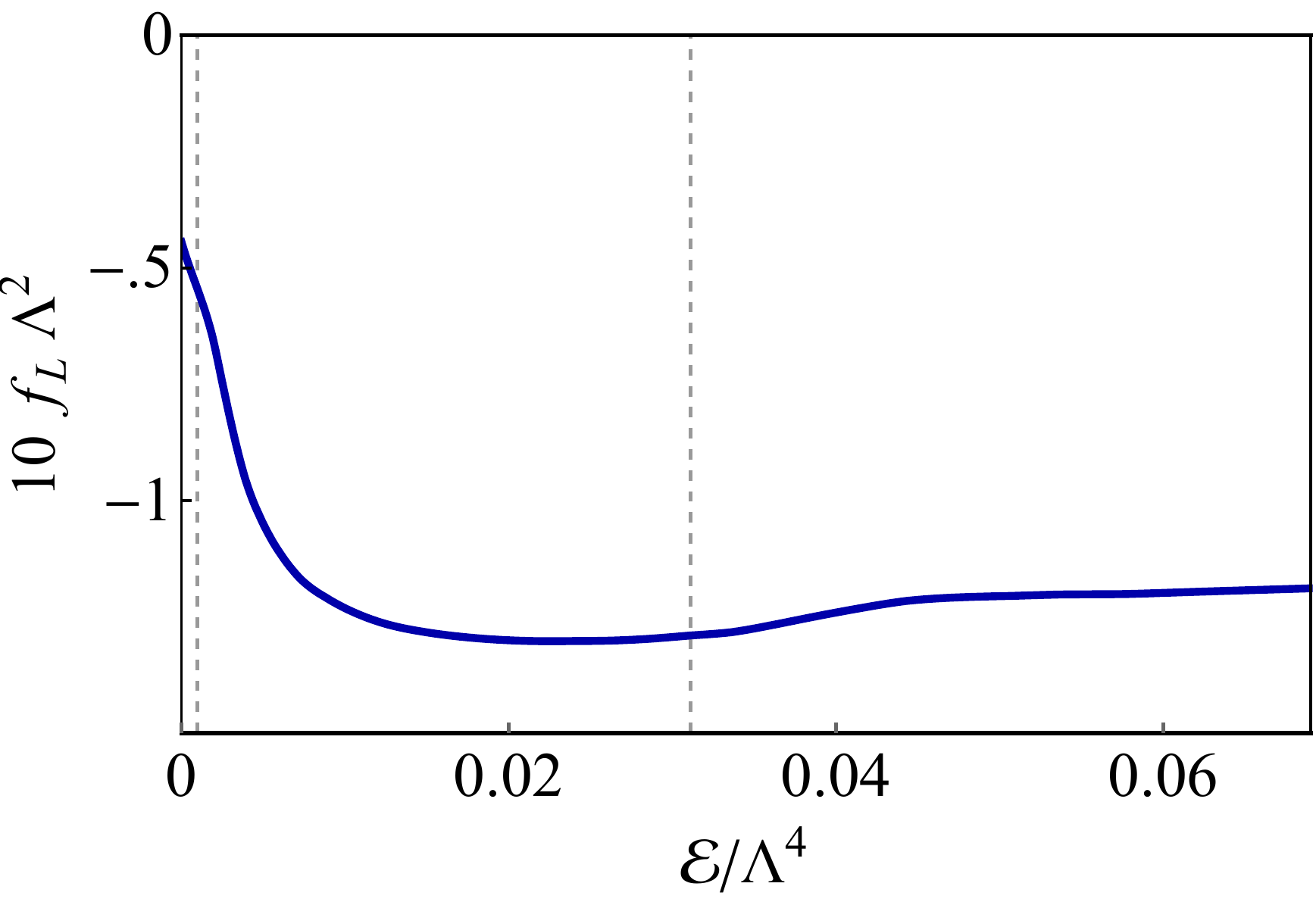}  
		\end{tabular}
	\end{center}
	\vspace{-5mm}
	\caption{\label{coef} 
		\small Second-order coefficients entering the purely-spatial formulation of hydrodynamics as a function of the energy density. The grey vertical  lines indicate the values at which the speed of sound crosses zero (see \fig{fig:cs2}(left)).  
	}
	\vspace{4mm}
\end{figure*}

 All the second-order terms are contained in $\Pi_{(2)}^{\mu\nu}$. For the case of interest here of fluid motion in flat space in 1+1 dimensions its tensor and scalar parts may be expanded as
\bse
\label{omitted}
\bal
\pi_{(2)}^{\mu\nu} &= \tilde c_1 \tilde{\mathcal{O}}_1 ^{\mu \nu}  + \tilde c_2 \tilde{\mathcal{O}}_2 ^{\mu \nu} + \tilde c_7 \tilde{\mathcal{O}}_7^{\mu\nu} \,, 
\\[2mm]
\Pi_{(2)} &= \tilde b_2 \tilde{\mathcal{S}}_2 + \tilde b_3 
\tilde{\mathcal{S}}_3 + \tilde b_4 \tilde{\mathcal{S}}_4 \,.
\end{align}
\ese
In order to make contact with \cite{Attems:2017ezz} we chose the basis of operators  to be
\bse
\label{basis1}
\bal
\tilde{\mathcal{O}}_1^{\mu \nu} &= 2 \nabla^{<\mu} \nabla^{\nu>} 
\mathcal{E}  \,, \\[2mm]
\tilde{\mathcal{O}}_2^{\mu \nu} &=  2  \nabla^{<\mu} \mathcal{E}  \nabla^{\nu>} 
\mathcal{E} \,, \\[2mm]
\tilde{\mathcal{O}}_7^{\mu\nu}&= \frac{\nabla \cdot u}{3} \sigma^{\mu\nu}\,, 
\\[2mm]
\tilde{\mathcal{S}}_2 &= 2 \nabla_\mu \mathcal{E} 
\nabla^\mu \mathcal{E} \,, \\[2mm]
\tilde{\mathcal{S}}_3 &=2  \nabla^2 \mathcal{E} \,, \\[2mm]
\tilde{\mathcal{S}}_4 &=  \left(\nabla \cdot u\right)^2\,.
\end{align}
\ese
Part of the notation above is chosen to make contact with 
\cite{Baier:2007ix,Romatschke:2009kr} below. The coefficients 
$\tilde c_1,  \tilde c_2,  \tilde b_2,  \tilde b_3$ are known because they are related to the coefficients $c_L, c_T, f_L, f_T$ of \cite{Attems:2017ezz} through
\bse
\bal
c_L&=\frac{2}{3} \left( 3\tilde b_2 + 2 \tilde c_2 \right) \,,
\\
\label{ape1}
f_L&= 2 \tilde b_3 + \frac{4}{3} \tilde c_1 \,,
\\
c_T&= \frac{2}{3} \left( 3\tilde b_2 -  \tilde c_2 \right) \,,
\\
f_T&= 2 \tilde b_3 - \frac{2}{3} \tilde c_1 \,.
\end{align}
\ese
These four coefficients are shown in \fig{coef} as a function of the energy density. Note that they are finite and smooth at the points where the speed of sound vanishes. We will come back to this point below.  
\begin{figure}[h!!!]
  \begin{center}
    \includegraphics[width=0.7 \textwidth]{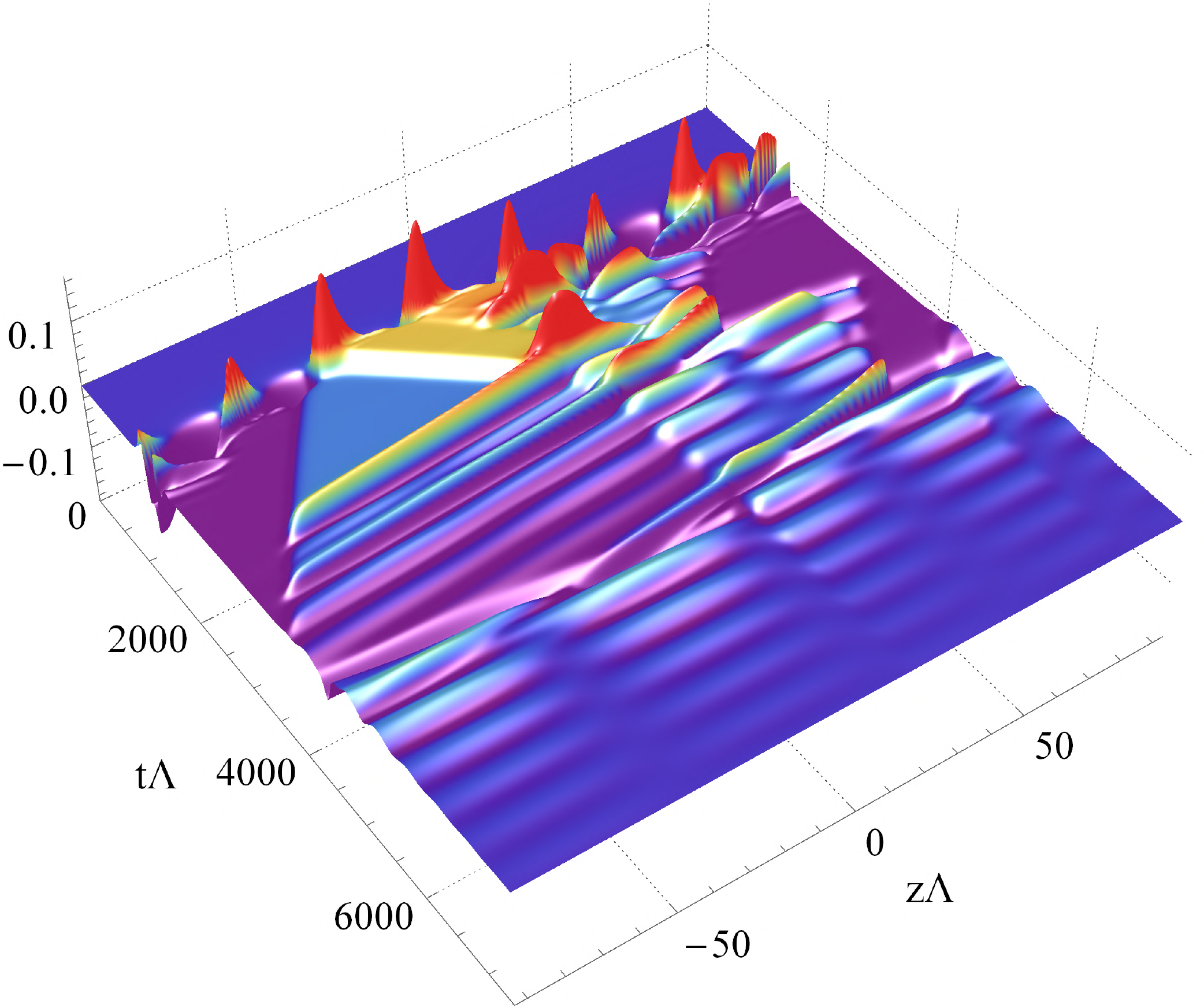}
    \put(-300,195){\Large $v$}
  \end{center}
  \vspace{-5mm}
\caption{\label{velo} 
\small Velocity of the fluid in the evolution  of Fig. \ref{3Denergy}(left). 
}
\end{figure}

We have not computed the coefficients $\tilde c_7, \tilde b_4$ but, as we will see,  they are not needed in order to obtain a good hydrodynamic description of our system. As in  \cite{Attems:2017ezz,Attems:2018gou}, the reason is that the operators 
$\tilde{\mathcal{O}}_7^{\mu\nu}, \tilde{\mathcal{S}}_4$ are suppressed in the dynamical situation under consideration because they are quadratic in the fluid velocity. In the case of the evolution shown in \fig{3Denergy} the absolute value of this velocity is everywhere smaller than 0.2  and typically no larger than 0.1,  as can be seen in \fig{velo}. 

\begin{figure*}[h!!!]
	\begin{center}
		\begin{tabular}{c}
	\includegraphics[width=0.85\textwidth]{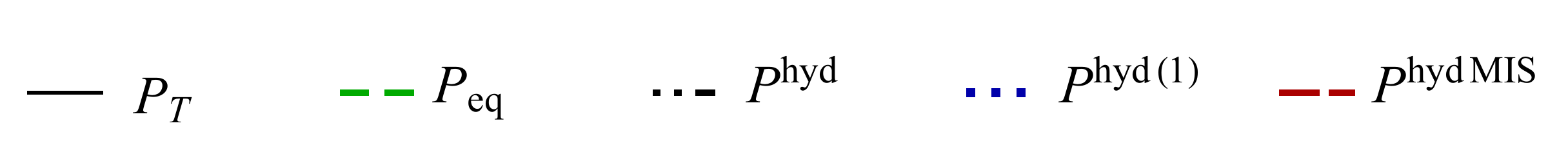} 
	\\
			\includegraphics[width=.87\textwidth]{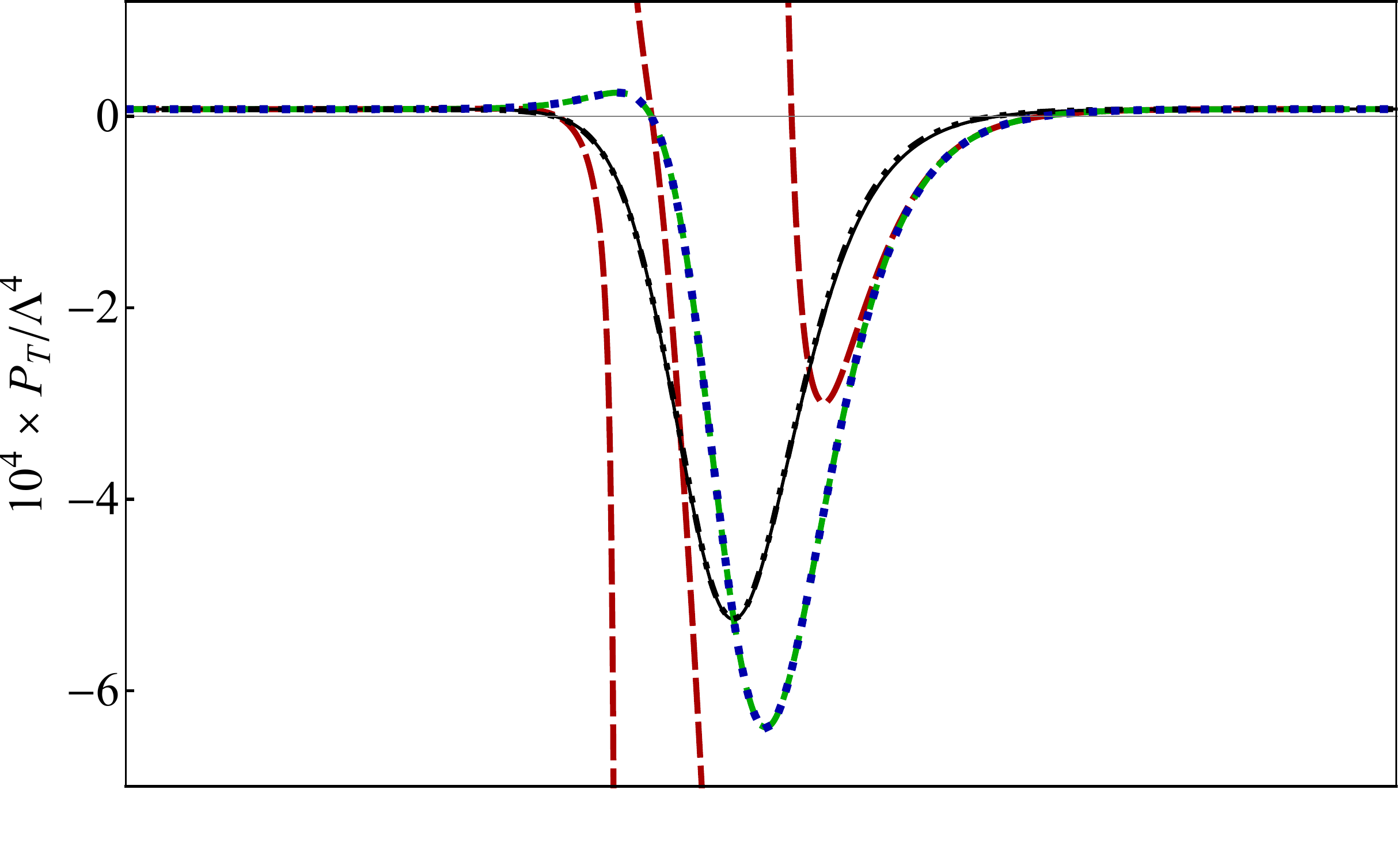}
			\\[-0.04\textwidth]
			\includegraphics[width=.87\textwidth]{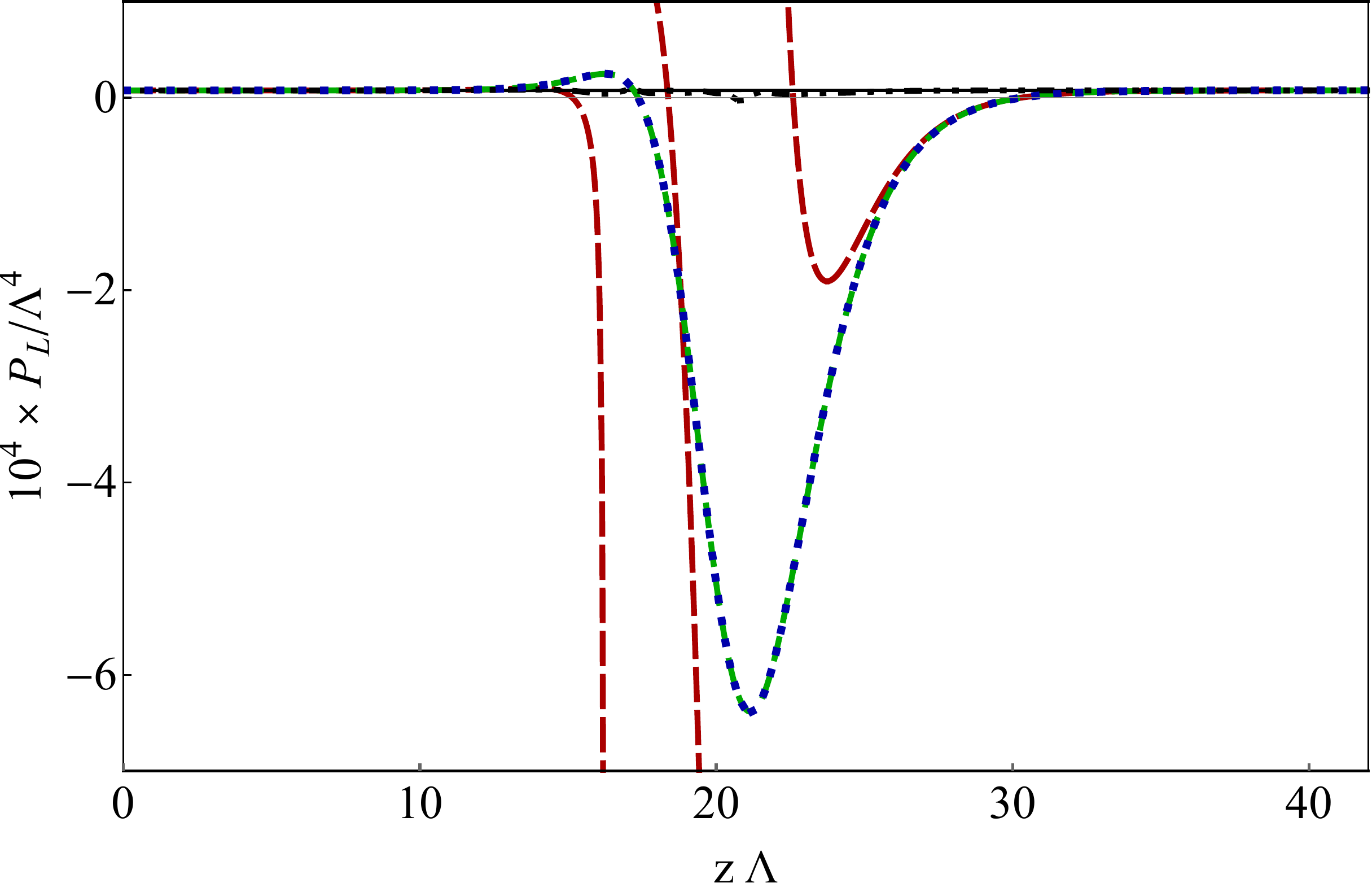} 
			\hspace{-0.0069\textwidth}
	\end{tabular}
	\end{center}
	\vspace{-5mm}
	\caption{\label{hydro1} 
		\small  Snapshots at asymptotically late times of the pressures for the evolution shown in \fig{3Denergy}(left). $P_L, P_T$ are the exact pressures  extracted from gravity. The plot of $P_T$ is the same as in 
\fig{AnalyticalApproxPlot}(bottom). $P_{\mt{eq}}$ is the equilibrium pressure. $P_L^{\mt{hyd(1)}}, P_T^{\mt{hyd(1)}}$ are the hydrodynamic pressures with only first-order viscous corrections included, i.e.~those in 
$T^{\mu \nu}_\mt{ideal} + \Pi_{(1)}^{\mu\nu} $.  $P_L^{\mt{hyd}}, P_T^{\mt{hyd}}$ are the  second-order hydrodynamic pressures in the purely spatial formulation omitting the contributions of 
$\tilde{\mathcal{O}}_7^{\mu\nu}$ and 
$\tilde{\mathcal{S}}_4$. 
$P_L^{\mt{hydMIS}}, P_T^{\mt{hydMIS}}$ are the  second-order hydrodynamic pressures in the MIS-type formulation. 
	}
	\vspace{4mm}
\end{figure*}

\begin{figure*}[h!!]
	\begin{center}
	\includegraphics[width=0.9\textwidth]{Leyenda1.pdf} 
			\includegraphics[width=0.99\textwidth]{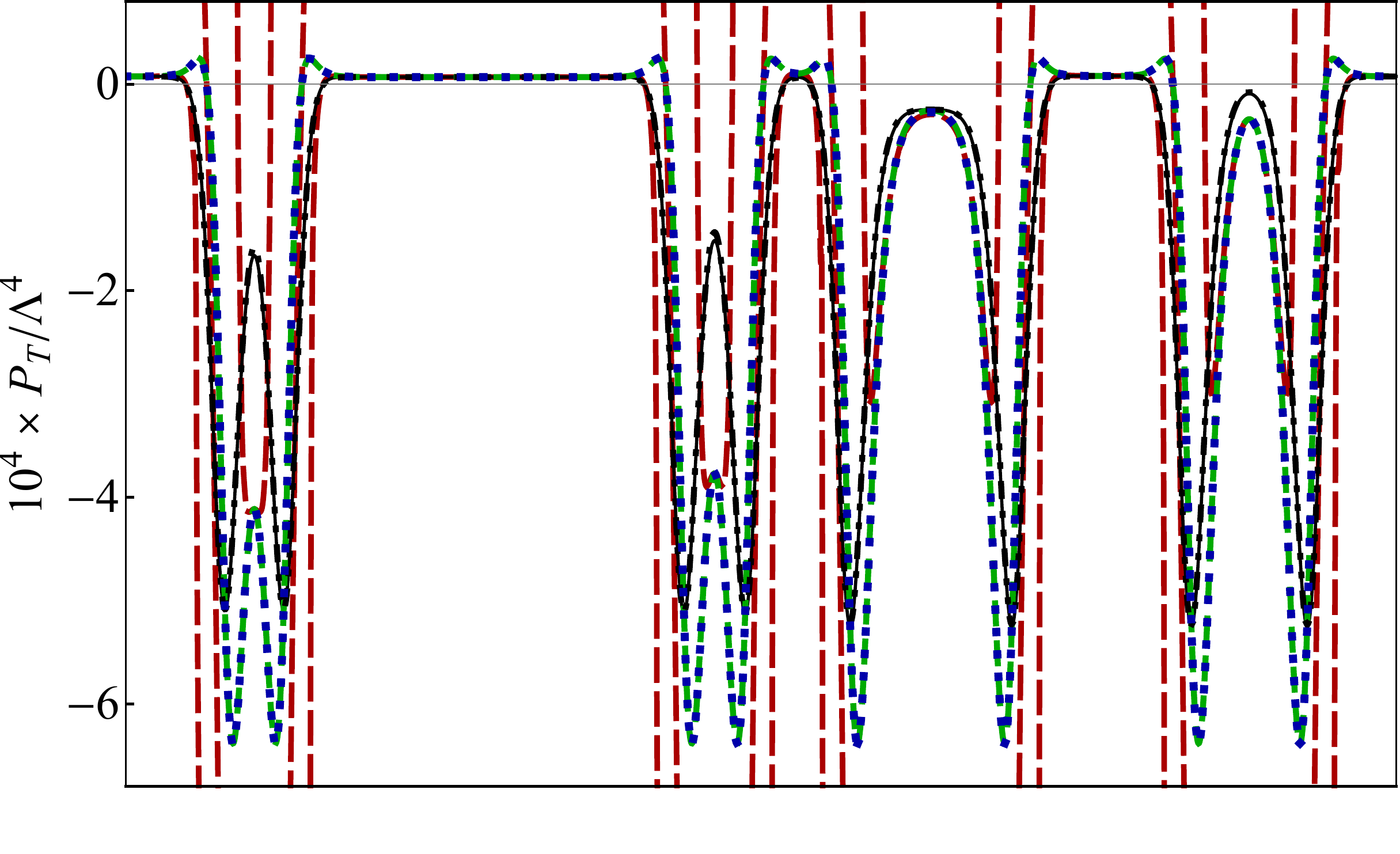} 
			\\[-0.02\textwidth]
			\includegraphics[width=0.99\textwidth]{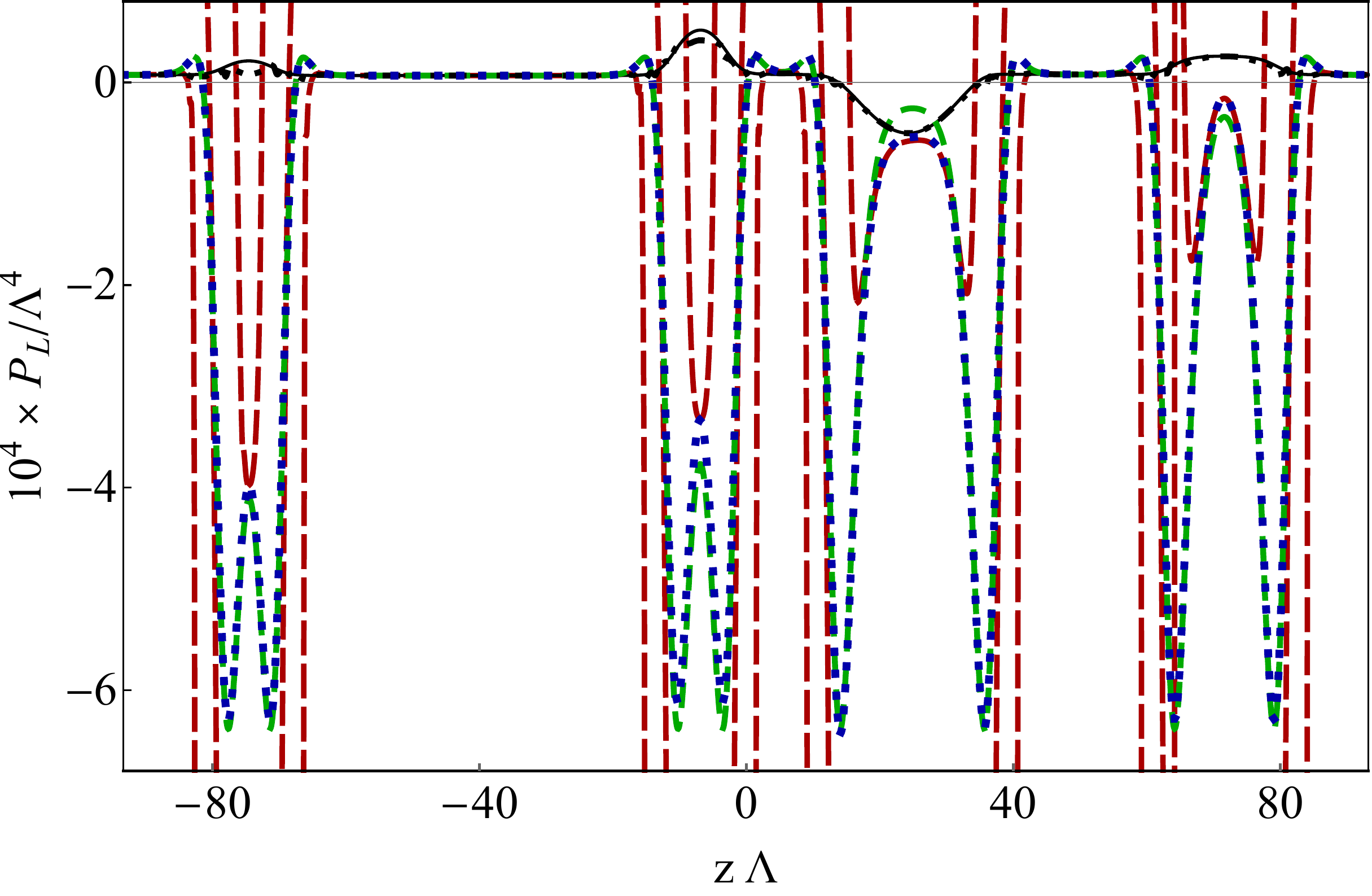} 
	\end{center}
	\vspace{-5mm}
	\caption{\label{hydro2} 
		\small Snapshots at $t\Lambda=2000$ of the pressures for the evolution shown in \fig{3Denergy}(left). See the caption of \fig{hydro1} for details.
	}
	\vspace{4mm}
\end{figure*}

\begin{figure*}[h!!]
	\begin{center}
	\includegraphics[width=0.9\textwidth]{Leyenda1.pdf} 
			\includegraphics[width=0.96\textwidth]{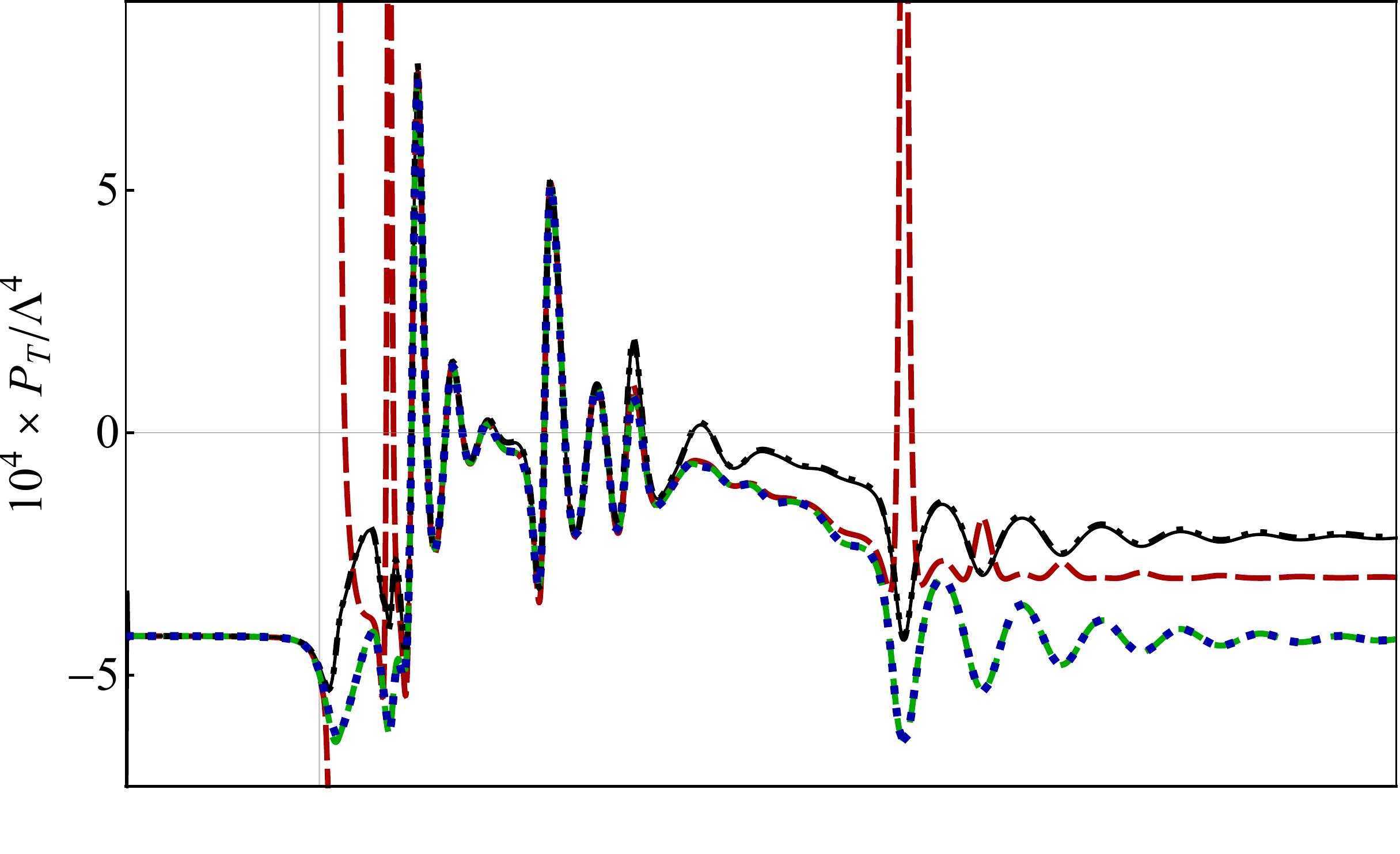} 
			\\[-0.02\textwidth]
			\includegraphics[width=0.96\textwidth]{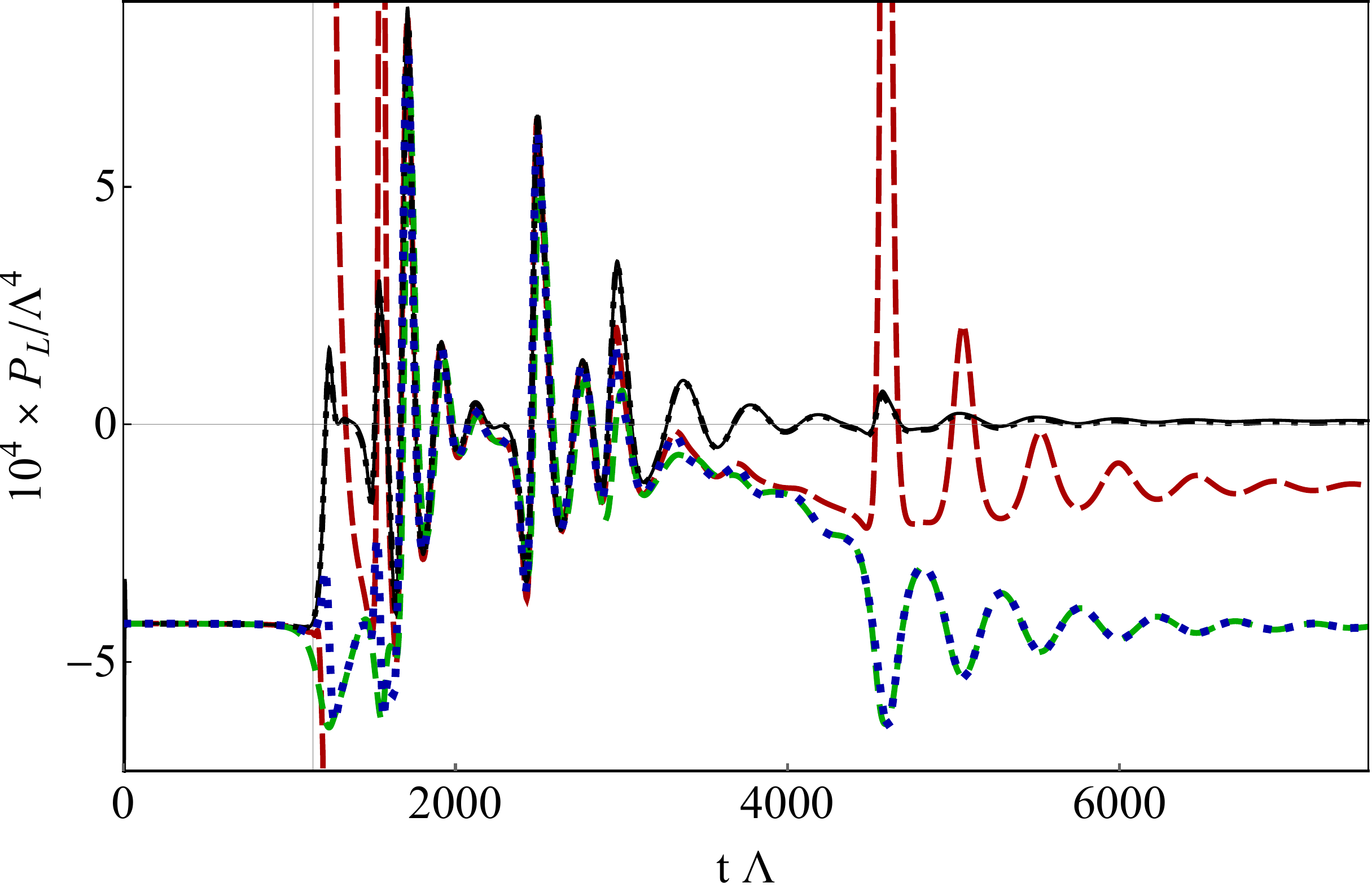} 
	\end{center}
	\vspace{-5mm}
	\caption{\label{hydro3} 
		\small Time evolution at the point $z\Lambda=23$ of the  pressures for the evolution shown in \fig{3Denergy}(left). See the caption of \fig{hydro1} for more details. The grey vertical lines indicate the end of the linear regime. 
	}
	\vspace{4mm}
\end{figure*}

In \figs{hydro1}, \ref{hydro2} and \ref{hydro3} we compare the longitudinal and transverse pressures 
$P_L$ and $P_T$ that we read off from the simulation on the gravity side with the second-order hydrodynamic pressures $P_L^{\mt{hyd}}, P_T^{\mt{hyd}}$. 
To obtain the latter we read off the energy density and the fluid velocity from gravity and we apply the constitutive relations \eqq{stresstensor} with 
$\Pi_{(2)}^{\mu\nu}$ given by \eqq{omitted} omitting the contributions of 
$\tilde{\mathcal{O}}_7^{\mu\nu}$ and  $\tilde{\mathcal{S}}_4$. 
In \fig{hydro1} we see an excellent agreement  with the exact pressures in the final, phase-separated configuration. 
Two aspects of this agreement are particular remarkable. First, the interface between the high- and the low-energy domains is well reproduced. Second, the tiny hydrodynamic longitudinal pressure, which is of order $10^{-6}\Lambda^4$, results from a huge cancellation between the equilibrium terms and the second-order gradient corrections, both of which are several orders of magnitude larger in the high-energy region.  Presumably, the small differences between $P_L^{\mt{hyd}}$ and the exact longitudinal pressure may be reduced by increasing the precision in the determination of the second-order transport coefficients.

Note that the configuration in  \fig{hydro1} cannot possibly be described by first-order hydrodynamics. The reason is that all first-order terms are  linear in the velocity, see \eqq{linear}. Since this vanishes in the final configuration, in this case first-order hydrodynamics reduces to ideal hydrodynamics. In turn, this implies that the profile of the longitudinal pressure follows that of the energy density through point-wise application of the equation of state, as is clear in \fig{hydro1}, in contradiction with the conservation of the stress-energy tensor, which for this case implies that $P_L$ must be constant. In contrast, in the static limit the constitutive relations for the pressures at second order become 
\bse
\label{reduce}
\bal
\label{ape2}
P_L^{\mt{hyd}} &= P_{\mt{eq}} (\E) + c_L(\E) (\partial_z \E)^2 
+ f_L(\E)(\partial_z^2 \E) \,, 
\\[2mm]
P_T^{\mt{hyd}} &= P_{\mt{eq}} (\E) + c_T(\E) (\partial_z \E)^2 
+ f_T(\E)(\partial_z^2 \E) \,.
\end{align}
\ese
These expressions motivated the definition of the  $c_L, c_T, f_L, f_T$ coefficients in \cite{Attems:2017ezz}. We conclude that it is  the contribution of the second-order terms with purely spatial gradients that brings about the agreement between the exact pressures and the hydrodynamic pressures in the phase-separated configuration. 

In fact, second-order hydrodynamics  describes well not just the final spatial profile but the entire spacetime evolution of the system. This is illustrated in  \figs{hydro2} and \ref{hydro3} which show, respectively, the spatial dependence at an intermediate time of the evolution, $t \Lambda=2000$, and the time dependence of the entire evolution at a particular point $z \Lambda= 21$. 

In \figs{hydro1}, \ref{hydro2} and \ref{hydro3} we have also plotted the ideal (equilibrium) pressure, as well as the hydrodynamic pressures obtained by including only the first-order viscous corrections. These agree rather well with one another almost everywhere but fail to describe the exact pressures. This shows that the first-order terms are suppressed not just in the final, phase-separated configuration but all along the evolution of the system, and also that the  second-order terms with purely spatial gradients are as large as the ideal terms. 

The purely spatial formulation of hydrodynamics is an acausal theory for which the initial-value problem is not well posed. For second-order hydrodynamics, a cure that is vastly used in hydrodynamic codes consists of  using the first-order equations of motion to exchange the terms with second-order purely spatial derivatives in the local rest frame for terms with one time and one spatial derivative (see \cite{Romatschke:2017ejr} for a review). This results in what we will call a M\"uller-Israel-Stewart-type (MIS) formulation. 
We emphasize that, strictly speaking, what is known as the MIS formulation is the phenomenological approach introduced in \cite{Muller:1967zza,Israel:1979wp,Israel:1976tn}, which is not second-order accurate. Building on it, different second-order-accurate formulations have been constructed \cite{Baier:2007ix,Romatschke:2009kr,Denicol:2012cn}, to which we will collectively refer as MIS-type formulations. The key point is that, while they differ from MIS and they may also differ from one another in certain details, all these formulations share the common property that, as a first step   to make the initial-value problem well posed, a second-order spatial derivative is replaced by one time and one spatial derivative. Since these two sets of second-order terms differ by higher-order terms, the purely spatial formulation and the MIS-type formulations are equivalent if all gradients are small. 
Different deviations from the MIS-type formulations were first reported
near the critical point of $\mathcal{N}=2^*$~\cite{Buchel:2009hv} and
for fluids with small viscosities~\cite{Florkowski:2013lya}.
Since second-order gradients are large in our situation, one may expect that the two formulations will differ, 
as we will now verify.  
We follow \cite{Romatschke:2009kr}, which is completely general for a non-conformal neutral fluid (see  \cite{Baier:2007ix} for the conformal case).

In 3+1 dimensions the tensor and the scalar parts of $\Pi_{(2)}^{\mu\nu}$ can be expanded in a basis of eight tensor operators $\mathcal{O}_i^{\mu\nu}$  and seven scalar operators $\mathcal{S}_j$, respectively \cite{Romatschke:2009kr}. For the case of fluid motion in flat space in 1+1 dimensions only the following operators of the basis chosen in \cite{Romatschke:2009kr} do not vanish identically:\footnote{Our definition of $\mathcal{O}_1$ differs from that in \cite{Romatschke:2009kr} but agrees with the one in \cite{Baier:2007ix}.}
\bse
\label{basis2}
\bal
\mathcal{O}_1^{\mu \nu} &= -2 c_s^2 \left(\nabla^{< \mu} \nabla^{\nu > } \log s - c_s^2 \nabla^{< \mu}  \log s \nabla^{\nu > }  \log s \right)\,, \,\,\,\,\,\,
\\[2mm]
\mathcal{O}_3^{\mu\nu} &= \sigma^{<\mu}_\lambda \sigma^{\lambda \nu>}\,,
\\[2mm]
\mathcal{O}_7^{\mu\nu}&=\tilde{\mathcal{O}}_7^{\mu\nu} \,,
\\[2mm]
\mathcal{O}_8^{\mu\nu} &=  \nabla^{< \mu}  \log s \nabla^{\nu > }  \log s \,,
\\[2mm]
\mathcal{S}_1 &= \sigma_{\mu\nu} \sigma^{\mu \nu} \,,
\\[2mm]
\mathcal{S}_3&= c_s^2 \nabla_\mu \nabla^\mu \log s + \frac{c_s^4}{2} \nabla_\mu \log s \nabla^\mu \log s + \frac{1}{6} \left(\nabla \cdot u\right)^2\,,
\\[2mm]
{\mathcal{S}}_4 &=\tilde{\mathcal{S}}_4 \,,
\\[2mm]
\mathcal{S}_6&= \nabla_\mu \log s \nabla^\mu \log s \,.
\end{align}
\ese
Note that the ${\mathcal{O}}_7^{\mu\nu}$ and $\mathcal{S}_4$ operators are the same in both basis.  Moreover, in 1+1 dimensions we have 
\be
\label{madeuse}
\mathcal{O}_3^{\mu\nu}=2 \mathcal{O}_7^{\mu\nu} \sac 
\mathcal{S}_1=\frac{8}{3} \mathcal{S}_4 \,,
\ee
showing that the number of independent operators is the same as in \eqq{omitted}. The first-order equations of motion imply the following identities 
\bse
\label{replace}
\bal
&D \sigma^{\mu\nu} + \mathcal{O}_7^{\mu\nu} = \mathcal{O}_1^{\mu\nu} - \frac{1}{2} \mathcal{O}_3^{\mu\nu} - 2 \frac{d c_s^2}{d \log s} \mathcal{O}_8^{\mu\nu} \,,
\\
&D \left( \nabla \cdot u \right) =  -\frac{1}{4} \mathcal{S}_1 - \mathcal{S}_3 -\frac{1}{6} \mathcal{S}_4 +\left(\frac{3}{2} c_s^4 -  \frac{d c_s^2}{d \log s}\right) \mathcal{S}_6 \,,
\end{align}
\ese
where $D=u^\mu \nabla_\mu$ is the time derivative in the local rest frame and the equal signs here mean equality up to third- and higher-order terms. These identities may be used to replace $\mathcal{O}_1^{\mu\nu}$ and $\mathcal{S}_3$ in the expansions of $\pi_{(2)}^{\mu\nu}$ and $\Pi_{(2)}$ in favor of the left-hand sides of \eqq{replace} \cite{Romatschke:2009kr}, thus replacing terms with two spatial derivatives in the local rest frame with terms with one time and one spatial derivative. Upon these replacement the expansions read 
\bse
\label{omitted2}
\bal
\pi_{(2)}^{\mu\nu} &= \eta \tau_\pi \left(D \sigma^{\mu\nu} + \mathcal{O}_7^{\mu\nu}\right) + 
\left( 2 \lambda_1  + \eta \tau_\pi^* \right) \mathcal{O}_7^{\mu\nu}+ \lambda_4 \mathcal{O}_8^{\mu\nu} \,,
 \\
\Pi_{(2)} &= \zeta \tau_\Pi  \, D \left( \nabla \cdot u \right)  + 
\left( \frac {8}{3} \xi_1 +\xi_2 \right) \mathcal{S}_4+  \xi_4 \mathcal{S}_6 \,,
\end{align}
\ese
where we have made use of \eqq{madeuse} and we have labelled the second-order coefficients as in \cite{Romatschke:2009kr}. The coefficients in the expansion \eqq{omitted2} can be related to those in \eqq{omitted} by changing from one basis of operators to the other, with the result
\bse
\label{relation}
\bal
\eta \tau_\pi &= - \frac{ w \tilde c_1}{c_s^2} \,,\\[2mm]
\lambda_4 &= \frac{2w}{c_s^4} \left[ w \tilde c_2 + \tilde c_1 \left(1 + 2 c_s^2 -  \frac{d \log |c_s^2|}{d \log s}\right)\right]
\,,\\[2mm]
2 \lambda_1 + \eta \tau_\pi^* &= \tilde c_7 -2 \frac{ w \tilde c_1}{c_s^2}
\,,\\[2mm]
\zeta \tau_\Pi &= - \frac{2 w \tilde b_3}{c_s^2}
\,,\\[2mm]
\xi_4 &= \frac{2 w}{c_s^4}  \left[ w \tilde b_2+ \tilde b_3 
\left(1 + 2 c_s^2 -  \frac{d \log |c_s^2|}{d \log s}\right)\right]
\,,\\[2mm]
\frac{8}{3} \xi_1 +\xi_2  &= \tilde b_4 - \frac{2 w \tilde b_3}{c_s^2}  \,,
\end{align}
\ese
where 
\be
w=\E + P=Ts
\ee
 is the enthalpy.  Upon this change the fact that in our dynamical situation 
 \be
 \tilde c_7 \tilde{\mathcal{O}}_7^{\mu\nu} \simeq \tilde b_4 \,\tilde{\mathcal{S}}_4 \simeq 0
 \ee
  translates into 
\bse
\label{reprep}
\bal
\left( 2 \lambda_1  +  \eta \tau_\pi^* \right) \mathcal{O}_7^{\mu\nu} &\simeq 
2\eta \tau_\pi \, \mathcal{O}_7^{\mu\nu} \,, \\[2mm]
\left( \frac {8}{3} \xi_1 +\xi_2 \right) \mathcal{S}_4 & \simeq 
\zeta \tau_\Pi \,\mathcal{S}_4 \,.
\end{align}
\ese
Using \eqq{reprep} in \eqq{omitted2} we finally arrive at the MIS-type constitutive relations 
\bse
\label{omitted3}
\bal
\pi_{(2)}^{\mu\nu} &= \eta \tau_\pi  \left(D \sigma^{\mu\nu} + 
\mathcal{O}_7^{\mu\nu}\right) + 2 \eta \tau_\pi \,  \mathcal{O}_7^{\mu\nu}+ \lambda_4 \mathcal{O}_8^{\mu\nu} \,,
 \\
\Pi_{(2)} &=\zeta \tau_\Pi \,  D \left( \nabla \cdot u \right)  + 
\zeta \tau_\Pi \, \mathcal{S}_4+  \xi_4 \mathcal{S}_6 \,.
\end{align}
\ese

\begin{figure*}[t!!!]
	\begin{center}
		\begin{tabular}{cc}
			\includegraphics[width=.46\textwidth]{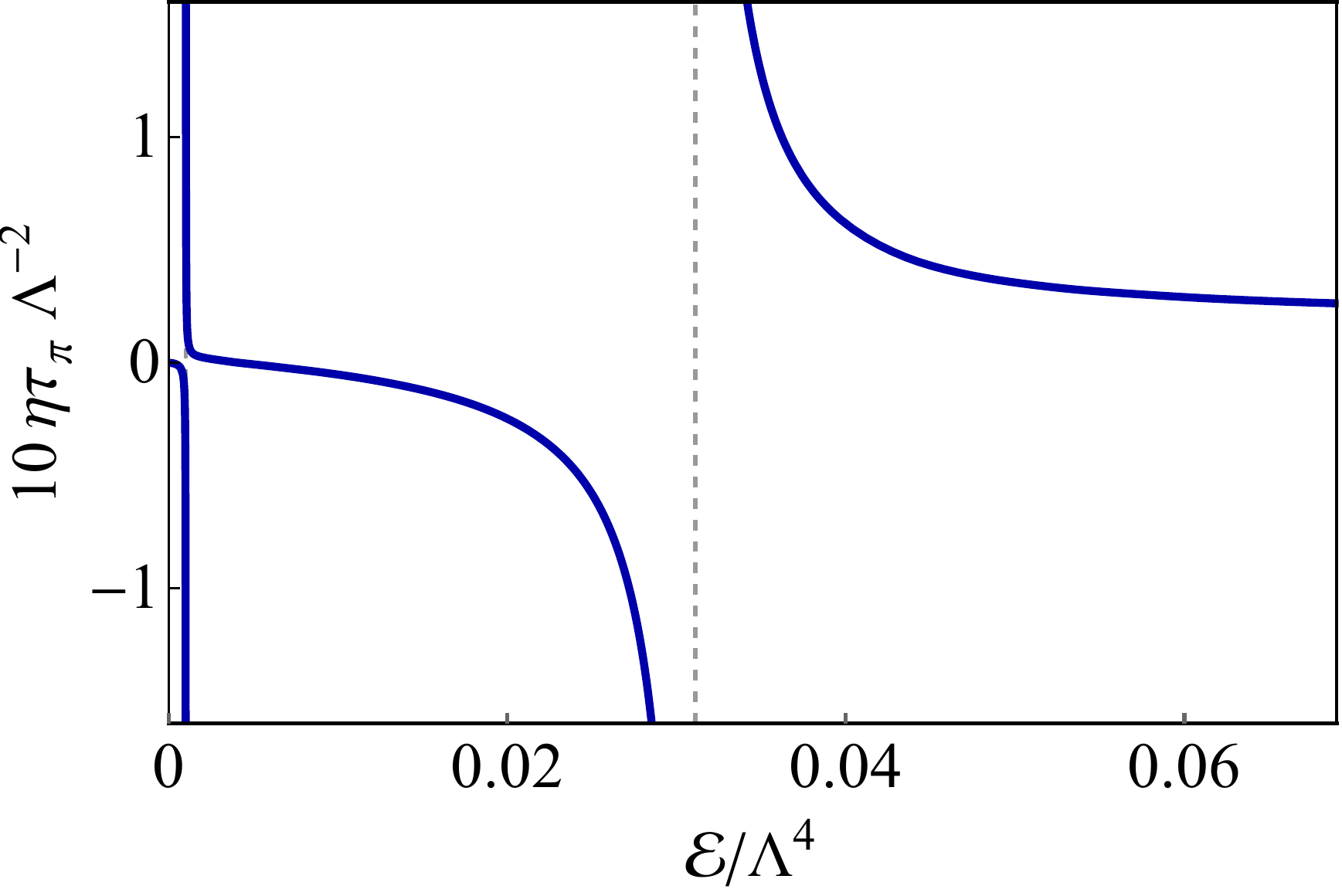} 
			\quad&\quad
			\includegraphics[width=.46\textwidth]{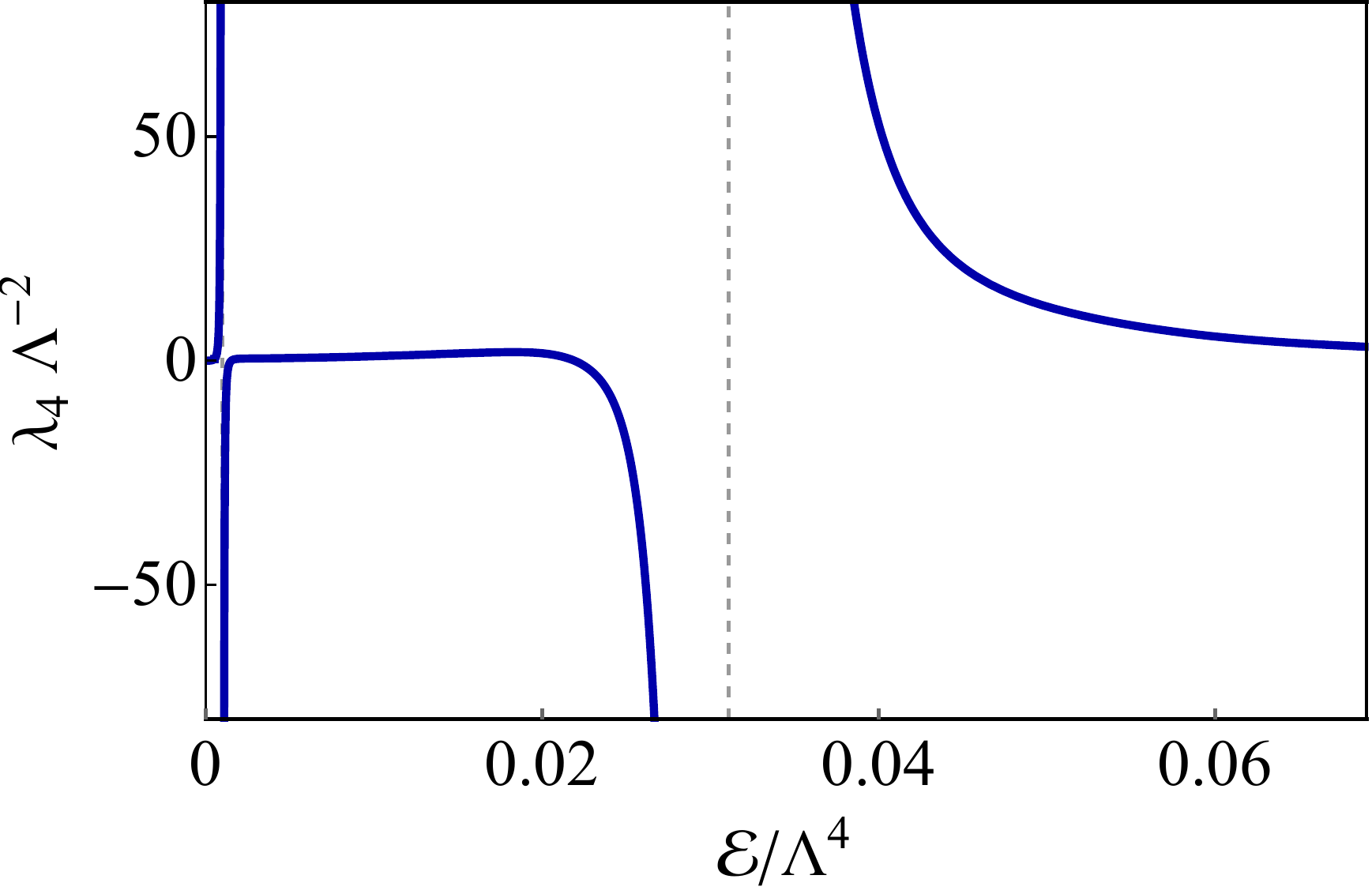} 
			\\[4mm]
			\includegraphics[width=.46\textwidth]{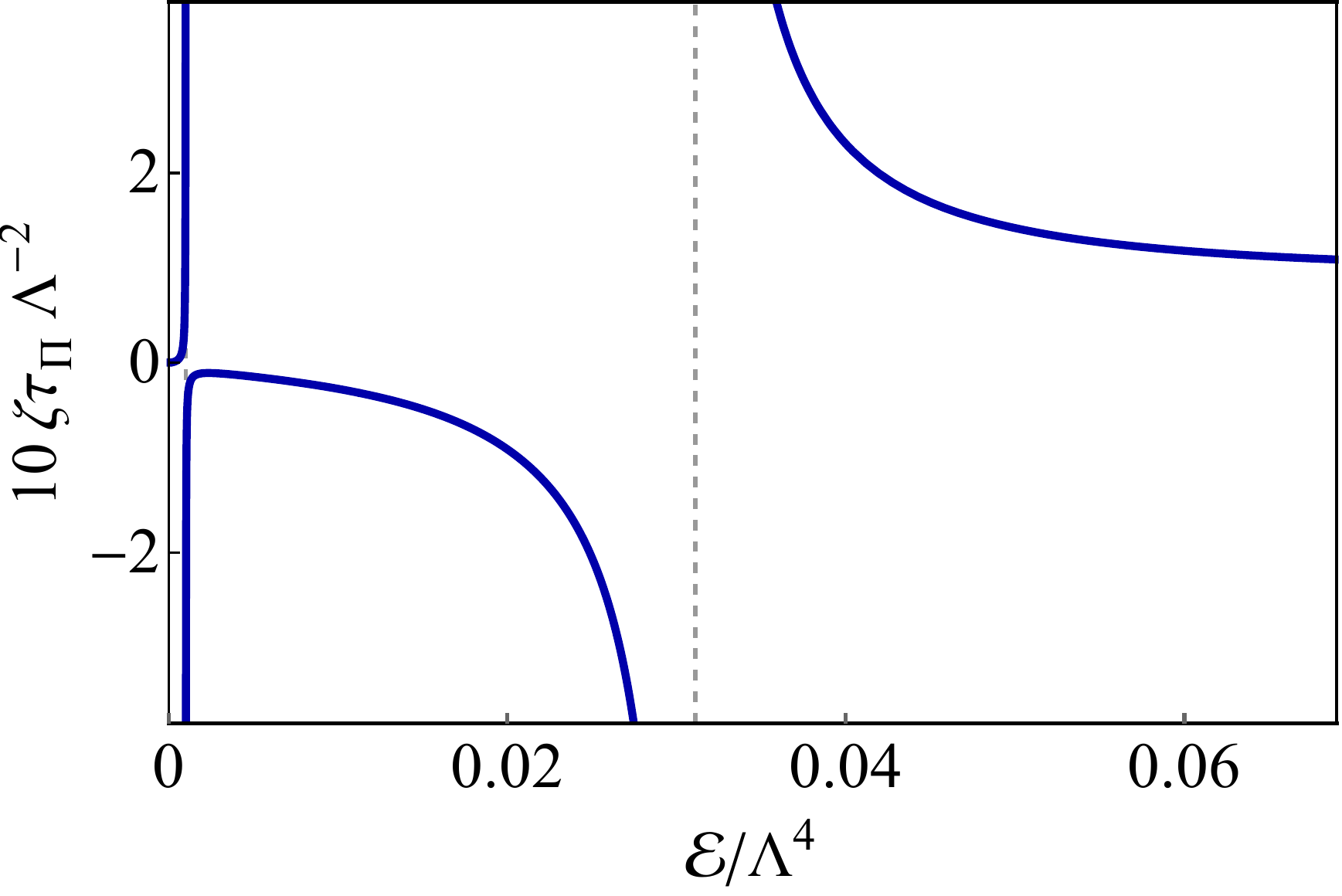} 
			\quad&\quad
			\includegraphics[width=.46\textwidth]{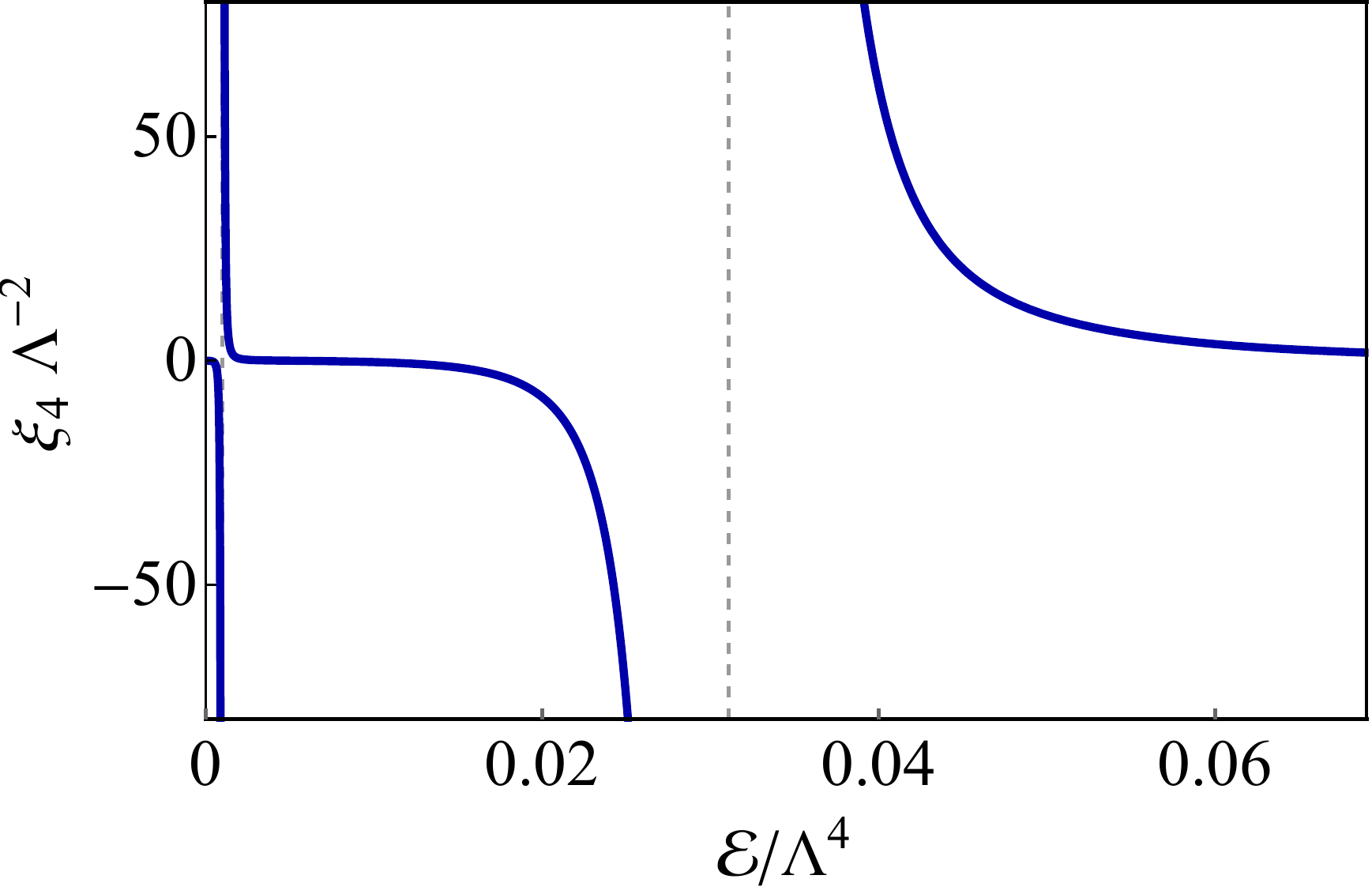}  
		\end{tabular}
	\end{center}
	\vspace{-5mm}
	\caption{\label{coefMIS} 
		\small Second-order coefficient entering the MIS-type formulation of hydrodynamics as a function of the energy density. The vertical grey lines indicate the values at which the speed of sound crosses zero.  
	}
	\vspace{4mm}
\end{figure*}

As shown in \figs{hydro1}, \ref{hydro2} and \ref{hydro3} the second-order hydrodynamic pressures  determined from these constitutive relations, $P_L^{\mt{hydMIS}}$ and  $P_T^{\mt{hydMIS}}$, fail to describe the exact pressures. It is interesting that there are two different reasons for this. The first one is the fact that the coefficients $\eta \tau_\pi, \lambda_4, \zeta \tau_\Pi, \xi_4$ entering the MIS-type formulation \eqq{omitted3} diverge at the points where the speed of sound vanishes, as can be seen in \fig{coefMIS}. 
This can be traced back to their relation \eqq{relation} with the \mbox{$\tilde c_1,  \tilde c_2 , \tilde b_2, \tilde b_3$} coefficients that enter the purely spatial formulation \eqq{omitted}. The fact that   
 the latter  are smooth and finite at the points where $c_s^2=0$ (see \fig{coef}) shows that $\eta \tau_\pi, \lambda_4, \zeta \tau_\Pi, \xi_4$ diverge as inverse powers of $c_s^2$ at those points. In turn, this results in the divergences of the MIS hydrodynamic pressures at the spacetime points in the evolution at which the energy density goes through a value such that $c_s^2 (\E)=0$, as can be seen in \figs{hydro1}, \ref{hydro2} and \ref{hydro3}. In contrast,  the hydrodynamic pressures predicted by the purely-spatial formulation are smooth and finite everywhere.

The second reason why the MIS-type formulation fails to reproduce the evolution of the system is that it does not include all the independent, spatial, second-order gradient corrections. Indeed, the operators in the purely spatial formulation \eqq{omitted}-\eqq{basis1} contain  $\nabla_z^2 \E$ and 
$(\nabla_z \E)^2$ terms. Both  types of terms are necessary in order to describe the evolution correctly. Instead, in the MIS-type formulation the 
$\nabla_z^2 \E$ terms are absent because the only operators that contain them,  $\tilde{\mathcal{O}}_1^{\mu \nu}$ and $\tilde{\mathcal{S}}_3$, have been eliminated in favor of the left-hand sides of \eqq{replace}, which contain crossed $\nabla_t \nabla_z$ derivatives. The effect of this replacement is most clearly illustrated by the late-time, static, phase-separated configurations. In these states the fluid velocity and all time derivatives vanish, so all first-order terms are zero and all second-order gradient corrections reduce to a combination of terms of the form $\partial_z^2 \E$ and $(\partial_z \E)^2$. Both of these are correctly captured by the purely-spatial formulation, as shown in \eqq{reduce}. In contrast, in the MIS-type formulation the only non-vanishing second-order operators are $\mathcal{O}_8^{\mu\nu}$ and 
$\mathcal{S}_6$ in \eqq{omitted3}, which only include $(\partial_z \E)^2$ terms. Incidentally, this also shows that the divergences in the MIS-type pressures in the phase-separated configuration at the points where $c_s^2=0$ are not due to the divergences of the relaxation times but to those of the $\lambda_4, \xi_4$ coefficients.



\section{Discussion}

We have used holography to develop a detailed physical picture of the real-time evolution of the spinodal instability of a four-dimensional, strongly-coupled, non-Abelian  gauge theory with a first-order, thermal phase transition. 

We have identified several characteristic stages in the dynamics of the system. In the first, linear stage the instability grows exponentially. In the second stage the evolution is non-linear and leads to the formation of peaks and/or domains. In the third stage these structures move towards each other with approximately constant shapes and slowly varying velocities until they merge, forming larger structures. In the fourth stage the system relaxes to equilibrium through damped oscillations which can be described in terms of linearised sound modes. For large enough boxes the final state after all mergers have taken place is a phase-separated configuration with one high- and one low-energy domain. As expected on general grounds, the interface separating them is universal in the sense that it is a property of the theory and does not depend on the initial conditions that led to the phase-separated configuration. We computed the surface tension of the interface in terms of the microscopic scale of the theory with the result \eqq{SurfaceTension}. We  noted that this interface moved with little deformation in the final relaxation stage towards the phase separated configuration. It would be interesting to understand in detail the precise conditions behind this ``rigidity'' property.

If the perturbation of the initial, homogenous state is dominated by a single mode with mode number $n\geq 2$ then the system evolves through an intermediate, almost static state with $n$ peaks or domains. Exactly static solutions with these number of structures do exist and, in principle, upon time evolution the system comes arbitrarily close to them provided that numerical noise is sufficiently suppressed. However, since these multi-structure static configurations are unstable, the evolution eventually deviates away.  We have shown that this instability precisely pushes the different peaks or domains  towards each other so that the final configuration at asymptotically late times is a phase-separated configuration if the box is large enough, or a configuration with a single peak otherwise. 

Remarkably, along the entire spacetime evolution of the system the pressures are well described by the constitutive relations of a formulation of second-order hydrodynamics in which all the gradient terms that are purely spatial in the local rest frame are included. In particular, the interface in the final phase-separated configuration is well described by this formulation. It is therefore interesting to place our system in the context of  the dynamics of fluids with boundaries. A good discussion of this topic in modern language can be found in \cite{Armas:2015ssd,Armas:2016xxg}. The general idea is that one can formulate hydrodynamics in the presence of an interface or  phase boundary. This  has an associated stress tensor that can be derivatively-expanded just like the stress tensor for the bulk of the fluid. At the zeroth, non-derivative order the time-time component of this stress tensor is the surface energy, namely the energy per unit area associated to the interface. Similarly, the diagonal space-space components give the pressure of the interface. General considerations imply that this equals minus the surface tension  that we computed in \eqq{SurfaceTension}. At higher orders in the derivative expansion the stress tensor of the interface is characterised by a set of transport coefficients called surface transport coefficients. In general, these coefficient are completely independent from those in the bulk of the fluid. Our case is an exception because the fact that the entire phase-separated configuration is well described by the hydrodynamics of the bulk fluid implies that  the surface transport coefficients could be computed in terms of the bulk transport coefficients. 

Purely spatial hydrodynamics is known to be acausal. This was not an issue for us since we did not evolve in time the hydrodynamic equations but simply verified the constitutive relations, but it is an issue in situations in which hydrodynamics is the only available description. For this reason we also investigated the validity of an MIS-type formulation, in which acausality is remedied by replacing terms with second-order spatial derivatives in the local rest frame by terms with one time and one space derivative. In the limit of small gradients this produces an equivalent formulation at long wavelengths. However, in our system the spatial gradients are large and the result is not equivalent. This is one reason why the MIS-type formulation fails to reproduce the correct pressures. The other reason is that, unlike in the purely spatial formulation, several  second-order coefficients in the MIS-type formulation, in particular some relaxation times, diverge at the points where the speed of sound vanishes, leading to a divergent prediction for the hydrodynamic pressures at those points.  

Since the conclusions in the paragraph above are important, let us state them in slightly different terms. Specifically, some readers may wonder what is the merit of using two different basis \eqq{basis1} and \eqq{basis2} given that  they are supposed to be equivalent and hence give the same result by construction. The point is that the last statement is only true provided two conditions hold: (1) that all the gradients are small, because the equivalence is only accurate up to third- and higher-order terms, and (2) that the transformation between the two basis is non-singular. In our case both conditions can be violated. The fact that the gradients are not small follows from the fact that second-order terms are as large as ideal terms, as we saw in \Sec{sectionhydro}. The failure of this condition means that the replacement \eqq{replace} is not justified, since it neglects third- and higher-order terms that are large. The discrepancy between the left- and the right-hand sides of \eqq{replace} becomes most extreme in the static, phase-separated configurations that we have considered, since in these the left-hand side vanishes identically whereas the right-hand side is generically non-zero. Condition (2) is violated at the points where $c_s^2=0$. At these points the two basis are not equivalent even if the gradients are small, because the transformation between the two basis becomes singular. This is most clearly exhibited by the fact the relations \eqq{relation} between the corresponding coefficients in each base become singular due to the inverse factors of $c_s^2$. This feature is the one that is ultimately responsible for the divergences seen in the hydrodynamic pressures in the MIS formulation.

Although our model is a specific bottom-up model our analysis suggests that the qualitative physics that we have just described may be quite universal in situations in which the gradients are large and/or the speed of sound is small. The latter property is guaranteed to hold near a critical point, so our results may have important implications for experimental searches of the QCD critical point.

It would be interesting to allow for dynamics in the $x_\perp$ directions. While some details will change, we expect that some of the qualitative lessons that we have extracted will remain true. For example, unstable modes in the initial homogeneous state will grow exponentially with the same growth rates as in our analysis. The details of the reshaping period will of course be more complicated as they will involve a shape adjustment in several dimensions. However, they will presumably lead to the formation of structures that will move with almost constant shapes and slowly varying velocities, since these features only depend on the large ratio between $\Ehigh$ and $\Elow$.  Finally, we expect that, in large enough boxes, the only stable configurations with average energy densities in the unstable region will be phase-separated configurations with a single high-energy domain.

It would be interesting to extend our analysis in several other directions, including a more systematic study of domain collisions or the inclusion of a conserved U(1) charge.

\section*{Acknowledgements}
We thank Tomas Andrade, Jean-Paul Blaizot, Oscar Dias, Roberto Emparan, Jaume Garriga, Thanasis  Giannakopoulos, Volker Koch, Guy Moore, Daniele Musso, Mikel Sanchez, Jorge Santos, Mikhail Stephanov and Benson Way for discussions. We thank the MareNostrum supercomputer at the BSC (project no.~UB65) and the FinisTerra II at CESGA for significant computational resources. 
MA would like to thank the Helsinki Institute of Physics for their hospitality during late stages of this work. MA acknowledges support through H2020-MSCA-IF-2014 FastTh 658574, FPA2017-83814-P, Xunta de Galicia (Conseller\'ia de Educaci\'on), MDM-2016-0692 and HPC17YDH55 by HPC-EUROPA3 (INFRAIA-2016-1-730897). MZ acknowledges support through the FCT (Portugal) IF programme, IF/00729/2015. We are also supported by grants FPA2016-76005-C2-1-P, FPA2016-76005-C2-2-P, 2014-SGR-104, 2014-SGR-1474, SGR-2017-754 and MDM-2014-0369.

\appendix

\section{Dispersion relation of sound modes}
\label{appendixA}
Let us assume that we start with a non-conformal fluid in equilibrium, such that its stress tensor is constant and characterised  by some energy density $\E_0$ and pressure $P(\E_0)$. We now excite the fluid with small fluctuations in the sound channel, such that we can express the fluctuations in terms of $\delta \E $ and $v^i$, both small. Since the fluid is isotropic we can choose both the momentum $k$ and the velocity $v$ (in the sound channel) to lie along  the $z$-direction. We can define the velocity and energy density fluctuations via the stress tensor components 
\be
\delta  T^{0z} = w_0 v^{z} \sac \delta T^{00} = \delta \E  \, ,
\ee
where 
\be
w_0=\E_0+ P(\E_0) = Ts
\ee
is the enthalpy.

Using the general expression for the second-order stress tensor, the fluctuations of all other (relevant) stress tensor components is given by
\be
\delta T^{zz}= c^2_s \, \delta \E - w_0 \, \Gamma\,  \del_z v^z + 
f_L \, \del_z^2 \, \delta \E \, ,
 \ee
 where we recall that $\Gamma$ is the sound attenuation constant \eqq{ate} and $f_L$ is the transport coefficient in \eqq{ape1} and \eqq{ape2}.

After performing a space Fourier transform we define
\be
 \delta \E_k (t) = \int dz \, e^{i k z} \, \delta \E (t,z ) \sac
 v_k (t) = \int dz \, e^{i k z} \, v(t,z ) \,.
\ee
In terms of these Fourier components the two conservation equations become
\bea
i \,k \,w_0 \, v_k(t)  + \E_k' (t)&=&0  \, ,
\\
 i \, k\left(c_s^2 - k^2 f_L\right) \E_k (t) + k\, w_0 \, \Gamma \, v_k(t) + 
 w_0 \, v_k'(t) &=&0  \,.
\eea
Differentiating the first equation and substituting in the second one we obtain a wave-like equation for the energy fluctuations: 
\be
\delta \E''_k (t) + \Gamma \, k^2 \,\delta \E'_k (t) + k^2 (c_s^2 - k^2 f_L)  \delta \E_k (t)=0  \,.
\ee
This equation leads to a non-trivial dispersion relation. When $c^2_s<0$ there is an unstable mode with frequency given by 
\be
\label{omegaunstable}
\omega_+=+ \sqrt{ c_s^2 k^2 - k^4 f_L - k^4 \frac{1}{4} \Gamma^2 } - \frac{i}{2} \Gamma k^2  \,.
\ee 
In the small-frequency limit this yields
\be
\omega_+ =  i\left |c_s \right | k - \frac{i}{2} \Gamma k^2 + i k^3 \frac{4 f_L + \Gamma^2 }{8 \left |c_s \right | } \, .
\ee
Note also that the intercept of the unstable mode with zero, namely the edge of the unstable dome, becomes
\be
k= \frac{c_s}{\sqrt{f_L}} \,.
\ee
This implies  that $f_L$ must be negative in order for the second-order hydrodynamic dispersion relation to become stable at some $k$. In addition to the unstable mode, there is a second, stable mode with frequency 
\be
\label{omegastable}
\omega_-=-\sqrt{ c_s^2 k^2 - k^4 f_L - k^4 \frac{1}{4} \Gamma^2 } - \frac{i}{2} \Gamma k^2  \, .
\ee

Equations (\ref{omegaunstable}) and (\ref{omegastable}) correspond to (\ref{disp}) in the main text.


\appendix

\bibliographystyle{JHEP}

\end{document}